\documentclass[12pt]{article}

\pdfoutput=1

\usepackage[utf8x]{inputenc}      
\usepackage{amsmath,amssymb,mathtools}
\usepackage[T1]{fontenc}          
\usepackage{booktabs,tabularx, multirow}
\usepackage{graphicx,subfig}
\usepackage{xspace}
\usepackage[usenames]{xcolor}\definecolor{fscolor}{RGB}{44,118,255}
\usepackage{geometry}
\usepackage{todonotes}
\usepackage{listings}
\usepackage[absolute]{textpos}
\usepackage[many]{tcolorbox}
\usepackage{xparse}
\usepackage[font=small,labelfont=bf,format=plain,margin=0.05\textwidth]{caption}
\usepackage{bbm}
\usepackage{tabularx}
\usepackage{cite}
\usepackage{soul}
\usepackage{hyperref}
\usepackage{cleveref}
\usepackage{lineno}
\usepackage{lmodern}
\usepackage{microtype}

\crefname{figure}{Fig.}{Figs.}
\crefname{equation}{Eq.}{Eqs.}
\crefname{section}{Sec.}{Secs.}
\crefname{appendix}{App.}{Apps.}
\crefname{table}{Tab.}{Tabs.}

\allowdisplaybreaks

\evensidemargin \oddsidemargin
\newcommand{\cp}{\ensuremath{{\cal CP}}\xspace}
\newcommand{\SM}{{\text{SM}}}

\newcommand{\tev}{\,\, \mathrm{TeV}}
\newcommand{\gev}{\,\, \mathrm{GeV}}

\newcommand{\fb}{\,\, \mathrm{fb}}

\definecolor{Darkgreen}{rgb}{0.,.7,0.2}
\definecolor{Darkblue}{rgb}{0.,.2,0.7}

\newcommand{\kgamma}{\ensuremath{\kappa_\gamma}\xspace}
\newcommand{\kg}{\ensuremath{\kappa_g}\xspace}
\newcommand{\kggzh}{\ensuremath{\kappa_{ggZH}}\xspace}

\newcommand{\cv}{\ensuremath{c_V}\xspace}
\newcommand{\ct}{\ensuremath{c_t}\xspace}
\newcommand{\cttilde}{\ensuremath{\tilde c_{t}}\xspace}
\newcommand{\cgamma}{\ensuremath{c_\gamma}\xspace}
\newcommand{\cgammatilde}{\ensuremath{\tilde c_{\gamma}}\xspace}
\newcommand{\cg}{\ensuremath{c_g}\xspace}
\newcommand{\cgtilde}{\ensuremath{\tilde c_{g}}\xspace}

\newcommand{\ttHmark}{$^{\color{magenta}\dagger}$}

\newcommand{\order}[1]{\ensuremath{{\cal O}(#1)}}

\begin{document}

\thispagestyle{empty}
\def\thefootnote{\fnsymbol{footnote}}

\begin{flushright}
\texttt{BONN-TH-2020-04, DESY 20-102, IFT--UAM/CSIC--20-090}
\end{flushright}
\vspace{2em}
\begin{center}
{\Large\bf Indirect \boldmath{\cp} probes of the Higgs--top-quark interaction:\\[1em] current LHC constraints and future opportunities}
\\
\vspace{3em}
{
Henning Bahl$^{1}$\footnotetext[0]{henning.bahl@desy.de, bechtle@physik.uni-bonn.de, sven.heinemeyer@cern.ch, judith.katzy@desy.de, klingl@physik.uni-bonn.de, krisztian.peters@desy.de, matthias.saimpert@cern.ch, tim.stefaniak@desy.de, georg.weiglein@desy.de},
Philip Bechtle$^{2}$,
Sven Heinemeyer$^{3,4,5}$,
Judith Katzy$^{1}$,
Tobias Klingl$^{2}$,\\[1em]
Krisztian Peters$^{1}$,
Matthias Saimpert$^{6}$,
Tim Stefaniak$^{1}$,
Georg Weiglein$^{1}$
}\\[2em]
{\sl $^1$Deutsches Elektronen-Synchrotron DESY, Notkestra{\ss}e 85, 22607 Hamburg, Germany}\\[0.2em]
{\sl $^2$Physikalisches Institut, Universit\"at Bonn, Nu{\ss}allee 12, 53115 Bonn, Germany}\\[0.2em]
{\sl$^3$Instituto de F\'isica Te\'orica, (UAM/CSIC), Universidad Aut\'onoma de Madrid,\\[0.2em]
Cantoblanco, E-28049 Madrid, Spain}\\[0.2em]
{\sl$^4$Campus of International Excellence UAM+CSIC, Cantoblanco, E-28049, Madrid, Spain}\\[0.2em]
{\sl$^5$Instituto de F\'isica de Cantabria (CSIC-UC), E-39005 Santander, Spain}\\[0.2em]
{\sl$^6$CERN, Geneva, Switzerland}

\def\thefootnote{\arabic{footnote}}
\setcounter{page}{0}
\setcounter{footnote}{0}
\end{center}
\vspace{2ex}
\begin{abstract}
{}

The \cp\ structure of the Higgs boson in its coupling to the particles of the Standard Model is amongst the most important Higgs boson properties which have not yet been constrained with high precision. In this study, all relevant inclusive and differential Higgs boson measurements from the ATLAS and CMS experiments are used to constrain the \cp-nature of the top-Yukawa interaction. The model dependence of the constraints is studied by successively allowing for new physics contributions to the couplings of the Higgs boson to massive vector bosons, to photons, and to gluons. In the most general case, we find that the current data still permits a significant \cp-odd component in the top-Yukawa coupling. Furthermore, we explore the prospects to further constrain the \cp\ properties of this coupling with future LHC data by determining $tH$ production rates independently from possible accompanying variations of the $t\bar{t}H$ rate. This is achieved via a careful selection of discriminating observables. At the HL-LHC, we find that evidence for $tH$ production at the Standard Model rate can be achieved in the Higgs to diphoton decay channel alone.

\end{abstract}

\newpage
\def\thefootnote{\arabic{footnote}}



\section{Introduction}
\label{sec:intro}

In 2012 the ATLAS and CMS collaborations have discovered a new particle that -- within current theoretical and experimental uncertainties -- is consistent with the predictions of a Standard-Model~(SM) Higgs boson at a mass of~$\sim 125 \gev$~\cite{Aad:2012tfa,Chatrchyan:2012xdj}. No conclusive signs of physics beyond the~SM have been found so far at the LHC. However, the measurements of Higgs-boson couplings, which are known experimentally to a precision of roughly $\sim 20\%$, leave room for Beyond Standard-Model (BSM) interpretations. Consequently, one of the main tasks of the LHC Run~3 and the high-luminosity LHC (HL-LHC) will be to determine the Higgs-boson coupling structures and quantum numbers with highest precision.

This experimental program has a direct link to cosmology. One of the most important questions connecting the two fields is related to the fact that the Cabbibo-Kobayashi-Maskawa (CKM) matrix, the only source of \cp violation in the SM, cannot explain the observed baryon asymmetry of the universe (BAU), as the SM prediction is off by many orders of magnitude~\cite{Gavela:1993ts,Huet:1994jb}. Consequently, additional sources of \cp violation must exist in nature. The main idea of this work is to investigate to which extent \cp violation beyond the CKM matrix can be present in the interactions of the detected Higgs boson. Specifically we investigate here the interaction between the Higgs boson and the top quark.

While the hypothesis that the Higgs boson at 125~GeV is a pure \cp-odd state was ruled out based on LHC Run~1 data~\cite{Khachatryan:2014kca,Aad:2015mxa}, only rather weak bounds exist on a possible admixture between a \cp-even and a \cp-odd component. The analyses so far were mainly based on observables involving the coupling of the observed Higgs boson, $H$, to two gauge bosons, $HVV$, where $V = W, Z$, in particular $H \to ZZ^* \to 4 \ell$, $H \to WW^* \to \ell\nu\ell\nu$, and Higgs production in weak vector boson fusion (VBF). Since in many BSM models only a small loop-induced coupling of a \cp-odd component to gauge bosons, $H^{\cp-{\rm odd}}VV$, is generated, the effects of the \cp-odd component are heavily suppressed compared to the tree-level contribution of $H^{\cp-{\rm even}}VV$, even if the \cp-odd component itself is large. Consequently, the couplings of the observed Higgs boson to fermions (in particular to the top quark) are crucial for investigating the \cp-nature of the observed state, since a \cp-odd component in the $Hf \bar f$ coupling may be unsuppressed and could be of similar magnitude as the \cp-even component. In addition to being a spectacular discovery by itself, the establishment of a non-zero \cp-odd component of the top-Yukawa coupling would also be a strong hint for an extended BSM Higgs sector.

Our investigations make use of the so-called ``Higgs characterization model'', a framework based on an effective field theory (EFT) approach. This framework allows one to take into account \cp-violating effects in the couplings and to perform studies in a consistent, systematic and accurate way, see e.g. Ref.~\cite{Artoisenet:2013puc}. The top-Yukawa coupling between the Higgs boson and the top quark in this approach is parametrized in terms of rescaling parameters for the \cp-even (SM-like) coupling and a \cp-odd (BSM) coupling. Alternatively, these can be expressed in terms of a magnitude and a \cp-violating phase. Such types of approaches are the basis of most published data analyses sensitive to \cp-violation and \cp-admixtures in the Higgs sector, briefly reviewed below.

\cp violation in the interactions between the Higgs boson and the top quark can be probed experimentally within this framework using two different strategies: In a \emph{direct} approach, a pure \cp-odd observable is constructed, typically from angular distributions of Higgs boson production~\cite{Aad:2020mnm,Sirunyan:2019nbs} or decay~\cite{Aad:2015mxa,Sirunyan:2017tqd,Sirunyan:2019twz}. Such a measurement can be used to constrain all models where both the \cp-odd and the \cp-even components of the Higgs boson couple with significant strength to the particle from which the \cp-odd observable is measured. For the experimental establishment of \cp\ violation in the Higgs boson sector, the measurement of a \cp-odd observable with a non-zero result would be crucial. For the Higgs--top-quark coupling, no experimental analysis of a pure \cp-odd observable~\cite{Faroughy:2019ird,Bortolato:2020zcg} exists yet. In an \emph{indirect} approach, the model parameters are fitted to \cp-even observables such as decay rates and kinematic distributions (e.g.\ $p_T$ spectra), and constraints on the \cp\ nature of the Higgs boson are derived indirectly from the constraints on the model parameters. This latter approach is the one followed in this paper. While the indirect approach is a very powerful test of possible deviations from the SM, in case a significant deviation is found it is not guaranteed that its origin can uniquely be associated with the presence of \cp-violating effects. Thus, in order to determine the \cp\ nature of the Higgs sector, the two approaches are complementary.

The properties of the Higgs boson under the pure \cp-even hypothesis and its coupling to top quarks are studied by ATLAS and CMS in great detail. An overview of early LHC Run~2 results on the constraints on the \cp-even top-Yukawa coupling $y_t$ can be found in Refs.~\cite{Aaboud:2018urx,Sirunyan:2018hoz}. Early fits to a possible \cp admixture to the observed Higgs state have been performed using Run~1 and partially early Run~2 data, either investigating all Higgs-boson couplings~\cite{Freitas:2012kw,Djouadi:2013qya}, or focusing on the Higgs--top-quark interaction~\cite{Boudjema:2015nda,Hou:2018uvr}. These analyses could set only very weak bounds on possible \cp violation in the Higgs-boson sector. Projections for future bounds on possible \cp-admixtures of the discovered Higgs-boson based on $gg \to H + 2$~jets data were obtained in Ref.~\cite{Anderson:2013afp,Dolan:2014upa,Gritsan:2020pib}, focusing on current and future Run~3 data. Experimental bounds on the top-Yukawa coupling under the \cp-even coupling assumption have been obtained more recently using an integrated luminosity of $36 \fb^{-1}$ in the $t \bar t\, t \bar t$ channel~\cite{Sirunyan:2017roi} and $t \bar t$ kinematic distributions~\cite{Sirunyan:2019nlw}.

Up to now, the most stringent experimental bounds on the \cp\ nature of the top-Yukawa coupling have been obtained by dedicated CMS~\cite{Sirunyan:2020sum} and ATLAS~\cite{Aad:2020ivc} analyses using full Run~2 data. Both analyses combine the direct and the indirect approach by fitting the rate of the $t\bar t H$, $tH$ and $tWH$ processes to data in the Higgs to di-photon decay channel in certain kinematic configurations. Under the assumptions of the considered model, a pure \cp-odd top-Yukawa coupling is excluded by $3.2\,\sigma$ (CMS) and $3.9\,\sigma$ (ATLAS), and, assuming a pure \cp-even coupling, the $t\bar t H$ signal is observed at the $5\,\sigma$ level. Combining the direct and the indirect approaches leads to the best sensitivity for the targeted model. Presenting the results of such a combined study in a fully model-independent format, however, is very challenging and would entail the public release of the likelihood as a function of all relevant parameters (including the relevant Higgs coupling scale factors). As this has not been done for the present analyses~\cite{Sirunyan:2020sum, Aad:2020ivc} we could not include these results as input for the \cp\ study carried out in the present paper, which is performed for several ``Higgs characterization models'' of different complexities. We will discuss the results of Refs.~\cite{Sirunyan:2020sum, Aad:2020ivc} further in \cref{sec:discussion}.

Another type of observables that are sensitive to \cp-violating interactions are electric dipole moment (EDM) measurements. These were used in Ref.~\cite{Brod:2013cka} to set upper bounds on the \cp-odd Higgs--top-quark coupling of \order{0.5}. The most recent EDM measurements (see e.g.~Refs.~\cite{Andreev:2018ayy,Abel:2020gbr}) which were not taken into account in Ref.~\cite{Brod:2013cka} may lead to stronger constraints (see Refs.~\cite{deVries:2017ncy,deVries:2018tgs,Fuchs:2020uoc}, in which the complementarity between collider and BAU constraints is also discussed). In the analysis of these constraints, however, it is assumed that the Higgs couples only to third-generation fermions.  While assuming SM values also for the first- and second-generation fermions leads to substantially stricter bounds, non-zero \cp-odd components in the couplings of the Higgs boson to leptons or first- and second-generation quarks, which are hardly experimentally accessible so far, could actually loosen the constraints on the \cp-odd component of the top-Yukawa interaction (see Refs.~\cite{Chien:2015xha,Cirigliano:2016nyn,Panico:2018hal}).

Besides the experimental publications, \cp violation in the Higgs--top-quark coupling has also been analysed in various phenomenological works focussing on $t\bar t H$, $tWH$ and $tH$ production. Tree-level analyses using Run~1 data can be found in Ref.~\cite{Ellis:2013yxa}. Corresponding studies at the next-to-leading (NLO) QCD level have been performed for $t\bar t H$~\cite{Demartin:2014fia}, $tH$~\cite{Demartin:2015uha} and $tWH$ production~\cite{Demartin:2016axk}. In these studies the different dependence of $t\bar tH$, $tWH$ and $tH$ production on a \cp-admixture of the Higgs boson was investigated. A study of the Higgs--top-quark interaction including partial Run~2 data has been presented in Ref.~\cite{Kobakhidze:2016mfx}. In Ref.~\cite{Cao:2019ygh}, future constraints from $t \bar t H$, $tH$ and $t \bar t\,t \bar t$ total rate measurements at the LHC Run~3 and the HL-LHC were investigated. Assuming that no deviations from the SM will be detected, prospective bounds on the \cp-admixture of the Higgs boson were derived. \cp-violating asymmetries involving $t \bar t H$ production were analyzed in Refs.~\cite{Buckley:2015vsa,Gritsan:2016hjl,Azevedo:2017qiz,Barger:2018tqn,Goncalves:2018agy}. These analyses correlated the top-quark spin and the four-momenta to obtain information about a possible \cp-admixture of the Higgs that is involved in the $t \bar t H$ vertex. However, no results based on the available data were presented.

In this work we will use all relevant inclusive and differential Higgs boson rate measurements (based on \cp-even observables) that are presently available to derive bounds on a possible \cp-odd coupling (or the aforementioned \cp-violating phase) using the \emph{indirect} approach discussed above. Based on the results of our analysis, we develop a possible strategy to measure the single top quark associated Higgs production mode with upcoming LHC data independently from possible \cp-structure-related effects in the $t \bar t H$ rate. This measurement would significantly enhance the sensitivity of the fit to the \cp\ nature of the top-Yukawa coupling in the future.

Our paper is organized as follows. The various LHC processes which depend on the top-Yukawa coupling at leading-order (LO) are briefly reviewed in \cref{sec:topYuk}. The effective model description we are using, i.e.\ the employed ``Higgs characterization model'', is defined in \cref{sec:model}. These definitions are applied to the various Higgs-boson production and decay modes in \cref{sec:xs_fits}, which also demonstrates the numerical dependences of the various Higgs-boson production modes on the \cp-even and \cp-odd couplings. In \cref{sec:results}, we define four parametrizations with increasing complexity, ranging from two free parameters in the Higgs--top-quark coupling to five free parameters, including also additional Higgs-boson production modes and the relevant Higgs-boson decay modes. The main results, using all relevant inclusive and differential Higgs boson rate measurements that are presently available, are obtained for these parametrizations. In \cref{sec:prospects}, we present a possible strategy to measure separately $tH$ and $t\bar tH+tWH$ production with more data and discuss the additional constraints on the \cp-structure that one would obtain from this new measurement. Our conclusions can be found in \cref{sec:conclusions}.


\section{Higgs--top-quark interaction at hadron colliders}
\label{sec:topYuk}

At hadron colliders the Higgs top-Yukawa coupling appears in multiple processes. For the present discussion we restrict ourselves to those processes in which the top-Yukawa coupling appears at the leading-order (LO). All of these processes are sensitive to the properties of the top-Yukawa coupling via the total rate and potentially via kinematic observables.

\begin{figure}\centering
\begin{minipage}[c]{.25\textwidth}
\includegraphics[width=\textwidth]{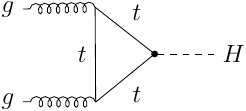}
\end{minipage}
\hspace{1cm}
\begin{minipage}[c]{.25\textwidth}
\includegraphics[width=\textwidth]{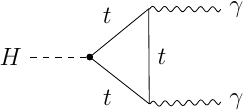}
\end{minipage}
\caption{Exemplary Feynman diagrams for $gg\rightarrow H$ and $H\rightarrow \gamma\gamma$.}
\label{fig:feynman_ggH_Hgamgamma}
\end{figure}

First, we discuss loop-induced processes which are mediated among others by a top-quark loop. One example is gluon fusion, the dominant LHC Higgs production mechanism for a SM-like Higgs boson. Its dominant leading-order contribution is mediated by a top-quark loop, shown in the left panel of \cref{fig:feynman_ggH_Hgamgamma}. One phenomenologically relevant Higgs decay mode -- the decay into two photons -- is also mediated by a top-quark loop, shown in the right panel of \cref{fig:feynman_ggH_Hgamgamma}, besides a dominantly contributing  $W$-boson loop (for the case of a SM-like Higgs boson) and other subdominant contributions.

\begin{figure}\centering
\begin{minipage}[c]{.17\textwidth}
\includegraphics[width=\textwidth]{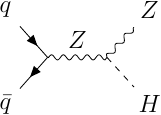}
\end{minipage}
\hspace{1cm}
\begin{minipage}[c]{.24\textwidth}
\includegraphics[width=\textwidth]{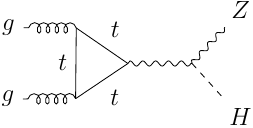}
\end{minipage}
\hspace{1cm}
\begin{minipage}[c]{.2\textwidth}
\includegraphics[width=\textwidth]{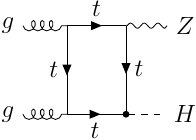}
\end{minipage}
\caption{Exemplary Feynman diagrams for $qq\rightarrow ZH$ and $gg\rightarrow ZH$.}
\label{fig:feynman_qqZH_ggZH}
\end{figure}

Another production mechanism sensitive to the top-Yukawa coupling is $Z$-boson associated production. While the dominant contribution to $ZH$ production involving a quark and an anti-quark in the initial state (see left diagram of \cref{fig:feynman_qqZH_ggZH}) does not depend on the top-Yukawa coupling at leading order, the subdominant gluon-induced channel has two LO contributions of which one involves the top-Yukawa coupling (see exemplary diagrams in the middle and right panel of \cref{fig:feynman_qqZH_ggZH}). In addition to the total rate, also the $p_T$-shape of the Higgs boson produced via $gg\rightarrow ZH$ is sensitive to the \cp-nature of the top-Yukawa coupling (alongside other kinematic distributions, see e.g.~\cite{Gritsan:2020pib}). This distribution can be studied in the simplified template cross-section (STXS) framework~\cite{deFlorian:2016spz}. In contrast, the $p_T$-shape of the Higgs boson produced via $gg\rightarrow H$ is not sensitive to the \cp-nature of the top-Yukawa coupling (see e.g.~Ref.~\cite{Demartin:2014fia}). If the Higgs is produced in association with two jets, the azimuthal correlations between the jets, however, offer sensitivity to the \cp-nature~\cite{Demartin:2014fia}.

\begin{figure}\centering
\begin{minipage}[c]{.2\textwidth}
\includegraphics[width=\textwidth]{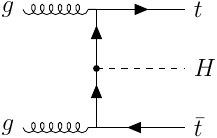}
\end{minipage}
\hspace{1cm}
\begin{minipage}[c]{.25\textwidth}
\includegraphics[width=\textwidth]{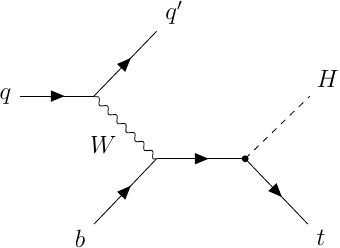}
\end{minipage}
\hspace{1cm}
\begin{minipage}[c]{.18\textwidth}
\includegraphics[width=\textwidth]{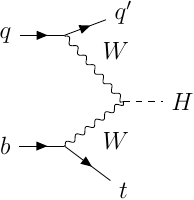}
\end{minipage}
\hspace{1cm}
\begin{minipage}[c]{.2\textwidth}
\includegraphics[width=\textwidth]{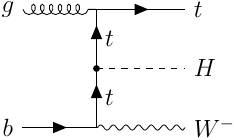}
\end{minipage}
\hspace{1cm}
\begin{minipage}[c]{.2\textwidth}
\includegraphics[width=\textwidth]{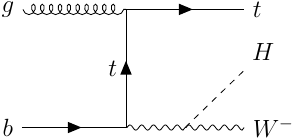}
\end{minipage}
\caption{Exemplary Feynman diagrams for $t\bar t H$, $tH$ and $tWH$ production.}
\label{fig:feynman_ttH_tH_tWH}
\end{figure}

So far we discussed only loop-induced  processes, where the top-Yukawa coupling dependence enters due to a virtual top quark appearing among other particles in the loop. At the LHC, however, we can also study channels which are sensitive to the top-Yukawa coupling already at the tree-level. These are Higgs production in association with one or two top-quarks. While all contributions to $t\bar tH$ production are proportional to the top-Yukawa coupling (see exemplary diagram in the upper left panel of \cref{fig:feynman_ttH_tH_tWH}), $tH$ production receives contributions proportional to the top-Yukawa coupling (see upper middle diagram of \cref{fig:feynman_ttH_tH_tWH}) and proportional to the electroweak gauge couplings (see upper right diagram of \cref{fig:feynman_ttH_tH_tWH}).\footnote{In addition to the $t$-channel $tH$ contributions, shown in \cref{fig:feynman_ttH_tH_tWH}, there is also a $s$-channel contribution mediated by a $W$ boson. The $s$-channel contribution is an order of magnitude smaller than the $t$-channel contribution~\cite{Demartin:2015uha}. Therefore, we neglect it in the present study.} Similar to $tH$ production, also $tWH$ production receives contributions proportional to the top-Yukawa coupling and to the electroweak gauge couplings (see bottom diagrams of \cref{fig:feynman_ttH_tH_tWH}). Experimentally, $tWH$ is challenging to distinguish from $t\bar tH$ production. At next-to-leading order in the five-flavor scheme or at leading-order in the four-flavor scheme, $tWH$ and $t\bar tH$ production even interfere with each other (see~\cite{Demartin:2016axk} for a detailed discussion). The distributions of the Higgs transverse momentum in $tH$, $t\bar tH$ and $tWH$ production offer additional sensitivity to the \cp-nature of the top-Yukawa coupling. Measurements of these shapes are not yet possible but are expected to become feasible in the future. STXS bins for the $t\bar tH$ Higgs $p_T$-shape have been defined already~\cite{STXSnewbinning}.

In addition to the processes discussed above, also the Higgs decay mode into a photon and a $Z$ boson, four leptons, as well as four-top-quark production~\cite{Chen:2015rha,Cao:2019ygh} can be used to constrain the \cp-nature of the top-Yukawa coupling. With the current experimental precision, these processes are, however, not competitive to the processes discussed above (but may become relevant after the high-luminosity upgrade of the LHC). Therefore, we do not include them into our analysis.


\section{Effective model description}
\label{sec:model}

For our analysis, we use a model similar to the Higgs-characterization model defined in Refs.~\cite{Artoisenet:2013puc,Maltoni:2013sma,Demartin:2014fia}. The top-Yukawa part of the Lagrangian is modified with respect to the SM,
\begin{align}\label{eq:topYuk_lagrangian}
\mathcal{L}_\text{yuk} = - \frac{y_t^\SM}{\sqrt{2}} \bar t \left(\ct + i \gamma_5 \cttilde\right) t H,
\end{align}
where $y_t^\SM$ is the SM top-Yukawa coupling, $H$ is used to denote the Higgs boson field and $t$ to denote the top quark field. The parameter \ct rescales the \cp-even coupling with respect to the SM prediction ($c_t = 1$). The \cp-odd coupling of the Higgs boson to top quarks is parametrized by \cttilde\
(in the SM, $\cttilde = 0$). In an EFT framework (e.g.~SMEFT~\cite{Demartin:2015uha}), this modification would be generated by dimension-six operators. For simplicity, we do not allow for any modification of the other Yukawa couplings in this work. The coupling rescaling parameters \ct and \cttilde in \cref{eq:topYuk_lagrangian} can furthermore be rewritten in terms of an absolute value $|g_t|$ and a \cp-violating phase $\alpha$, defined as
\begin{align}
|g_t| \equiv \sqrt{\ct^2 + \cttilde^2}, \qquad \qquad \tan\alpha = \frac{\cttilde}{\ct}.
\label{eq:phase}
\end{align}
Note that $|g_t|$ corresponds to the $\kappa_t$ parameter used in Ref.~\cite{Aad:2020ivc}.

In addition to the modification of the top-Yukawa sector we shall consider an $SU(2)_L$ preserving modification of the Higgs-gauge-boson coupling strength,
\begin{align}\label{eq:HVV}
\mathcal{L}_V = \cv H \left(\frac{M_Z^2}{v} Z_\mu Z^\mu + 2\frac{M_W^2}{v} W_\mu^+ W^{-\mu}\right),
\end{align}
where $Z$ and $W$ are the massive vector boson fields ($M_Z$ and $M_W$ are their respective masses). The SM Higgs vacuum expectation value, $v$, is $\simeq 246\gev$. The parameter \cv rescales the SM-Higgs interaction strength with the massive vector bosons. In addition to rescaling the SM-Higgs--gauge interaction, we could also include additional non-SM like operators of the form $Z_{\mu\nu}Z^{\mu\nu}H$ or $Z_{\mu\nu}\widetilde Z^{\mu\nu}H$  (and analogously for the $W$~boson), where $Z_{\mu\nu}$ and $\widetilde Z_{\mu\nu}$ are the $Z$~boson's field strength and its dual, respectively. Since the focus of this study lies mainly on the top-Yukawa interaction, we do not take into account these additional couplings, which are not expected to have a significant impact on our analysis.

We furthermore include a set of operators that couple the Higgs boson directly to photons and gluons,
\begin{align}\label{eq:dim5}
\mathcal{L}_{Hgg,H\gamma\gamma} ={}& - \frac{1}{4 v} H \left(- \frac{\alpha_s}{3\pi}\cg G_{\mu\nu}^a G^{a,\mu\nu} + \frac{\alpha_s}{2\pi}\cgtilde G_{\mu\nu}^a \widetilde{G}^{a,\mu\nu} \right)  \nonumber\\
& - \frac{1}{4 v} H\left( \frac{47\alpha}{18\pi}\cgamma A_{\mu\nu}A^{\mu\nu} + \frac{4\alpha}{3\pi}\cgammatilde A_{\mu\nu}\widetilde{A}^{\mu\nu}\right),
\end{align}
where $\alpha_s = g_3^2/(4\pi)$, with $g_3$ being the strong gauge coupling, and $\alpha = e^2/(4\pi)$, with $e$ being the elementary electric charge. $A_{\mu\nu}$ and $G^a_{\mu\nu}$ (with $a$ being the color index) are the field strengths of the photon and the gluons.
The couplings \cg, \cgtilde, \cgamma and \cgammatilde can be generated by heavy undiscovered BSM particle(s), while the SM limit corresponds to $\cg = \cgtilde = \cgamma = \cgammatilde = 0$. The prefactors are chosen as in the SM for the case where the top quark and the $W$-boson are integrated out. These additional couplings take into account the possibility that new-physics contributions could decorrelate the tight constraints from Higgs production via gluon fusion and the Higgs decay to photons from the Higgs--top-quark interactions. In the analysis, we will always assume that $\cgtilde = \cgammatilde = 0$.

The parametrization of the adopted effective Lagrangian with up to five free parameters, \ct, \cttilde, \cv, \cg and \cgamma, has been chosen for demonstrating the impact of the experimental constraints on possible \cp-violating effects in the top quark sector. More general parametrizations allowing additional sources of \cp violation or additional free couplings would in general lead to somewhat increased allowed ranges of the top quark Yukawa coupling parameters \ct and \cttilde.


\section{Coupling dependence of cross-sections and decays}
\label{sec:xs_fits}

In order to constrain the effective model described in \cref{sec:model}, we need to obtain theoretical predictions for the relevant Higgs production and decay channels in dependence of the various coupling parameters. In this section we derive these theory predictions. Following the structure of \cref{sec:topYuk}, we first discuss Higgs production via gluon fusion and the Higgs to di-photon decay, and then turn to $Z$- and top-associated Higgs production. The cross section of processes involving a Higgs--vector-boson interaction but not the Higgs top-Yukawa coupling (e.g.\ weak-vector-boson fusion) can be obtained by multiplying the corresponding SM cross section by $\cv^2$ (according to \cref{eq:HVV}).


\subsection{Cross-section calculation and event generation}
\label{sec:xs_fits_calc}

We calculate the cross sections for different parameter choices in order to derive numerical fit formulas which we will use as input for the global fit in \cref{sec:results}. While for gluon fusion and the Higgs decay into two photons such fit formulas are available at leading-order (LO) in analytic form including the full dependence on the generalized top-Yukawa coupling, \cref{eq:topYuk_lagrangian}, this is not the case for Higgs boson production in association with one or two top-quarks or a $Z$~boson.

For calculating the cross-sections for these processes, we use \texttt{MadGraph5\_aMC@NLO~2.7.0}~\cite{Alwall:2014hca} with \texttt{Pythia~8.244}~\cite{Sjostrand:2007gs} as parton shower (PS) employing the A14 set of tuned parameters~\cite{ATL-PHYS-PUB-2014-021}. As model file, we use the ``Higgs charaterization model''~\cite{Artoisenet:2013puc,Maltoni:2013sma,Demartin:2014fia}. We derive the cross-sections at LO in the five-flavor scheme. For the parton-distribution functions (PDF) we use the \texttt{MSTW2008LO}~\cite{Martin:2009iq} fit evaluated through the \texttt{LHAPDF} interface~\cite{Whalley:2005nh}. The derived results are then rescaled to the state-of-the-art SM predictions reported in Ref.~\cite{deFlorian:2016spz}.\footnote{In the case of Higgs production via gluon fusion, this procedure could be improved by taking into account the full NLO predictions available for scalar and pseudoscalar production. However, we cross-checked our predictions with the ones of the code \texttt{SusHi}~\cite{Harlander:2012pb,Harlander:2016hcx}, finding differences which are negligible within the theoretical and experimental uncertainties.}

Note that the ``Higgs characterization model'' allows the derivation of results at next-to-leading order in the strong coupling constant only if the Higgs--gluon--gluon BSM operators of \cref{eq:dim5} are zero (or in case of the infinite top-mass limit). Therefore, we include the BSM effects only at the LO and then rescale the predictions by the ratio of the state-of-the-art SM predictions to the SM LO predictions. Since the contributions of the Higgs--gluon--gluon operators and the explicit loop contributions from top quarks that are coupled to the external Higgs boson should largely compensate each other if $\kg\sim1$, we expect that neglecting BSM NLO corrections leads to a negligible effect in the phenomenologically relevant parameter region.
For the analysis of the kinematic properties of $ZH$ and Higgs plus jets production, we use \texttt{Rivet}~\cite{Buckley:2010ar} to distribute the generated events into bins (using the STXS analysis script provided in Ref.~\cite{RivetSTXSscript}). In the case of $ZH$ production, we cross-checked the results obtained using \texttt{Rivet} against an independent analysis script written in the \texttt{MadAnalysis} framework~\cite{Conte:2012fm,Conte:2014zja,Dumont:2014tja,Conte:2018vmg}.


\subsection{Higgs boson production in gluon fusion}
\label{sec:xs_fits_gluon_fusion}

Following the notation of~Ref.~\cite{Freitas:2012kw}, the modification of the SM gluon fusion production cross-section at leading order (LO) is given by
\begin{align}\label{eq:ggH}
\kg^2 \equiv \frac{\sigma_{gg\rightarrow H}}{\sigma_{gg\rightarrow H}^\SM} = \frac{\left|\ct H_{1/2}(\tau_t) + \frac{2}{3}\cg + \ldots\right|^2 + \left|\cttilde A_{1/2}(\tau_t) + \cgtilde\right|^2}{\left|H_{1/2}(\tau_t) + \ldots\right|^2},
\end{align}
where $\tau_t = M_H^2/(4 M_t^2)$. The ellipses (i.e., the ``$+\dots$'') denotes the contributions involving the remaining SM quarks. They are included with their SM values in our numerical analysis, but their numerical effect is negligible (and we omit those contributions in the following discussions). The loop functions $H_{1/2}$ (\cp-even scalar attached to fermion loop) and $A_{1/2}$ (\cp-odd scalar attached to fermion loop) are given by
\begin{align}
H_{1/2}(\tau) = \frac{(\tau - 1)f(\tau) + \tau}{\tau^2}, \hspace{.5cm} A_{1/2}(\tau) = \frac{f(\tau)}{\tau}
\end{align}
with
\begin{align}
f(\tau)=
\begin{cases}
\arcsin^2(\sqrt{\tau}) & \text{for } \tau\le 1,\\
-\frac{1}{4}\left(\log\frac{1+\sqrt{1-1/\tau}}{1-\sqrt{1-1/\tau}}-i\pi\right)^2 & \text{for } \tau> 1.
\end{cases}
\end{align}
In the heavy top-quark mass limit the dependence simplifies to\footnote{For our numerical results, we employ \cref{eq:ggH} including the full top-quark mass dependence.}
\begin{align}\label{eq:ggH_HET}
\frac{\sigma_{gg\rightarrow H}}{\sigma_{gg\rightarrow H}^\SM}\bigg|_{M_t\rightarrow\infty} = (\cg + \ct)^2 + \frac{9}{4}(\cgtilde + \cttilde)^2,
\end{align}
in correspondence to the equation employed in Ref.~\cite{Demartin:2015uha}. According to Refs.~\cite{Demartin:2014fia,Demartin:2015uha}, also in the total rates for Higgs plus up to three jets no interference effects between the scalar and pseudoscalar components appear. Therefore, we employ \cref{eq:ggH} also for the numerical evaluation of gluon fusion in association with jets.

In addition, we investigated the azimuthal correlations between the jets in $gg\rightarrow H + 2j$, $\Delta\phi(j_1,j_2)$, as a function of \ct and \cttilde. This observable is known to have some sensitivity to the \cp character of the Higgs boson~\cite{DelDuca:2001fn,Klamke:2007cu,Dolan:2014upa,Demartin:2015uha}. However, numerically, we find modifications due to the \cp-properties at the sub-percent level, which cannot be resolved with the current experimental and theoretical precision. Nevertheless, we include $\Delta\phi(j_1,j_2)$ observables and their \cp-dependence in our fit (more details are given in \cref{app:ggH2j}). As mentioned in \cref{sec:topYuk}, the $p_T$-shape of Higgs bosons produced via gluon fusion is not sensitive to the \cp-nature of the top-Yukawa coupling. Therefore, we do not include corresponding observables into our fit.


\subsection{Higgs decay to two photons}

The decay width of the Higgs boson into photons is modified as follows,
\begin{align}
\kgamma^2 \equiv \frac{\Gamma_{H\rightarrow \gamma\gamma}}{\Gamma_{H\rightarrow \gamma\gamma}^\SM} = \frac{ \left|\frac{4}{3}\ct H_{1/2}(\tau_t) - \cv H_1(\tau_W) - \frac{47}{18}\cgamma +\ldots\right|^2 + \left|\frac{4}{3}\cttilde A_{1/2}(\tau_t) + \frac{4}{3}\cgammatilde\right|^2 }{\left|\frac{4}{3}H_{1/2}(\tau_t) - H_1(\tau_W) +\ldots\right|^2},
\end{align}
where $\tau_W = M_H^2/(4 M_W^2)$. The ellipses denotes subdominant SM contributions (e.g.\ from $c$ and $b$ quarks and $\tau$ leptons), which we take into account in our numerical analysis with their SM values. The loop function $H_1$ (\cp-even scalar attached to vector boson loop) is defined as
\begin{align}
H_1(\tau) = \frac{3 (2\tau - 1) f(\tau) + 3\tau + 2\tau^2}{2\tau^2}.
\end{align}
With the exception of $H\rightarrow gg$, which is modified in accordance with $gg\rightarrow H$ as outlined in \cref{sec:xs_fits_gluon_fusion}, the partial widths of all other Higgs decay channels are unmodified with respect to the SM or trivially rescaled by $\cv^2$ (see explanation above).


\subsection{$Z$-boson associated Higgs production}

\begin{figure}\centering
\includegraphics[width=.49\textwidth, height=0.4\textwidth]{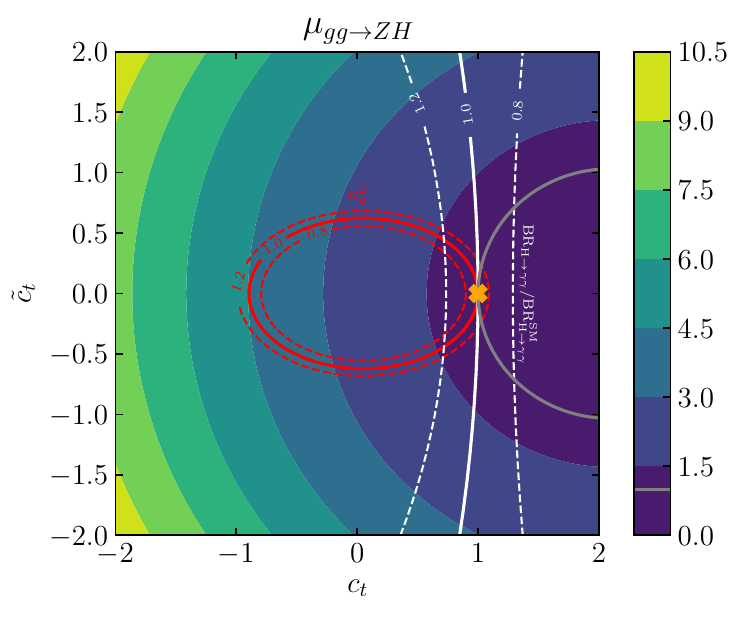}
\caption{Ratio of the $gg\rightarrow ZH$ cross-section over the SM cross-section in dependence of \ct and \cttilde for $\cv = 1$. \textit{Red contours}: $\kappa_g = 1.0 \pm 0.2$ for $\cg = 0$. \textit{White contours}: $\text{BR}_{H\rightarrow\gamma\gamma}/\text{BR}^\SM_{H\rightarrow\gamma\gamma} = 1.0 \pm 0.2$ for $\cgamma = 0$. The parameter point $(\ct = 1, \cttilde = 0)$ corresponding to the SM case is marked by an orange cross, while the SM value for the plotted observable is indicated by the gray curve.}
\label{fig:ggZH}
\end{figure}

The largest contribution to $Z$-boson associated Higgs production is the $q\bar q$-initiated process. The $gg$-initiated process plays a subdominant role ($\sigma_\text{SM}(q\bar q \to ZH) = 761$ fb and $\sigma_\text{SM}(gg \to ZH) = 123$ fb at $13\tev$).

The $q\bar q$-initiated process scales trivially with $\cv^2$. For the signal strength of gluon-initiated $Z$-boson associated Higgs production, we obtain
\begin{align}
\mu_{gg\rightarrow ZH}&\equiv\kappa_{ggZH}^2\equiv\frac{\sigma_{gg\rightarrow ZH}}{\sigma_{gg\rightarrow ZH}^\SM} = 0.45 \ct^2 + 0.50 \cttilde^2 + 2.44 \cv^2 - 1.89 \cv\ct \label{eq:mu_ggZH}
\end{align}
with $\sigma$ and $\sigma^\text{SM}$ denoting the $13\tev$ inclusive cross sections at the parameter point and for the SM, respectively, for the given process. We cross-checked this formula against the corresponding fit function implemented in \texttt{HiggsBounds-5}~\cite{Bechtle:2020pkv} (based on \texttt{vh@nnlo-v2}~\cite{Harlander:2018yio}) and found very good agreement. Note that there can be contributions from additional operators affecting the $gg\rightarrow ZH$ channel. We do not include the effect of such operators in the parametrization given in \cref{eq:mu_ggZH}. However, in order to assess the possible impact of these operators, we treat $\kappa_{ggZH}$ as a free parameter instead of using \cref{eq:mu_ggZH} in some of the fits presented in \cref{sec:results}.

\cref{fig:ggZH} shows the numerical dependence of the $gg\rightarrow ZH$ signal strength, as obtained via \cref{eq:mu_ggZH}, on \ct and \cttilde for $\cv=1$. As additional contour lines, we display $\kg^2$ (red solid: $\kg^2 = 1$; red dashed: $\kg^2 = 0.8, 1.2$) and $\text{BR}_{H\rightarrow\gamma\gamma}/\text{BR}^\SM_{H\rightarrow\gamma\gamma}$ (white solid: $\text{BR}_{H\rightarrow\gamma\gamma}/\text{BR}^\SM_{H\rightarrow\gamma\gamma}=1.0$; white dashed: $\text{BR}_{H\rightarrow\gamma\gamma}/\text{BR}^\SM_{H\rightarrow\gamma\gamma} = 0.8, 1.2$) for illustration setting $\cg = \cgamma = 0$. Furthermore the contour corresponding to $\mu_{gg\rightarrow ZH} = 1$ is indicated in gray. It forms an ellipse that is roughly centered around the point $(\ct,\cttilde) \simeq (2,0)$, while the contours for constant \kg are centered at the origin, $(\ct,\cttilde) = (0,0)$.

\begin{figure}\centering
\includegraphics[width=.49\textwidth]{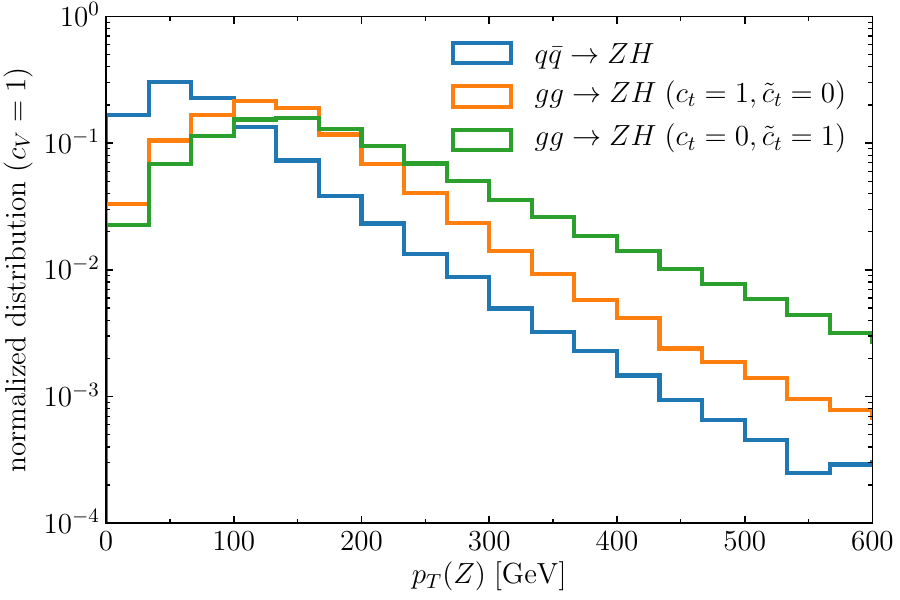}
\includegraphics[width=.49\textwidth]{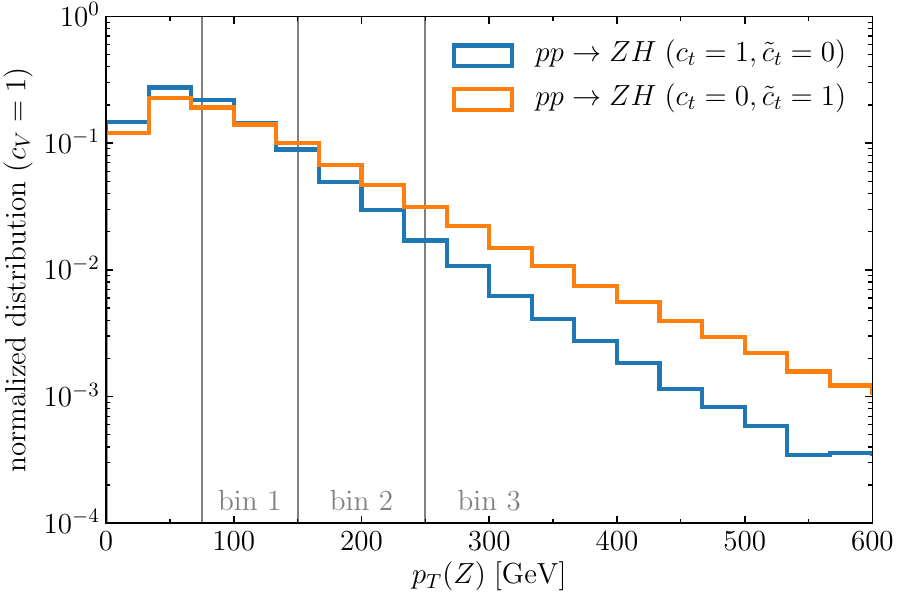}
\caption{\textit{Left}: $Z$-boson $p_T$-shape for $gg\rightarrow ZH$ and $q\bar q\rightarrow ZH$. \textit{Right:} $Z$-boson $p_T$-shape for $pp\rightarrow ZH$. The gray vertical lines mark the boundaries of the bins used in Ref.~\cite{Aaboud:2019nan}.}
\label{fig:ZH_ptshape}
\end{figure}

As additional input, we take into account modifications of the Higgs-boson $p_T$-distribution in $Z$-associated Higgs production. Experimentally, though, measurements are performed in $p_T$ bins of the leptonically decaying $Z$-boson which recoils against the Higgs boson. The $Z$-boson $p_T$ is strongly correlated to the Higgs boson $p_T$ and provides an experimentally cleaner observable. The transverse momentum distribution of $gg\rightarrow ZH$ is sensitive to the \cp-structure of the top-Yukawa coupling while it is insensitive for the $q\bar q\rightarrow ZH$ process. Exemplary $Z$-boson $p_T$-shapes are shown in the left panel of \cref{fig:ZH_ptshape}. The $p_T$-distribution for $gg\to ZH$ is harder in the pure \cp-odd case. As mentioned above, the $q\bar q$- and $gg$-initiated processes are difficult to distinguish experimentally. The $p_T$-shape of the $Z$ boson produced in the combined $pp\rightarrow ZH$ channel is shown in the right panel of \cref{fig:ZH_ptshape}, which features a visible increase in the high-$p_T$ range in the case of a large \cp-odd coupling due to the behavior of the $gg\to ZH$ subprocess.

\begin{figure}\centering
\includegraphics[width=.48\textwidth,height=0.4\textwidth]{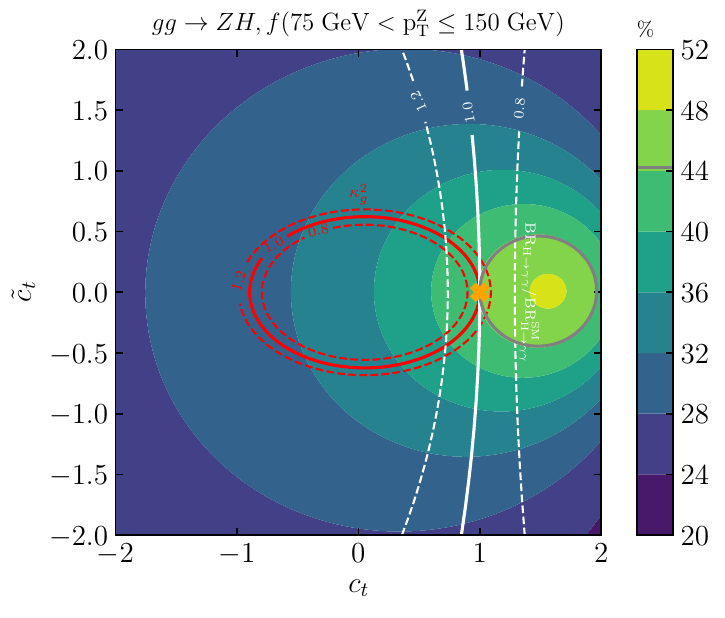}
\includegraphics[width=.48\textwidth,height=0.4\textwidth]{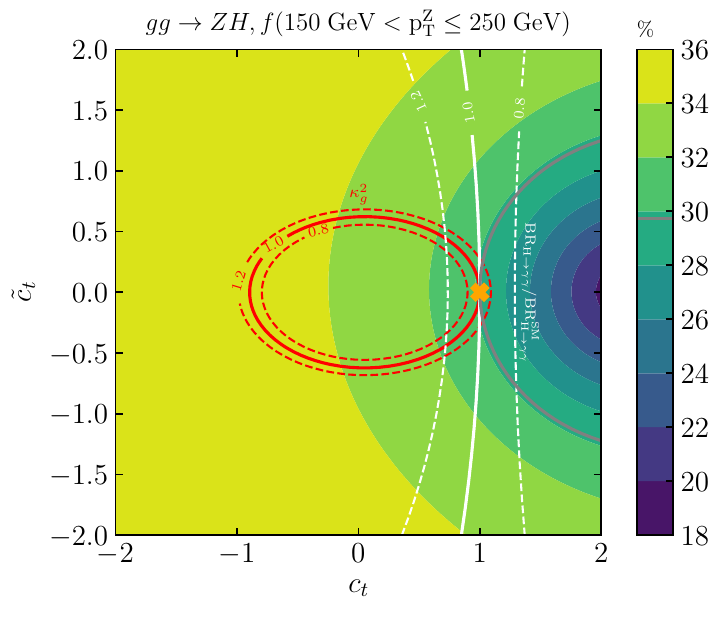}
\includegraphics[width=.48\textwidth,height=0.4\textwidth]{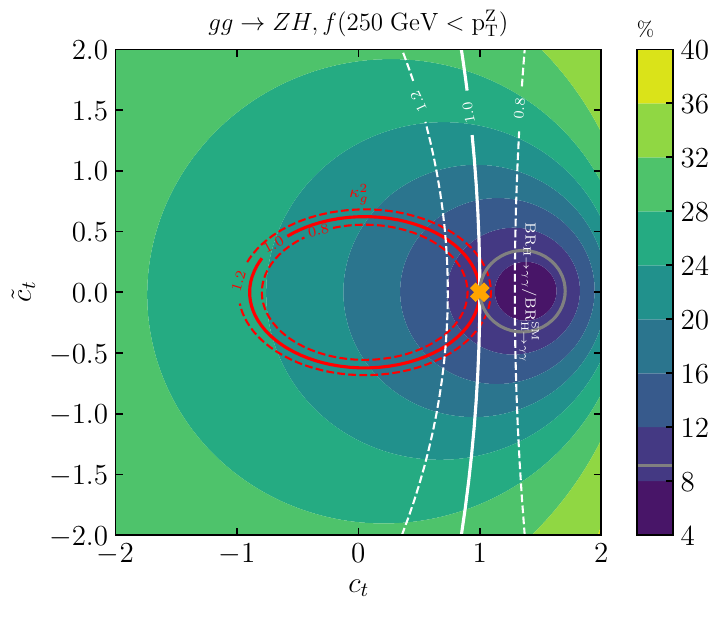}
\caption{Fraction of $gg\rightarrow ZH$ events in the specific $p_T$ bins in dependence of \ct and \cttilde for $\cv = 1$. \textit{Red contours}: $\kappa_g = 1.0 \pm 0.2$ for $\cg = 0$. \textit{White contours}: $\text{BR}_{H\rightarrow\gamma\gamma}/\text{BR}^\SM_{H\rightarrow\gamma\gamma} = 1.0 \pm 0.2$  for $\cgamma = 0$. The parameter point $(\ct = 1, \cttilde = 0)$ corresponding to the SM case is marked by an orange cross, while the SM value for the plotted observable is indicated by the gray curve.}
\label{fig:ggZH_bins}
\end{figure}

Measurements of the kinematic shape of $ZH$ production employ the STXS framework, in which the $p_T$-bins are defined. The current ATLAS analysis~\cite{Aaboud:2019nan,Aad:2020jym} uses the reduced $VH, V\to \text{leptons}$ stage-1.2\ STXS region scheme, given by~\cite{Badger:2016bpw,Berger:2019wnu,STXSnewbinning}:
\begin{itemize}
\item  bin~1: $75\gev <  p_T^Z \le 150\gev$,
\item  bin~2: $150\gev <  p_T^Z \le 250\gev$,
\item  bin~3: $250\gev <  p_T^Z$,
\end{itemize}
and illustrated in the right panel of \cref{fig:ZH_ptshape}. For the \cp-dependent predictions of the signal strength in these $p_T$-bins we obtain
\begin{align}
\mu_{gg\rightarrow ZH}^\text{bin-1} &= 0.20 \ct^2 + 0.22 \cttilde^2 + 1.96 \cv^2 - 1.13 \cv\ct, \\
\mu_{gg\rightarrow ZH}^\text{bin-2} &= 0.53 \ct^2 + 0.61 \cttilde^2 + 2.76 \cv^2 - 2.29 \cv\ct, \\
\mu_{gg\rightarrow ZH}^\text{bin-3} &= 2.02 \ct^2 + 2.22 \cttilde^2 + 5.12 \cv^2 - 6.14 \cv\ct.
\end{align}
In order to confirm the results shown in \cref{fig:ZH_ptshape}, the fraction of events, $f$, falling into the respective bins is plotted in \cref{fig:ggZH_bins}. As in \cref{fig:ZH_ptshape}, the $Z$-boson $p_T$-shape becomes harder if the \cp-odd component of the top-Yukawa coupling is increased.

In addition to the analysis performed in Refs.~\cite{Aaboud:2019nan,Aad:2020jym}, a specialized analysis targeting the high $p_T$-range was presented in Ref.~\cite{ATLAS:2020udg}. Since a significant statistical correlation between these analyses is expected (and the corresponding correlations are not available), we only take into account the most sensitive analyses -- i.e., Refs.~\cite{Aaboud:2019nan,Aad:2020jym} -- in our global fit (see \cref{sec:results}).


\subsection{Top-quark associated Higgs production}

For the signal strengths for inclusive top-quark associated Higgs production, we obtain the following fit formulas,
\begin{align}
\mu_{t\bar tH}&\equiv\frac{\sigma_{pp\rightarrow t\bar tH}}{\sigma_{pp\rightarrow t\bar tH}^\SM} = 1.00 \ct^2 + 0.42 \cttilde^2, \label{eq:mu_ttH} \\
\mu_{tH}&\equiv\frac{\sigma_{pp\rightarrow tH}}{\sigma_{pp\rightarrow tH}^\SM} = 3.28 \ct^2 + 1.00 \cttilde^2 + 3.82 \cv^2 - 6.10 \cv\ct, \label{eq:mu_tH} \\
\mu_{tWH}&\equiv\frac{\sigma_{pp\rightarrow tWH}}{\sigma_{pp\rightarrow tWH}^\SM} = 2.73 \ct^2 + 2.07 \cttilde^2 + 2.01 \cv^2 - 3.74 \cv\ct. \label{eq:mu_tWH}
\end{align}
The $tH$, $tWH$ and $t\bar t H$ production, channels are difficult to disentangle experimentally. Therefore, so far an inclusive quantity is measured, with possible contributions from all three processes. Correspondingly, we define the combined signal strength for $t\bar{t}H$, $tWH$ and $tH$ production as
\begin{align}
\mu_{t\bar{t}H+tWH+tH} \equiv \frac{\sigma(pp \to t\bar{t}H) + \sigma(pp \to tH)+\sigma(pp \to tWH)}{\sigma_\text{SM}(pp \to t\bar{t}H) + \sigma_\text{SM}(pp \to tH)+\sigma_\text{SM}(pp \to tWH))}.
\label{eq:mutthth_sum}
\end{align}
Since a distinction of $tH$ production from $ttH$ and $tWH$ production may be feasible at the LHC in the future (see \cref{sec:prospects}), we in addition define the SM-normalized $tH$ over $t\bar{t}H+tWH$ cross section ratio as
\begin{align}
\mu_{tH/(t\bar{t}H+tWH)} \equiv \frac{\sigma(pp \to tH)/\left[\sigma(pp \to t\bar{t}H)+\sigma(pp \to tWH)\right]}{\sigma_\text{SM}(pp \to tH)/\left[\sigma_\text{SM}(pp \to t\bar{t}H)+\sigma_\text{SM}(pp \to tWH)\right]}
\label{eq:mutthth_ratio}
\end{align}
and the combined signal strength for $t\bar{t}H$ and $tWH$ as
\begin{align}
\mu_{t\bar{t}H+tWH} \equiv \frac{\sigma(pp \to t\bar{t}H) + \sigma(pp \to tWH)}{\sigma_\text{SM}(pp \to t\bar{t}H) + \sigma_\text{SM}(pp \to tWH)}.
\end{align}

\begin{figure}\centering
\includegraphics[width=.49\textwidth,height=0.4\textwidth]{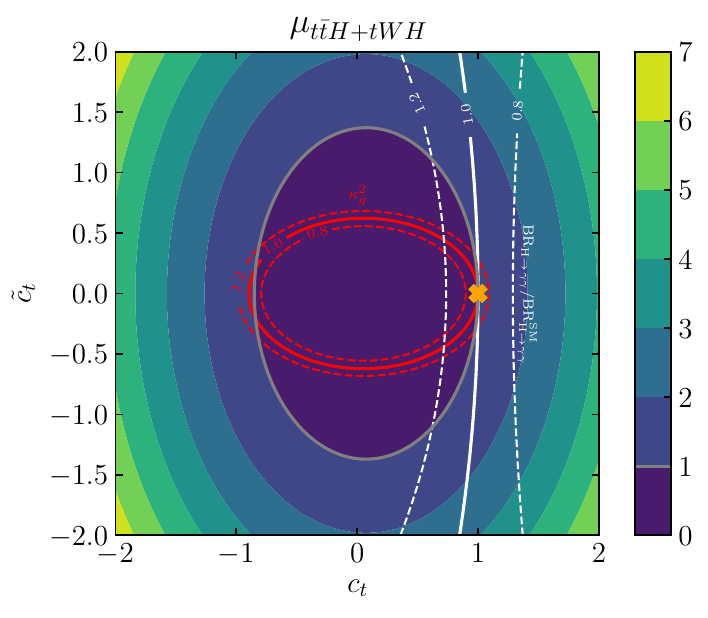}
\includegraphics[width=.49\textwidth,height=0.4\textwidth]{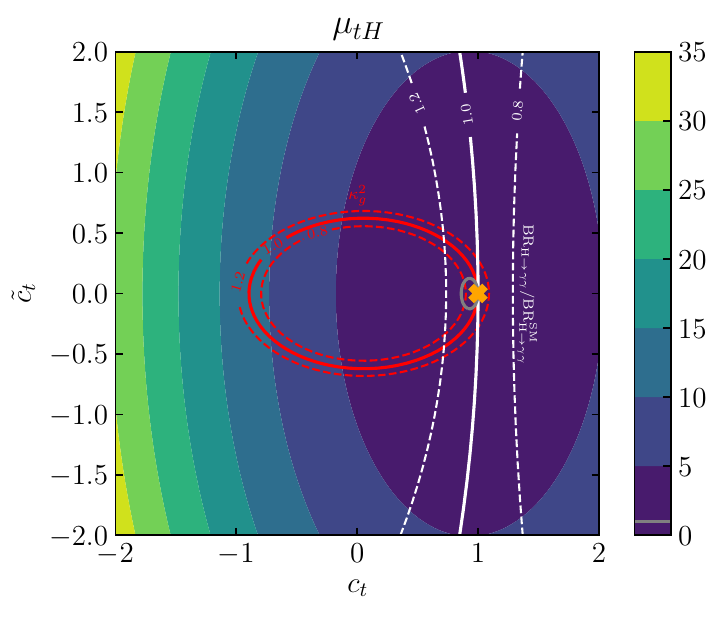}
\includegraphics[width=.495\textwidth,height=0.4\textwidth]{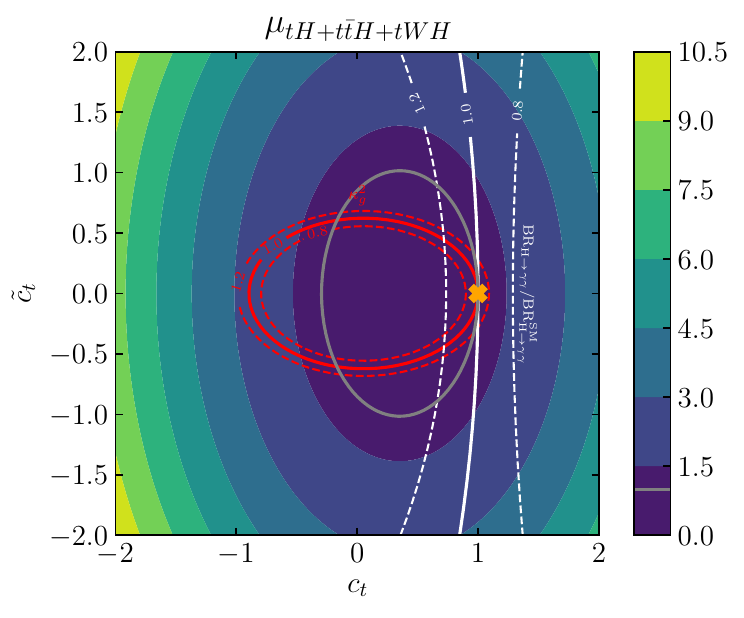}
\includegraphics[width=.485\textwidth,height=0.4\textwidth]{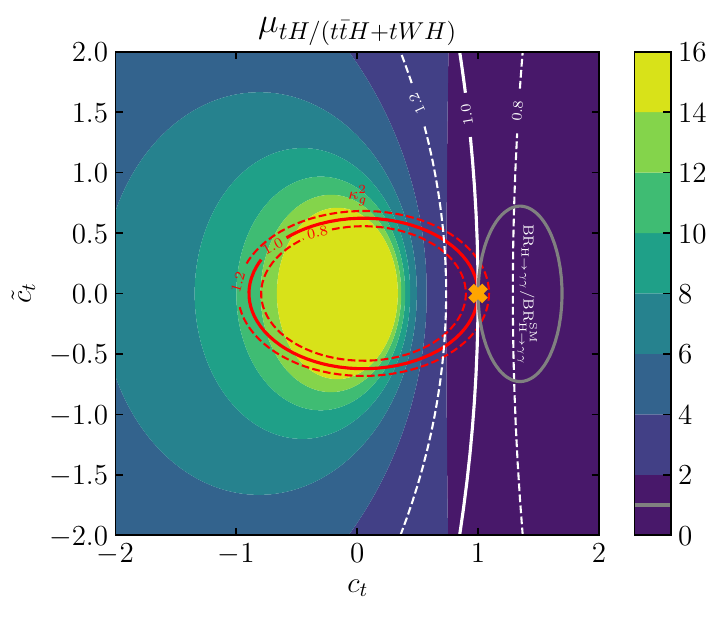}
\caption{Signal strengths of $tH$, $ttH$ and $tWH$ (and combinations) in dependence of \ct and \cttilde for $\cv = 1$. \textit{Red contours}: $\kappa_g = 1.0 \pm 0.2$ for $\cg = 0$. \textit{White contours}: $\text{BR}_{H\rightarrow\gamma\gamma}/\text{BR}^\SM_{H\rightarrow\gamma\gamma} = 1.0 \pm 0.2$  for $\cgamma = 0$. The parameter point $(\ct = 1, \cttilde = 0)$ corresponding to the SM case is marked by an orange cross, while the gray curve marks $\mu = 1$.}
\label{fig:tH-ttH}
\end{figure}

We plot $\mu_{t\bar tH + tWH}$ and $\mu_{tH}$ in dependence of \ct and \cttilde (for $\cv = 1$) in the upper left and right panel of \cref{fig:tH-ttH}, respectively. The contours of constant $\mu_{t\bar tH+tWH}$ are centered around the origin, $(\ct,\cttilde) = (0,0)$. This result originates from the fact that $\mu_{t\bar tH+tWH}$ is dominated by $t\bar{t}H$ production (in the SM case, $\sigma_\text{SM}(pp \to t\bar{t}H) = 506$ fb and $\sigma_\text{SM}(pp \to tWH) = 15$ fb).  This is different for $tH$ production. The interference of the top-Yukawa contribution with the $W$-boson contribution shifts the center of the contours for constant $\mu_{tH}$ from the origin towards positive \ct values.

We show the signal strength for the sum of the three production channels, \cref{eq:mutthth_sum}, in the lower left plot of \cref{fig:tH-ttH}. Since $t\bar tH$ has the highest cross-section in most parts of the considered parameter space ($\sigma_\text{SM}(pp \to tH) = 74$ fb), $t\bar tH$ production is the dominant contribution to the combined signal strength.

The lower right panel of \cref{fig:tH-ttH} shows the ratio of the signal strengths of $tH$ and the combined $t\bar tH$ and $tWH$ production, \cref{eq:mutthth_ratio}. The comparably large variation of this ratio along the contours of constant \kg indicates that this ratio is a promising future observable for disentangling the effects of \ct and \cttilde (see \cref{sec:prospects}).


\section{Constraints from current LHC measurements}
\label{sec:results}

In this section we discuss the fit of the model presented in \cref{sec:model} to experimental data, including all relevant inclusive and differential Higgs boson rate measurements available, and using the theory predictions outlined in \cref{sec:xs_fits}.\footnote{The fitting code is publicly available at \url{https://gitlab.com/timstefaniak/higgs_cp_fitting_code}.}


\subsection{Fit setup}
\label{sec:fit_setup}

In order to study the impact of current Higgs boson rate measurements from the LHC on the \cp-nature of the Higgs--top-quark interaction we performed numerical scans of the parameter space using four different model parametrizations:\footnote{Note that we do not use the parameters \cg, \cgtilde, \cgamma and \cgammatilde directly, but instead we fit the coupling modifiers \kg and \kgamma, see \cref{sec:xs_fits}.}
\begin{enumerate}
  \item (\ct, \cttilde) free [\emph{2D parametrization}]

  Only the \cp-even and \cp-odd components of the Higgs--top-quark coupling are allowed to vary freely; all remaining tree-level Higgs couplings (including $\cv$) are fixed to their SM value; the loop-induced coupling scale factors \kg and \kgamma\ are derived in terms of \ct and \cttilde, i.e. we assume $\cg = \cgamma = 0$ (see \cref{sec:xs_fits});

  \item (\ct, \cttilde, \cv) free [\emph{3D parametrization}]

  In addition to the \cp-even and \cp-odd components of the Higgs--top-quark coupling, we allow a freedom in the Higgs coupling to weak vector bosons ($V = W^\pm, Z$) via $c_V$; again, \kg and \kgamma\ are derived in terms of \ct and \cttilde, i.e. we assume $\cg = \cgamma = 0$ (see \cref{sec:xs_fits});

  \item (\ct, \cttilde, \cv, \kgamma) free [\emph{4D parametrization}]

  In addition to the three free parameters of the previous setup, the loop-induced Higgs coupling to photons, $\kappa_\gamma$, is allowed to vary freely, parametrizing our ignorance about possible contributions to the $H\to \gamma\gamma$ decay from new, undiscovered color-neutral but electrically charged particles; the loop-induced Higgs coupling to gluons, parametrized by $\kappa_g$, is still derived in terms of \ct and \cttilde, i.e. we assume $\cg = 0$ (see \cref{sec:xs_fits});

  \item (\ct, \cttilde, \cv, \kgamma, \kg) free [\emph{5D parametrization}]

  Both loop-induced Higgs coupling parameters -- $\kappa_\gamma$  and $\kappa_g$ -- are allowed to vary freely, accounting for the possibility of additional contributions to gluon fusion and the ${H\to \gamma\gamma}$ decay from undiscovered particles carrying color or electric charge. This five-dimensional fit setup is the most general parametrization that we consider.
\end{enumerate}

For brevity, we shall refer to these parametrizations by stating the number of dimensions (D), i.e.~the number of free fit parameters, as denoted above.
In the baseline fits for these four parametrizations, $\kappa_{ggZH}$ is derived from the other coupling modifiers following \Cref{eq:mu_ggZH}. However additional fits are performed in order to address specific issues in the analysis. Specifically, we performed fits where the 5D parametrization is extended by $\kappa_{ggZH}$ as an additional free fit parameter so that a fit with six free parameters is performed, as well as fits to a reduced set of observables (without including the dedicated $t\bar{t}H$ analyses) and/or without taking Higgs-$p_T$-shape effects due to \cp-violation into account. More details on those fits are given below in the discussion of the results.

Technically, we sample the parameter space in each fit setup randomly with $\sim\mathcal{O}(10^7-10^8)$ points. We use uniform priors for \ct, \cttilde, \cv (and $\kappa_{ggZH}$ when applicable), and Gaussian priors with mean value at the SM prediction for \kg and \kgamma. Note that the details of the sampling technique do not influence the fit results, but are only chosen to ensure a sufficiently dense sampling of the relevant parameter space.

At each parameter point we evaluate the predicted cross sections and decay rates, as detailed in \cref{sec:xs_fits}, and include them into the \texttt{HiggsBounds}/\texttt{HiggsSignals} framework~\cite{Bechtle:2008jh,Bechtle:2011sb,Bechtle:2013xfa,Bechtle:2013wla,Bechtle:2014ewa,Bechtle:2020pkv}. \texttt{HiggsSignals}~(version 2.5.0)~\cite{Bechtle:2013xfa,Bechtle:2013wla,Bechtle:2014ewa,HS25} incorporates the latest available Higgs rate measurements from ATLAS and CMS from Run~2~\cite{Aaboud:2018gay,Aaboud:2017rss,ATLAS:2019nvo,Aaboud:2018jqu,Aad:2020mkp,ATLAS:2019jst,Aaboud:2018pen,Aad:2019lpq,Aad:2020jym,Sirunyan:2020sum,Sirunyan:2018hbu,Sirunyan:2017elk,Sirunyan:2017dgc,CMS:2019lcn,Sirunyan:2018mvw,Sirunyan:2018shy,CMS:2018dmv,CMS:2019kqw,CMS:2019chr,CMS:1900lgv,CMS:2019pyn}, as well as the combined measurements from Run~1~\cite{Khachatryan:2016vau}. In total, \texttt{HiggsSignals} includes 81 Run-2 measurements and 20 Run-1 measurements. A detailed list of the included observables is given in \cref{app:expinput}. Note, in particular, that we take into account the shape modification of the Higgs transverse momentum distribution expected in the presence of \cp-violation for the available $p_T$-binned STXS measurements of $pp\to ZH$, $H\to b\bar{b}$ by ATLAS~\cite{Aaboud:2019nan,Aad:2020jym},
as well as in the $\Delta\Phi_{j_1j_2}$ distribution for the $pp \to H+jj,H\to\gamma\gamma$ channel, for which ATLAS provided measurements in four $\Delta\Phi_{j_1j_2}$ bins~\cite{ATLAS:2019jst}.

 Based on this large collection of measurements, \texttt{HiggsSignals} is used to calculate a total $\chi^2$ value, which includes various information on correlations of systematic uncertainties as provided by the experiments. In the $\chi^2$ evaluation, we assume the theoretical uncertainties on Higgs boson production and decay rates, as well as their correlations, to be the same as in the SM across the parameter space. We then determine the favored regions in parameter space by calculating the $\chi^2$ difference, $\Delta \chi^2 \equiv \chi^2 - \chi^2_\text{min}$, with $\chi^2_\text{min}$ being the minimal $\chi^2$ value found within the parameter scan at the so-called best-fit point. In the following we mainly present the fit results projected onto binned two-dimensional parameter planes. In each bin we display the minimal $\chi^2$ for the subset of scan points within the bin. In the frequentist approach, this corresponds to the profile likelihood for the two-dimensional plane of the parameters of interest. We then derive the $1\,\sigma$, $2\,\sigma$ and $3\,\sigma$ confidence regions from the profile $\Delta\chi^2$, corresponding (in a two-dimensional projection of the parameter space) to $\Delta\chi^2 < 2.3, 6.18$ and $11.83$, respectively, assuming the Gaussian limit is approximately realized. In the case of a profile likelihood in a one-dimensional parameter space (used for the \cp-violating phase interpretation) the $1\,\sigma$, $2\,\sigma$  and $3\,\sigma$  corresponding $\Delta\chi^2$ values  are $1$, $4$ and $9$, respectively.



In all four parametrizations we find the best-fit point to be remarkably close to the SM. The minimal $\chi^2$ improves only insignificantly with respect to the SM $\chi^2$, despite the up to five additional free parameters. A more quantitative discussion about how well the models describe the data, as well as implications about which model may be favored, can be found in \cref{app:bfpoint}.  


\subsection{\cp-violating Higgs--top-quark interactions}
\label{sec:res_ctcttilde}

\begin{figure}[!bt]
  \centering
\includegraphics[width=0.49\textwidth]{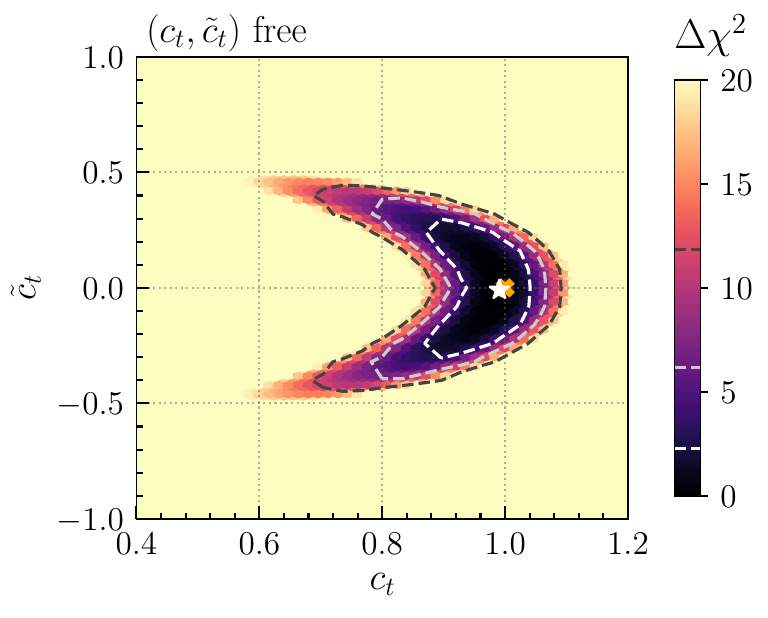}\hfill
\includegraphics[width=0.49\textwidth]{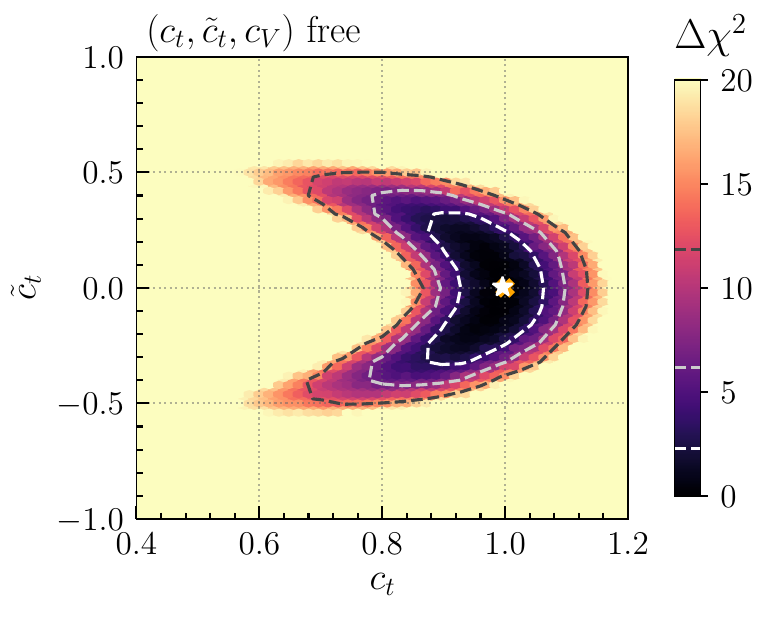}\\
\includegraphics[width=0.49\textwidth]{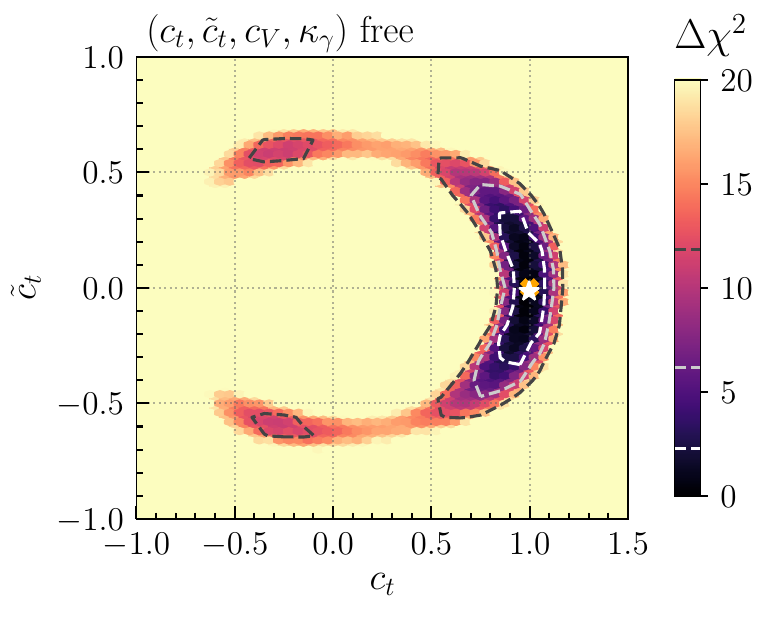}\hfill
\includegraphics[width=0.49\textwidth]{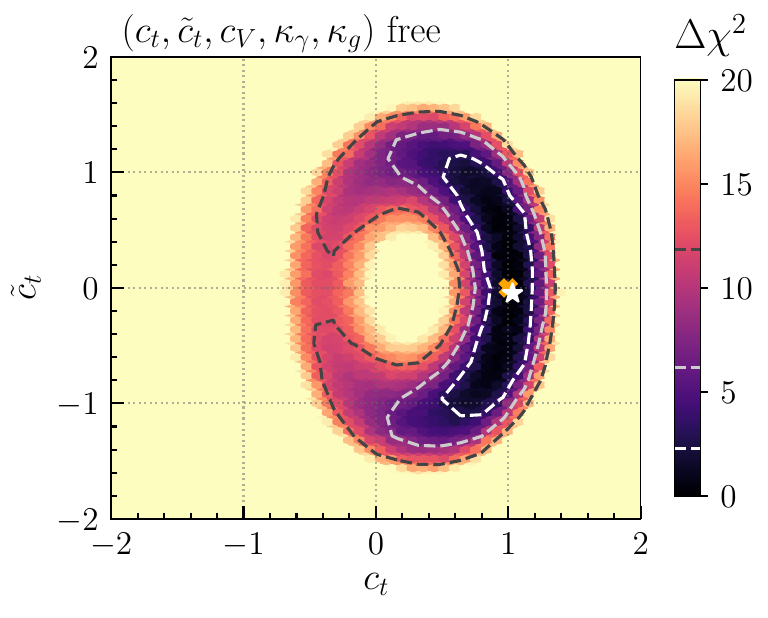}
\caption{Fit results in the (\ct, \cttilde) parameter plane for the four considered models (see header labels), using all currently available measurements from the LHC. The color corresponds to the profile $\Delta\chi^2$ of the global fit, and the $1\,\sigma$, $2\,\sigma$ and $3\,\sigma$ confidence regions are shown as white, light-gray and dark-gray dashed contours, respectively. The best-fit point and the SM case are marked by a white star and an orange cross, respectively. Note the larger scale required to display the bottom plots.}
\label{fig:ctcttilde}
\end{figure}

First, we investigate the two-dimensional plane of the \cp-even and \cp-odd Higgs--top-quark coupling parameters, \ct and \cttilde, respectively. The corresponding fit results for all four model parametrizations are shown in \cref{fig:ctcttilde}. The profile $\Delta\chi^2$ distribution is shown in color (the maximal value includes all values of $\Delta\chi^2 \ge 20$) and the $1\,\sigma$, $2\,\sigma$ and $3\,\sigma$ confidence regions as defined above are indicated by the white, light-gray and dark-gray dashed contours, respectively.

In the first three parametrizations (see above) the gluon fusion cross section (or equivalently \kg) is derived from \ct and \cttilde. The fact that the experimental results are well compatible with a gluon fusion cross section that is close to the SM value gives rise to the feature that the favored region of parameter space is located around the ellipse in the (\ct, \cttilde) parameter space where $\kappa_g \approx 1$ (see also \cref{fig:tH-ttH,fig:ggZH}). However, the favored region is restricted to the part where \ct is around $+1$. In the 2D and 3D parametrizations -- (\ct, \cttilde) free and (\ct, \cttilde, \cv) free -- this constraint originates from the fact that the preferred value for the $H\to\gamma\gamma$ partial width modifier, \kgamma, which is derived in these parametrizations from the fit parameters, is close to $+1$ (see also the discussion in \cref{sec:xs_fits}). As the 3D parametrization has additional freedom in the Higgs-vector boson coupling, \cv, which influences the $H\to \gamma\gamma$ partial width, the allowed region in the (\ct, \cttilde) parameter space is slightly larger than in the 2D parametrization (which has fixed $c_V = 1$). In contrast, the 4D parametrization treats \kgamma as a free parameter, hence the lower bound on \ct is weakened here, as shown in the \emph{bottom left panel} of \cref{fig:ctcttilde} (note the different scale of the \ct\ axis in comparison to the previous plots in the figure). In fact, two local minima of the likelihood appear in the negative \ct region, at $(\ct,\cttilde) \sim (-0.3,0.6)$ and $(-0.3,-0.6)$, which are still allowed at the $3\,\sigma$ level. In these regions, the predictions for Higgs production in gluon fusion, $gg\to H$, as well as the combined top-quark associated Higgs production, $pp\to t\bar{t}H, tWH$~and $tH$, are SM-like [see the intersection of the gray  and red solid contour in \cref{fig:tH-ttH} (\emph{bottom left panel})]. However, the process $gg\to ZH$ is strongly enhanced here, which excludes this region at the $2\,\sigma$ level. Overall, the interplay of Higgs measurements in channels containing the top quark associated and $gg\to ZH$ Higgs production modes restricts the $2\,\sigma$ favored region to the positive \ct range, as will be discussed in more detail in the following subsections.

The shape of the favored region in the 5D parametrization is different, as \kg and \kgamma are both treated as free parameters. Here, the favored region in the (\ct, \cttilde) parameter space extends to much larger values of \cttilde\ than in the previous parametrizations. It should be noted that the scale on the \cttilde-axis has been enlarged by a factor two compared to the other plots in this figure, while the scale on the \ct-axis has been much enlarged and displays also negative values. As Higgs production in gluon fusion is now governed by an individual fit parameter, the fit does not constrain the allowed parameter space to the region near the $\kappa_g = 1$ ellipse derived from \ct and \cttilde (see \cref{sec:xs_fits}). Instead, again, the Higgs measurements in channels containing the top quark associated and $gg\to ZH$ Higgs production modes become important and impose the constraints on the (\ct,~\cttilde) parameter space observed in \cref{fig:ctcttilde} (bottom right).
In particular, the allowed region aligns with the ellipse in the parameter space where SM-like rates are obtained for inclusive top quark associated Higgs production, see \cref{fig:tH-ttH},
and allow substantially smaller \ct values as compared to the other parametrizations. However, they are constrained at the $2\,\sigma$ level to the region $c_t > 0$ by measurements that include $gg\to ZH$ production, see \cref{fig:ggZH}, as well as by dedicated top-quark associated Higgs measurements that feature different signal acceptances among the various top quark associated Higgs production modes (see \cref{app:expinput} and discussion below).

\medskip

\begin{figure}[!tb]
  \centering
\includegraphics[width=0.49\textwidth]{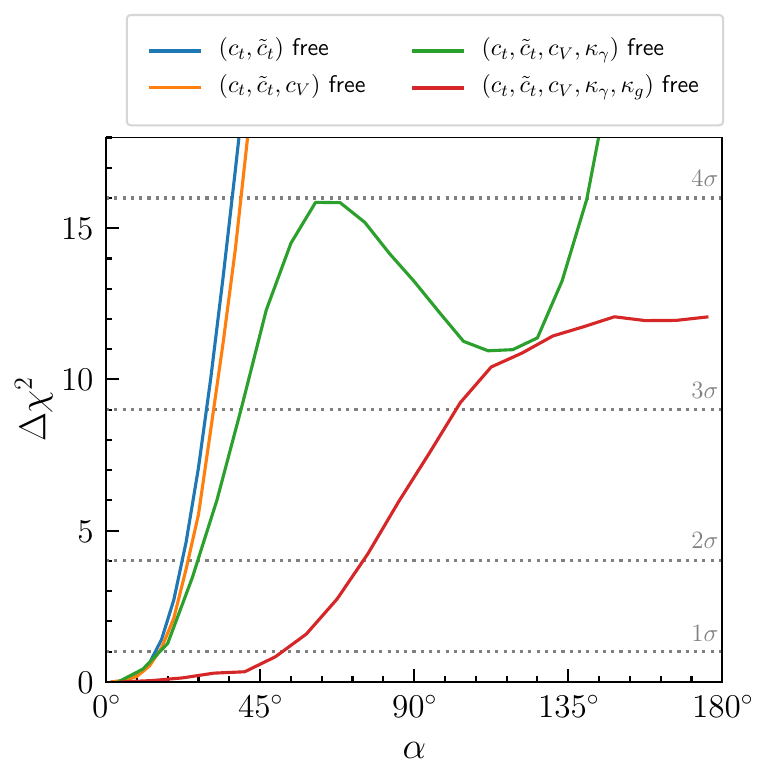}
\caption{Fit results for the \cp-violating phase $\alpha$ in the four different model parametrizations.}
\label{fig:alpha}
\end{figure}

The constraints derived in the various parametrizations can also be interpreted in terms of the \cp-violating phase $\alpha$, defined in \cref{eq:phase}.
The fit results for $\alpha$ are shown in \cref{fig:alpha}, where we again profile over the other parameters. Since \cttilde enters the theoretical predictions only in squared form (see \cref{sec:xs_fits}), the constraints on the \cp-violating phase $\alpha$ are symmetric around zero. The constraints are similar in the 2D, 3D and 4D parametrizations, with $|\alpha|$ being bounded to be $\lesssim 22.5 - 27$ degrees at the $2\,\sigma$ level.
In the 4D parametrization, a second minimum appears around $\alpha \sim 110$ degrees, which corresponds to the local minima in the negative \ct\ region discussed above [see \cref{fig:ctcttilde} (\emph{bottom left panel})].
In the 5D parametrization, the $2\,\sigma$ upper limit on the \cp-violating phase is $\lesssim 72$ degrees.


\subsection{$pp\to ZH$ subprocesses}
\label{sec:res_ppZH}

  We now study the behavior of the rate of Higgs production in association with a $Z$ boson in our fit. In particular, we focus on the gluon-initiated subprocess, $gg\to ZH$, which depends non-trivially on the \cp-nature of the Higgs--top-quark interaction, as discussed in \cref{sec:xs_fits}. In contrast, the quark-initiated subprocess, $q\bar{q}\to ZH$, depends dominantly on the Higgs--$Z$-boson interaction, i.e.\ on the coupling modifier $c_V$, which is either set to the SM prediction (in the 2D parametrization) or treated as an individual fit parameter (in the other parametrizations). For the latter, in an inclusive measurement of $pp\to ZH$, the $c_V$-dominated $q\bar{q}\to ZH$ process can at least partially compensate a non-trivial modification of the $gg\to ZH$ rate from the \cp-violating Higgs--top-quark interaction. Consequently, non-inclusive measurements, where the subprocesses $q\bar{q}\to ZH$ and $gg\to ZH$ can be separated, as well as Higgs transverse momentum ($p_T$) dependent measurements are crucial to increase the sensitivity to the \cp-nature of the Higgs--top-quark interaction in $Z$-boson associated Higgs production.

\begin{figure}[!htbp]
  \centering
\includegraphics[width=0.49\textwidth]{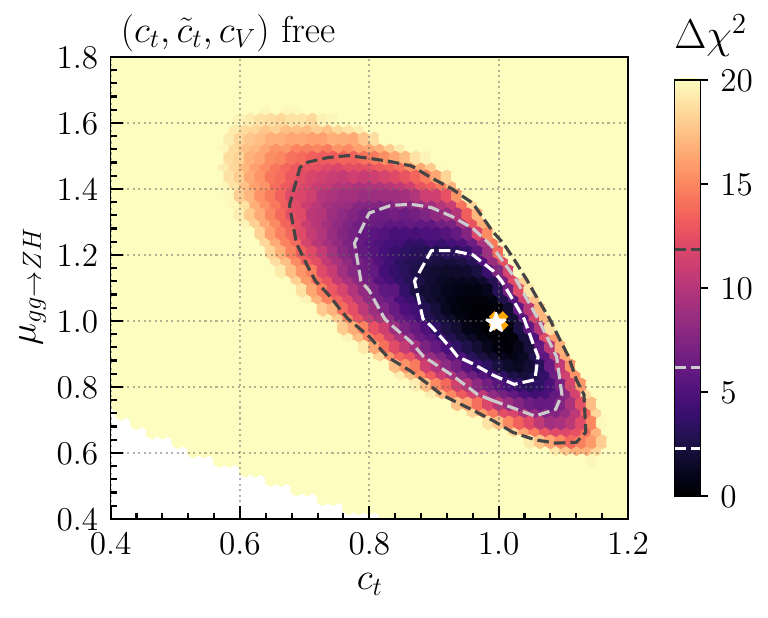}\hfill
\includegraphics[width=0.49\textwidth]{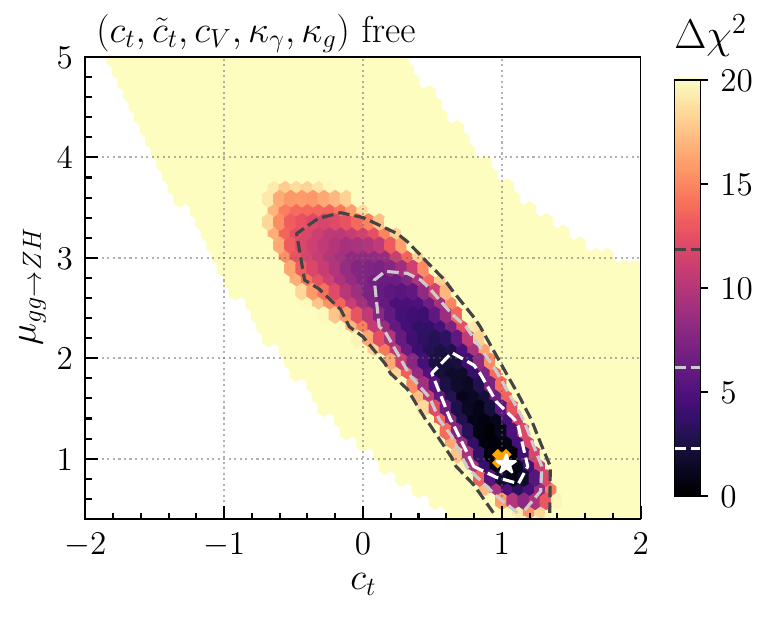}
\includegraphics[width=0.49\textwidth]{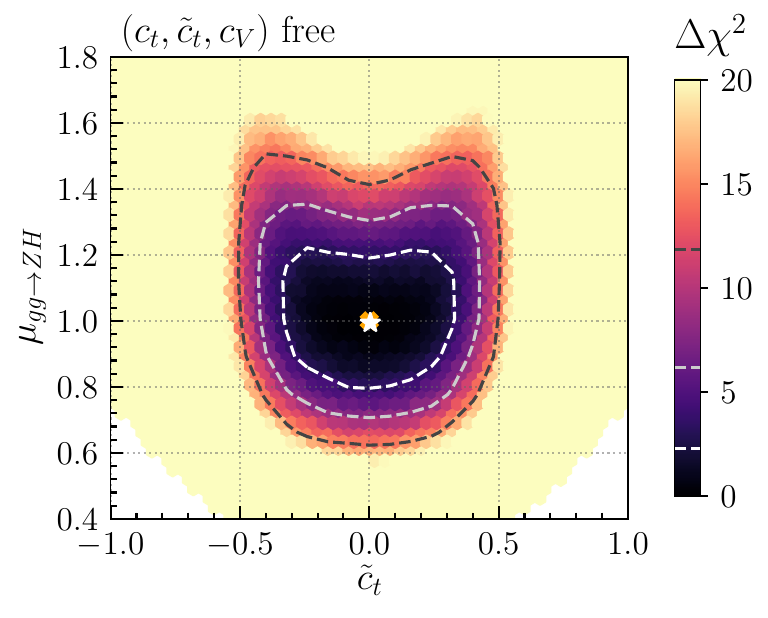}\hfill
\includegraphics[width=0.49\textwidth]{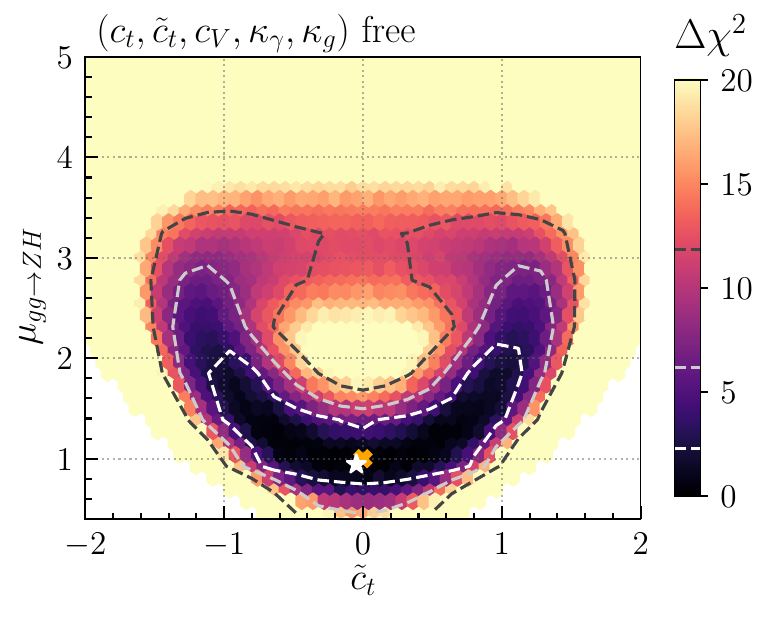}
\includegraphics[width=0.49\textwidth]{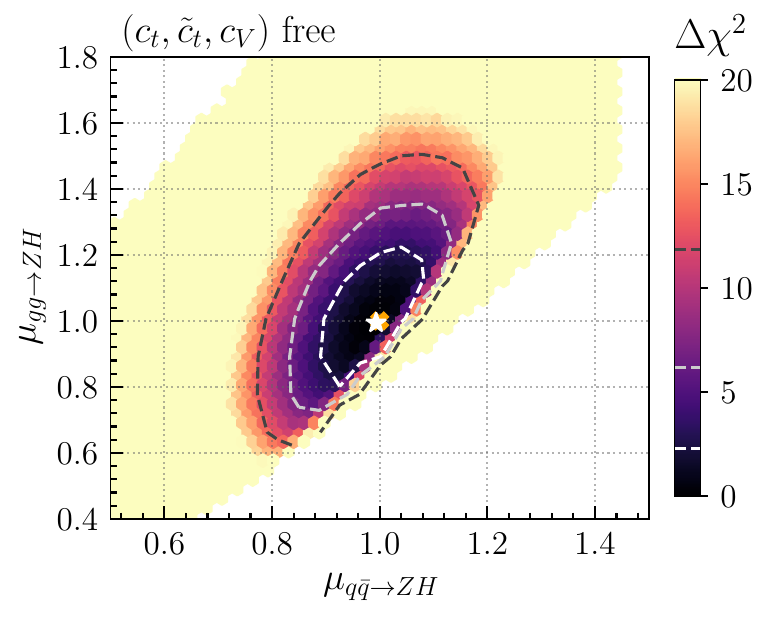}\hfill
\includegraphics[width=0.49\textwidth]{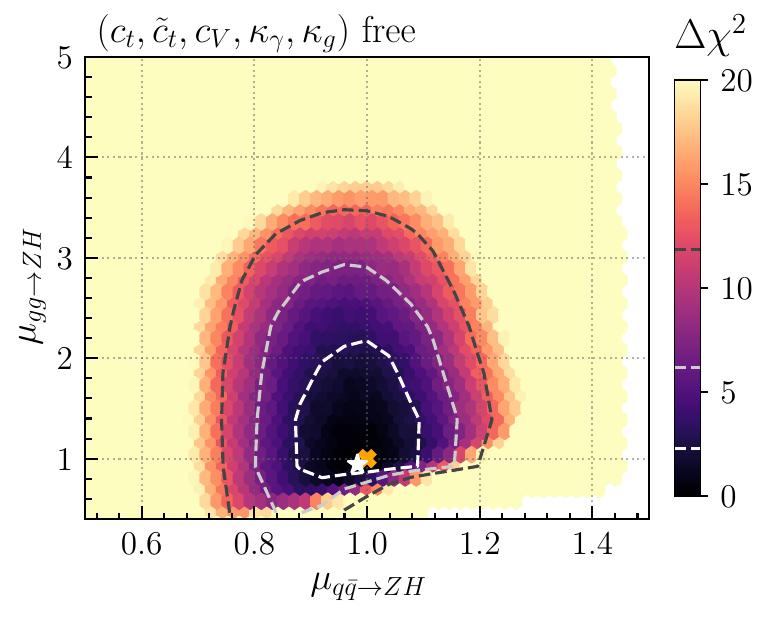}
\caption{Predicted signal strength of gluon-initiated $ZH$ production, $\mu_{gg\to ZH}$,
in dependence of $c_t$ (\emph{top row}) and $\tilde{c}_t$ (\emph{middle row}) and its correlation with the $q\bar{q} \to ZH$ signal strength, $\mu_{q\bar{q}\to  ZH}$, (\emph{bottom row}) for the 3D parametrization (\emph{left panels}) and the 5D parametrization (\emph{right panels}). The color corresponds to the profile $\Delta\chi^2$ of the global fit, and the $1\,\sigma$, $2\,\sigma$ and $3\,\sigma$ confidence regions are shown as white, light-gray and dark-gray dashed contours, respectively. The best-fit point and the SM case are marked by a white star and an orange cross, respectively. Note the larger scale required to display the plots in the right panels.}
\label{fig:res_ggZH}
\end{figure}

We display in \cref{fig:res_ggZH} the inclusive cross section of the $gg\to ZH$ process normalized to the SM prediction, $\mu_{gg\to ZH}$, in dependence of \ct (\emph{top panels}) and \cttilde (\emph{bottom panels}) for the 3D parametrization -- (\ct, \cttilde, \cv) free -- (\emph{left panels}) and the 5D parametrization -- (\ct, \cttilde, \cv, \kg, \kgamma) free -- (\emph{right panels, note the larger scale required to display the plots}). In both models, this rate is derived via \cref{eq:mu_ggZH} from the fit parameters \ct, \cttilde and \cv. The differences in the allowed $\mu_{gg\to ZH}$ ranges in the two models therefore entirely originate from differences in the allowed ranges of the fit parameters \ct, \cttilde and \cv.\footnote{The corresponding results for the 2D and 4D parametrization not shown here are qualitatively very similar to those for the 3D parametrization.} We find the following allowed ranges for $\mu_{gg\to ZH}$ at the $2\,\sigma$ level:
\begin{align}
\mu_{gg\to ZH} \in [0.70, 1.36]& \qquad \text{in~the~3D parametrization -- (\ct, \cttilde, \cv) free}, \\
\mu_{gg\to ZH} \in [0.44, 2.90]& \qquad \text{in~the~5D parametrization -- (\ct, \cttilde, \cv, \kg, \kgamma) free}.
\end{align}
The upper limits of the allowed ranges are reached only for a non-zero \cp-odd component of the Higgs--top-quark interaction, i.e. $\cttilde \ne 0$. This feature is particularly pronounced in the 5D parametrization, see \cref{fig:res_ggZH} (\emph{bottom right panel}), but is also visible in the 3D parametrization (\emph{bottom left panel}). From the upper panels in \cref{fig:res_ggZH} we again observe how $\mu_{gg\to ZH}$ increases for decreasing \ct, as shown in \cref{fig:ggZH} in \cref{sec:xs_fits}, thus limiting the allowed \ct range in the 5D parametrization to positive values at the $2\,\sigma$ level.

Concerning the $q\bar{q} \to ZH$ subprocess, we found in all parametrizations with \cv as a fit parameter that the $q\bar{q}\to ZH$ cross section is restricted at the $2\,\sigma$ level to be within $20\%$ of the SM prediction, as can be seen in the bottom panels of \cref{fig:res_ggZH} for the 3D and 5D parametrization. While we observe a mild linear correlation of the $gg\to ZH$ and $q\bar{q}\to ZH$ signal strengths in the 3D and 4D (not shown) parametrization, the signal strengths are rather uncorrelated in the 5D parametrization. Hence, observing a strongly enhanced $gg\to ZH$ component in Higgs channels targeting $ZH$ production may indicate a \cp-violating Higgs--top-quark interaction. 

We now analyze to what extent measurements of differential distributions already help to constrain the parameter space. As noted earlier, the only measurements that allow us to account for kinematic modifications due to the \cp-odd component of the Higgs--top-quark coupling are the STXS measurements in three Higgs-$p_T$ bins of the $pp\to ZH$, $H\to b\bar{b}$ channel, as well as four  $\Delta\Phi_{j_1j_2}$ bins of the $pp\to H + 2j$, $H\to \gamma\gamma$ channel, both provided by ATLAS. The modifications of the $\Delta\Phi_{j_1j_2}$ distribution in $pp\to H + 2j$ are very small compared to current experimental and theoretical uncertainties, see \cref{sec:xs_fits} and \cref{app:ggH2j}. The impact of these effects on the fit to current data is therefore negligible.

\begin{figure}[!tb]
  \centering
\includegraphics[width=0.49\textwidth]{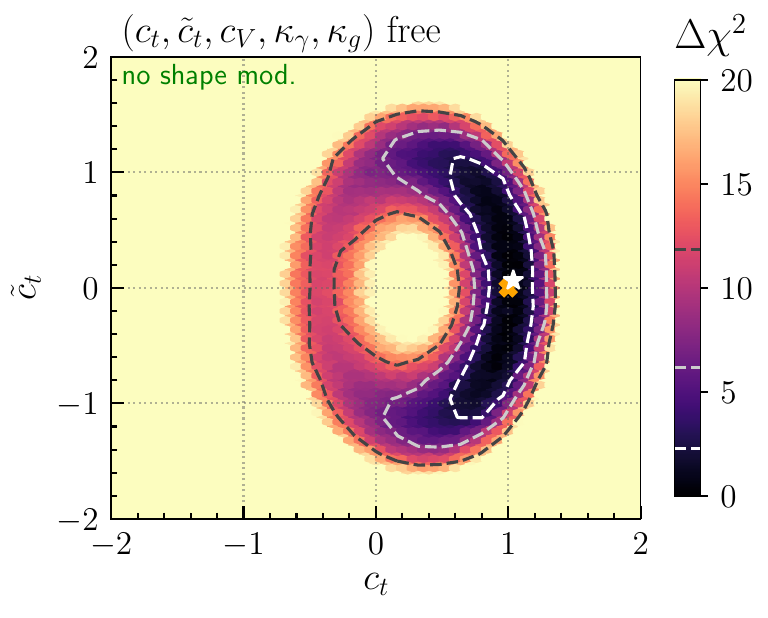}\hfill
\includegraphics[width=0.49\textwidth]{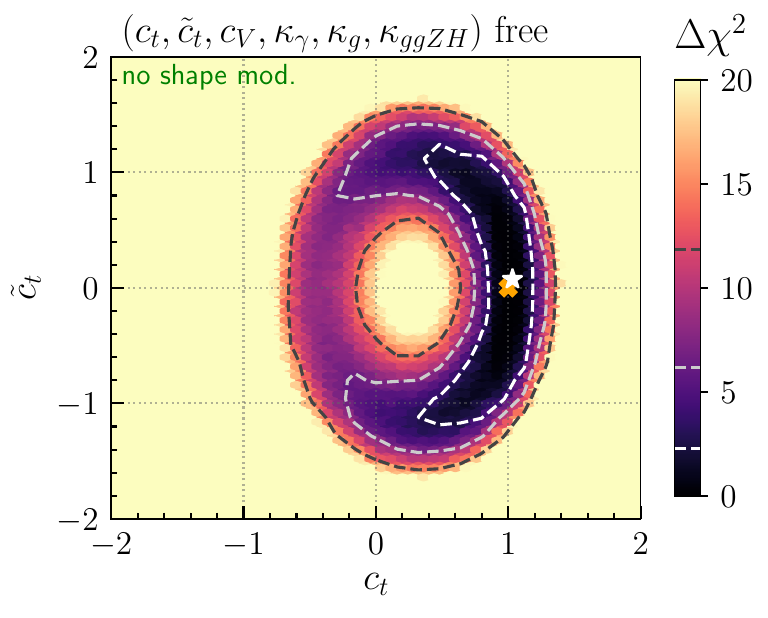}
\caption{Fit results in the (\ct, \cttilde) plane if Higgs-$p_T$-shape modifications in $gg\to ZH$ due to a modified top-Yukawa coupling are ignored. Results are shown for the 5D parametrization (\emph{left panel}) and for a parametrization where the $gg\to ZH$ cross section is additionally modified by an individual fit parameter, $\kggzh$ (\emph{right panel}). The color corresponds to the profile $\Delta\chi^2$ of the global fit, and the $1\,\sigma$, $2\,\sigma$ and $3\,\sigma$ confidence regions are shown as white, light-gray and dark-gray dashed contours, respectively. The best-fit point and the SM case are marked by a white star and an orange cross, respectively.
}
\label{fig:ctcttilde_shapecompare}
\end{figure}

In order to assess the impact of the Higgs-$p_T$-shape modifications in the $pp\to ZH$, $H\to b\bar{b}$ channel, we repeated the fit in the 5D parametrization -- (\ct, \cttilde, \cv, \kg, \kgamma) free -- while ignoring the modifications of the kinematic Higgs boson properties
that could be induced by \cp-violating effects.\footnote{Technically, the predicted signal strength in each $pp\to ZH, H\to b\bar{b}$ $p_T$-bin is set to the signal strength of the inclusive $pp\to ZH$ rate.} In \cref{fig:ctcttilde_shapecompare} we show the fit result in the (\ct, \cttilde) plane using the full observable set without these kinematic shape modifications (\emph{left panel}). Comparing these results with the result when taking these effects into account, see \cref{fig:ctcttilde}
(\emph{bottom right panel}), we find the impact of the additional information provided through the current kinematic shape measurements to be moderate, but clearly visible. The impact is most pronounced in the regions of large $|\tilde{c}_t|$ and small \ct, as well as in the negative \ct region at small \cttilde\ values. This can be seen, for instance, by comparing the $2\,\sigma$ and $3\,\sigma$ contours. In particular, if the shape modifications are neglected, the $3\,\sigma$ allowed region closes in the negative \ct parameter space.  We conclude that accounting for the Higgs-$p_T$ modifications in the $gg\to ZH$ process, even at the current stage, already helps to constrain the parameter space. We therefore strongly encourage the ATLAS and CMS collaborations to (continue to) provide such measurements, e.g.~by employing the STXS framework.

So far, the discussion and results were based on the assumption that the $gg\to ZH$ cross section can be derived entirely from the modified Higgs--top-quark and Higgs--vector-boson interactions, i.e.~from the parameters \ct, \cttilde and \cv. In other words, genuine contributions to $gg\to ZH$ from higher-dimensional operators were assumed to be absent. If we relax this assumption, both the inclusive and Higgs-$p_T$ differential cross sections can be altered by new operators. While a detailed study of these effects would require to consider all relevant SMEFT operators\footnote{At dimension-six, three operators directly contribute to $gg\to ZH$ production~\cite{Bylund:2016phk}. In addition, there are three dimension-six operators modifying the Higgs $Z$-boson coupling.} which is beyond the scope of the present work, we nevertheless assess the possible effects of additional loop contributions to $gg\to ZH$ production using a single scale factor ($\kappa$) parametrization. For this purpose, we consider higher-dimensional operator contributions that --- in combination with the modifications from our model parameters \ct, \cttilde and \cv\ --- change the inclusive $gg\to ZH$ cross section by $\kappa_{ggZH}^2$ with respect to the SM prediction, see \cref{eq:mu_ggZH}. We can then treat $\kappa_{ggZH}$ as an additional free floating parameter, which decorrelates the $gg \to ZH$ cross section from the other fit parameters. However in this approach we have to neglect Higgs-$p_T$-shape modifications in $gg \to ZH$, as this would require knowledge of all underlying relevant operators in the Lagrangian.

The resulting constraints in the (\ct, \cttilde) parameter plane for this fit are displayed in the right panel of \cref{fig:ctcttilde_shapecompare}. We find these to be slightly weaker than those from the previous parametrization, \cref{fig:ctcttilde_shapecompare} (\emph{left panel}), i.e.~when performing the fit without the additional freedom in $gg\to ZH$ and without accounting for the Higgs-$p_T$-shape modifications. From this we conclude that, in the model parametrization(s) considered here, both the inclusive and Higgs-$p_T$ differential rate measurements of the $gg\to ZH$ process have a sizable impact on the allowed parameter space.\footnote{As noted earlier, in an inclusive $ZH$ rate measurement, the $gg\to ZH$ rate modification can be partially compensated by modifications of the $q\bar{q} \to ZH$ rate if $c_V$ is treated as a free parameter in the fit. In contrast, if $c_V$ were fixed, the inclusive $gg\to  ZH$ measurement would have an even larger impact.} 


\subsection{Top quark associated Higgs production}
\label{sec:res_ttH}

 We now discuss the rates for Higgs production in association with top quarks. In the SM the dominant process is $pp\to t\bar{t}H$ with a $13\tev$ cross section of $\sim 0.5~\mathrm{pb}$, while the rates for  Higgs production in association with a single top quark, $pp\to tH$ and $pp\to tWH$, are smaller by roughly a factor of $7$ and $33$, respectively. However this picture can be strongly affected by the presence of a non-zero \cp-odd component. Experimental measurements of top-quark and Higgs associated production have been so far inclusive in the Higgs boson $p_T$, as well as in all three processes, due to the limited size of the LHC dataset. The measured rate is then composed of $t\bar{t}H$, $tH$ and $tWH$ events. The $tH$ contribution is sometimes subtracted assuming the SM prediction for the $tH$ rate.
Unfortunately, information about the acceptance of $t\bar{t}H$, $tH$ and $tWH$ events is rarely provided by the experiments. If no such information is given for a measurement, we have to make assumptions about the relative contributions of these processes. In these cases, we assumed equal acceptances for $t\bar{t}H$ and $tWH$ events. The acceptance of $tH$ events is also assumed to be equal to the $t\bar{t}H$ acceptance, except in measurements requiring two leptons originating from the Higgs production process, where the acceptance of $tH$ events is set to zero. Furthermore, we assume all acceptances to be independent of the \cp-nature of the Higgs--top-quark coupling. More details and explanations are given in \cref{app:expinput}.

While detailed information on the $tH$ and $tWH$ signal acceptance may appear irrelevant for the SM, it is very relevant when testing models where these processes can be strongly enhanced. The impact of our assumptions on the signal acceptances on the fit is demonstrated in \cref{app:expinput},  where we show fit results obtained for a variation of the $tH$ acceptance in all Higgs measurements. The associated uncertainty of this assumption on the fit is sizable in parameter regions of enhanced $tH$ production. We therefore strongly encourage the experimental collaborations to publicly provide more detailed information about the $t\bar{t}H$, $tWH$ and $tH$ signal acceptances in future measurements. In fact, in case experimental analyses feature significantly different acceptances for these processes, it should be possible to determine the rate of each process separately in a global fit including several measurements of top quark associated Higgs production, which in turn can constrain the \cp-character of the Higgs--top-quark interaction.

\begin{figure}[!thbp]
  \centering
\includegraphics[width=0.49\textwidth]{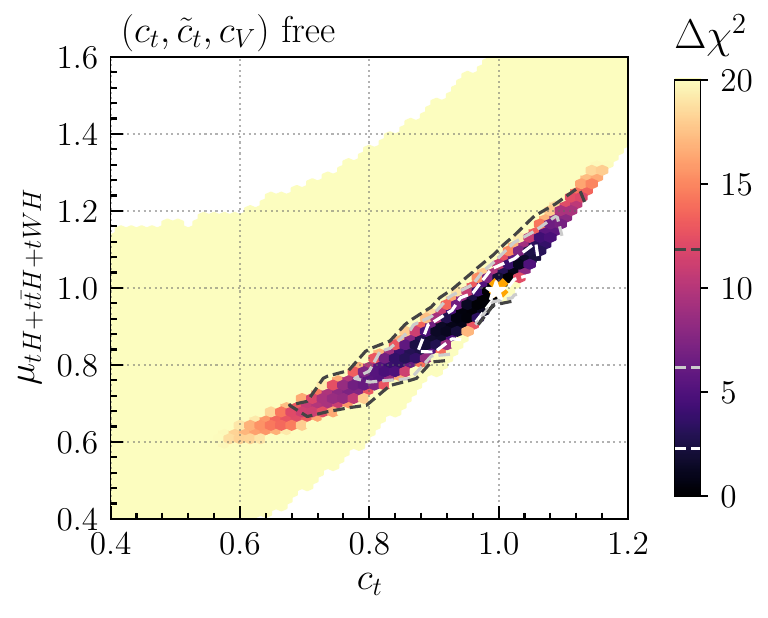}\hfill
\includegraphics[width=0.49\textwidth]{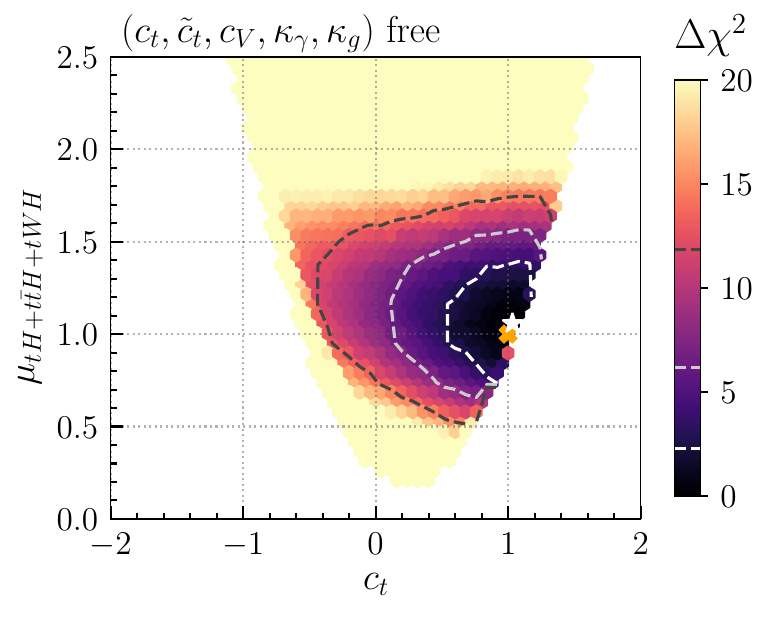}
\includegraphics[width=0.49\textwidth]{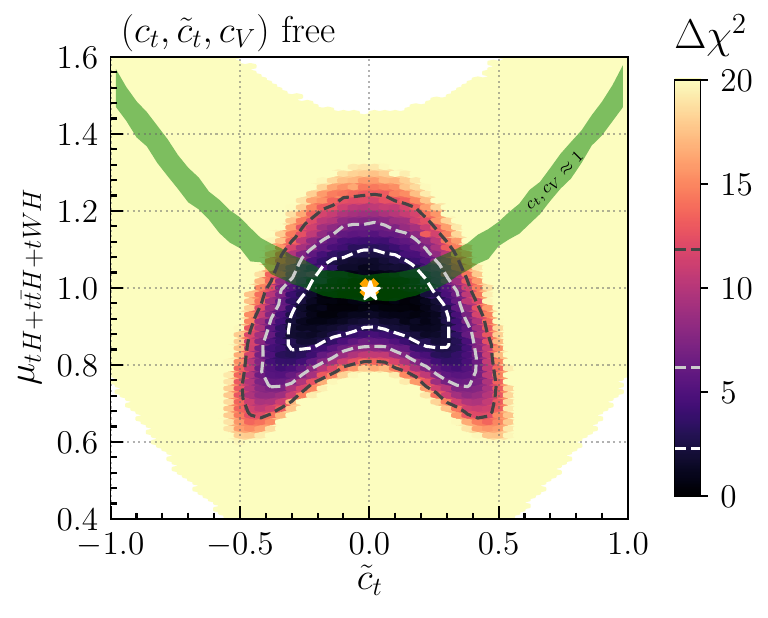}\hfill
\includegraphics[width=0.49\textwidth]{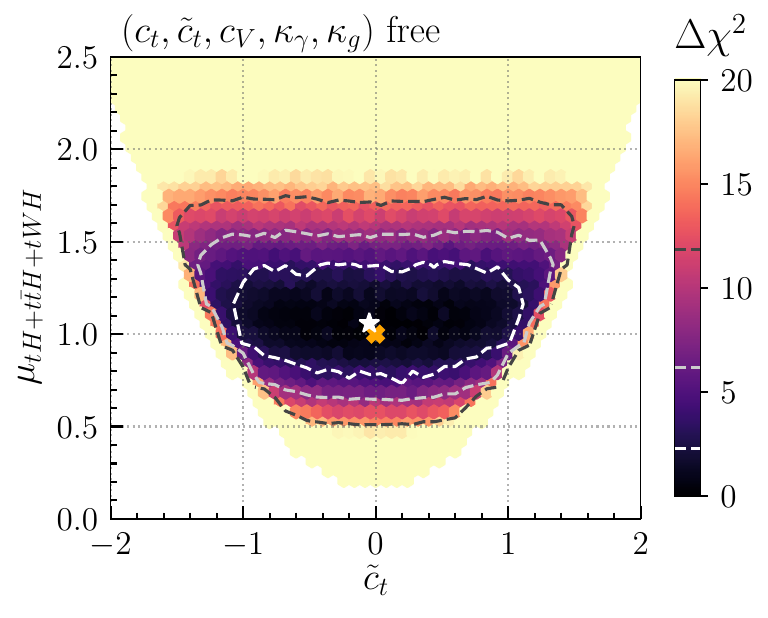}
\caption{Predicted signal strength for combined $t\bar{t}H+tH+tWH$ production, $\mu_{t\bar{t}H+tH+tWH}$, in dependence of $c_t$ (\emph{top panels}) and $\tilde{c}_t$ (\emph{bottom panels}) for the 3D parametrization (\emph{left panels}) and the 5D parametrization (\emph{right panels}). The color corresponds to the profile $\Delta\chi^2$ of the global fit, and the $1\,\sigma$, $2\,\sigma$ and $3\,\sigma$ confidence regions are shown as white, light-gray and dark-gray dashed contours, respectively. The best-fit point and the SM case are marked by a white star and an orange cross, respectively. The green band in the bottom left panel indicates the parameter points with $c_V$ and $c_t$ being within $2\%$ of the SM prediction. Note the larger scale required to display the plots in the right panels.}
\label{fig:mu_ttHtH}
\end{figure}

The combined $t\bar{t}H+tH + tWH$ signal strength is shown in \cref{fig:mu_ttHtH} in dependence of \ct (\emph{top panels}) and \cttilde  (\emph{bottom panels}) for the 3D parametrization -- (\ct, \cttilde, \cv) free -- (\emph{left panels}) and 5D parametrization -- (\ct, \cttilde, \cv, \kg, \kgamma) free -- (\emph{right panels, note the larger scale required to display the plots}). We find the $2\,\sigma$ allowed ranges to be
\begin{align}
  \mu_{t\bar{t}H+tH+tWH} \in [0.76, 1.19]& \qquad \text{in~the~3D parametrization -- (\ct, \cttilde, \cv) free}, \\
  \mu_{t\bar{t}H+tH+tWH} \in [0.64,1.56]& \qquad \text{in~the~5D parametrization -- (\ct, \cttilde, \cv, \kg, \kgamma) free}.
\end{align}
In the 3D parametrization\footnote{The phenomenology of the $t\bar{t}H$, $tH$ and $tWH$ production modes in the 2D and 4D parametrization (not shown here) is qualitatively very similar to the behavior found in the 3D parametrization.} the constraints on $\mu_{t\bar{t}H+tH+tWH}$ are mainly induced through model correlations. Measurements of Higgs production in gluon fusion and the $H\to \gamma\gamma$ decay rate strongly constrain the model parameters \ct and \cttilde, which, in turn, restricts the allowed range for $\mu_{t\bar{t}H+tH+tWH}$. From \cref{fig:mu_ttHtH} (\emph{top left panel}) we see that, in this parametrization, the allowed values of the combined signal strength when $c_t \ge 1$ can be approximated by $\mu_{t\bar{t}H+tH+tWH} \approx \ct^2$, as \cttilde is small and a strong $tH$ enhancement is not possible here. For $c_t < 1$, larger values of \cttilde are allowed (compare with~\cref{fig:ctcttilde}). This can also be observed by comparing the areas above ($c_t \gtrsim 1$) and below ($c_t \lesssim 1$) the green band in the \emph{bottom left panel} of  \cref{fig:mu_ttHtH}. Thus an enhancement in $tH$ production leads to $\mu_{t\bar{t}H+tH+tWH}$ values slightly above the naive $c_t^2$ expectation.  Still, in the $c_t < 1$ region, the combined signal strength cannot significantly exceed the SM expectation ($\mu_{t\bar{t}H+tH+tWH} = 1$).

In contrast, in the 5D parametrization, where the gluon fusion cross section and the $H\to \gamma\gamma$ decay rate are governed by individual fit parameters, the signal strength is much less constrained (we stress again that the scale in the right panels is enlarged as compared to the left panels), due to the absence of correlations with the sensitive gluon fusion and  $H\to \gamma\gamma$ observables.  Here, dedicated measurements of top quark associated Higgs production directly restrict the allowed range of $\mu_{t\bar{t}H+tH+tWH}$. This, in turn, yields constraints on the underlying model parameters \ct and \cttilde, in an important interplay with measurements sensitive to $gg\to ZH$, as explained in \cref{sec:res_ctcttilde}. It is interesting to note that the correlation between $c_t$ and $\mu_{t\bar{t}H+tH+tWH}$ is much weaker in the 5D parametrization, \cref{fig:mu_ttHtH} (\emph{upper right panel}), than in the 3D parametrization (\emph{upper left panel}). Here, even for $c_t < 1$, an enhancement of the $t\bar{t}H+tH+tWH$ inclusive cross section by more than $+50\%$ is possible at the $2\,\sigma$ level.

\begin{figure}[!thbp]
  \centering
\includegraphics[width=0.49\textwidth]{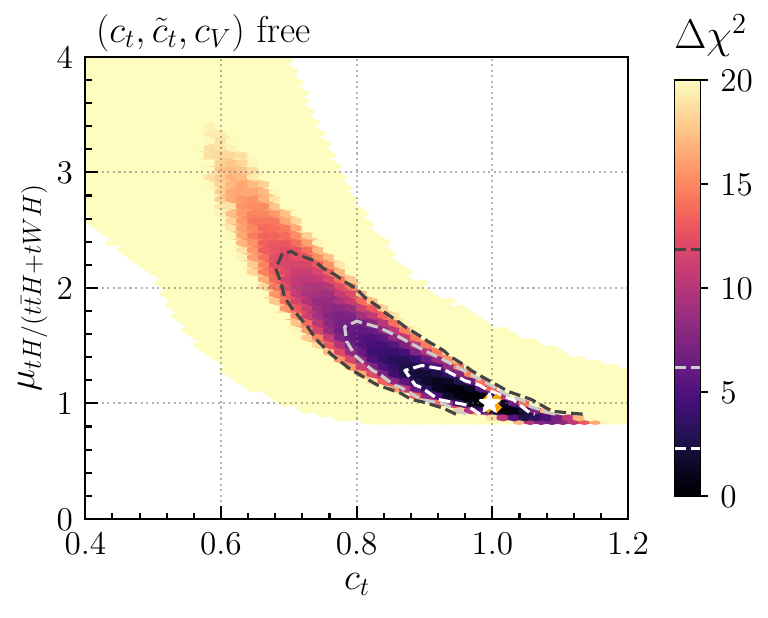}\hfill
\includegraphics[width=0.49\textwidth]{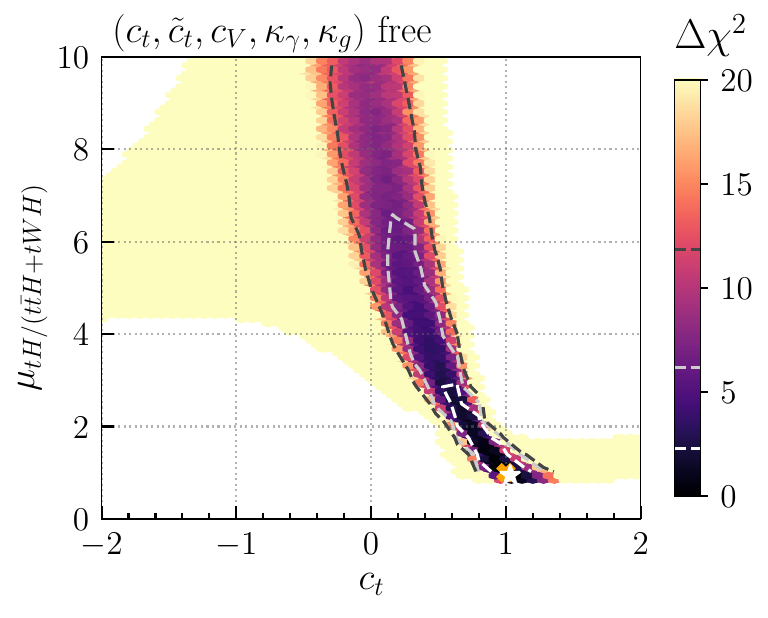}
\includegraphics[width=0.49\textwidth]{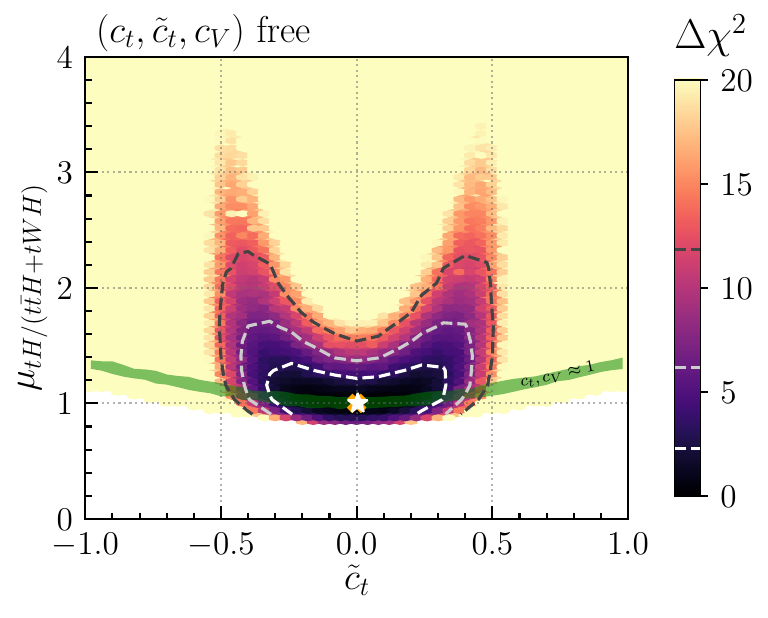}\hfill
\includegraphics[width=0.49\textwidth]{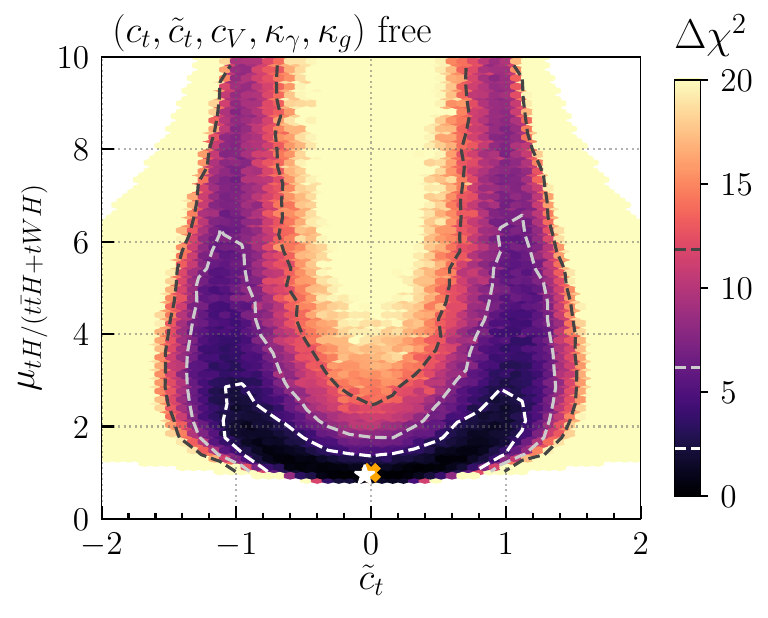}
\caption{Predicted ratio between the SM-normalized rates of $tH$ and $t\bar{t}H$ production, $\mu_{tH}/\mu_{t\bar{t}H}$, in dependence of $c_t$ (\emph{top panel}) and $\tilde{c}_t$ (\emph{bottom panel}) for the 3D parametrization (\emph{left panels}) and the 5D parametrization (\emph{right panels}). The color corresponds to the profile $\Delta\chi^2$ of the global fit, and the $1\,\sigma$, $2\,\sigma$ and $3\,\sigma$ confidence regions are shown as white, light-gray and dark-gray dashed contours, respectively. The best-fit point and the SM case are marked by a white star and an orange cross, respectively. The green band in the bottom left panel indicates the parameter points with $c_V$ and $c_t$ being within $2\%$ of the SM prediction. Note the larger scale required to display the plots in the right panels.}
\label{fig:mu_tHoverttH}
\end{figure}

Nevertheless, from the lower panels in \cref{fig:mu_ttHtH} we can conclude that, independently of the specific model parametrization, the measurement of the combined inclusive signal strength of top quark associated production does not feature sensitive discrimination power regarding the \cp-odd Higgs--top-quark coupling \cttilde. Instead, one may aim at a separate determination of the $t\bar{t}H+tWH$ and $tH$ cross sections. We will discuss some strategies for this in \cref{sec:prospects}. In order to illustrate the possible impact of separate experimental information on the $t\bar{t}H+tWH$ and the $tH$ cross section, we show in \cref{fig:mu_tHoverttH} the corresponding plots where the SM-normalized ratio of the $tH$ rate over the $t\bar{t}H+tWH$ rate, $\mu_{tH/(t\bar{t}H+tWH)}$, is displayed (see \cref{eq:mutthth_ratio}). This quantity is shown in \cref{fig:mu_tHoverttH} in dependence of \ct (\emph{top panels}) and \cttilde (\emph{bottom panels}), for the same two model parametrizations as above (note the larger scale required to display the plots in the right panels). We find that the cross section ratio can be significantly enhanced with respect to the SM. Specifically, within the $2\,\sigma$ region, we find
\begin{align}
  \mu_{tH/(t\bar{t}H+tWH)} \in [0.8, 1.7]& \qquad \text{in~the~3D parametrization -- (\ct, \cttilde, \cv) free}, \\
  \mu_{tH/(t\bar{t}H+tWH)} \in [0.8, 6.7]& \qquad \text{in~the~5D parametrization -- (\ct, \cttilde, \cv, \kg, \kgamma) free}.
\end{align}
As can be seen in \cref{fig:mu_tHoverttH}, this enhancement increases with decreasing \ct and increasing $|\cttilde|$, and is therefore a sensitive probe of the \cp-nature of the Higgs--top-quark coupling.

\begin{figure}[!t]
  \centering
\includegraphics[width=0.49\textwidth]{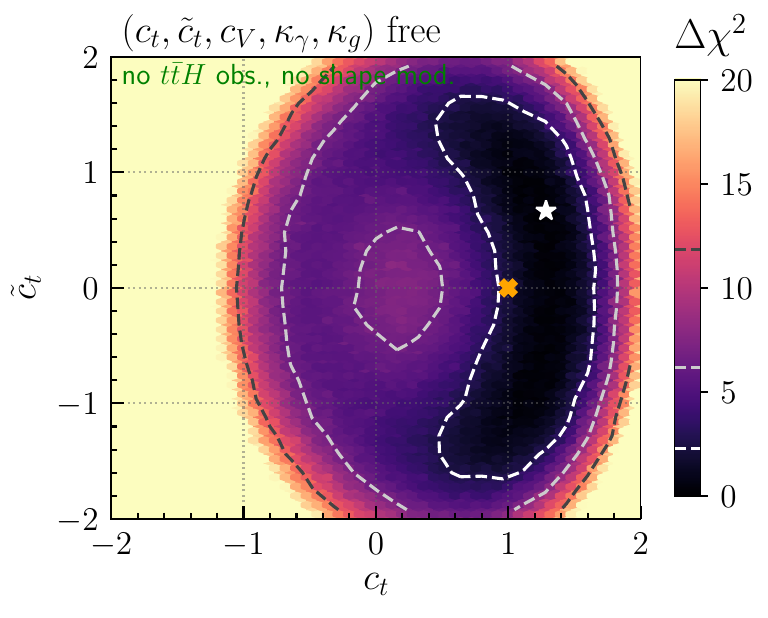}\hfill
\includegraphics[width=0.49\textwidth]{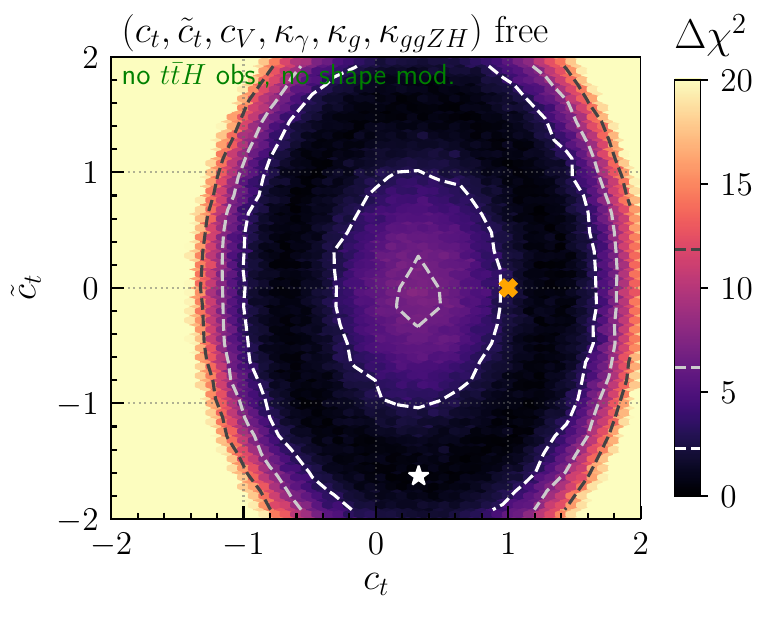}
\caption{Fit results in the (\ct, \cttilde) plane if Higgs-$p_T$-shape modifications in $gg\to ZH$ due to \cp-violation are ignored and measurements from dedicated $t\bar{t}H$ analyses are excluded from the fit. Results are shown for the 5D parametrization (\emph{left panel}) and for a parametrization where the $gg\to ZH$ cross section is additionally modified by an individual fit parameter, $\kggzh$ (\emph{right panel}). The color corresponds to the profile $\Delta\chi^2$ of the global fit, and the $1\,\sigma$, $2\,\sigma$ and $3\,\sigma$ confidence regions are shown as white, light-gray and dark-gray dashed contours, respectively. The best-fit point and the SM case are marked by a white star and an orange cross, respectively.
}
\label{fig:ctcttilde_ttHcompare}
\end{figure}

In order to assess the impact of the dedicated experimental analyses targeting top quark associated Higgs production on the parameter space, we repeated the fit in the 5D parametrization -- (\ct, \cttilde, \cv, \kg, \kgamma) free -- with a reduced observable set in \texttt{HiggsSignals} which excludes all specific $t\bar{t}H$  analyses (see \cref{app:expinput} for details).\footnote{In this fit we furthermore ignored Higgs-$p_T$-shape modifications in $pp \to ZH$, as done in \cref{sec:res_ppZH}.} The fit result for the (\ct, \cttilde) parameter plane is given in the left panel of \Cref{fig:ctcttilde_ttHcompare}. The impact of the $t\bar{t}H$ measurements can be clearly seen in the comparison with the left panel of \Cref{fig:ctcttilde_shapecompare}, which includes these measurements. Obviously, leaving out the information from the dedicated $t\bar{t}H$ measurements largely opens up the parameter space in the (\ct, \cttilde) plane for this model parametrization. In the fit without the $t\bar{t}H$ measurements the favored region still aligns with the contour ellipses of constant $\mu_{t\bar{t}H+tH+tWH}$, see \cref{fig:tH-ttH} (\emph{bottom left panel}), where --- in contrast to the fit with all observables included, \Cref{fig:ctcttilde_shapecompare} (\emph{left)} --- the $2\,\sigma$ allowed parameter space covers the whole ellipse, i.e.~it closes in the negative \ct region. Within the $2\,\sigma$ region we find the combined rate and the ratio of $tH$ and $t\bar{t}H$ production to be
\begin{align}
  \mu_{t\bar{t}H+tH+tWH} \in [0.3, 3.5] \qquad\mbox{and}\qquad   \mu_{tH/(t\bar{t}H+tWH)} \in [0.8 ,38.0].
\end{align}
The fit still favors positive values of $c_t$ at the $1\,\sigma$ level. This constraint originates from $ZH$-sensitive measurements, as the $gg\to ZH$ rate is enhanced for decreasing $c_t$ in this parametrization, see \cref{fig:ggZH}.

In order to explicitly prove this influence of the $gg\to ZH$ process, we repeated the fit to the same observable set (i.e.\ without dedicated $t\bar{t}H$ analyses) in the six-dimensional parametrization from the preceding subsection, treating \ct, \cttilde, \cv, \kg, \kgamma and \kggzh as free parameters, i.e.\ allowing the $gg\to ZH$ rate to be freely adjusted by $\kappa_{ggZH}^2$. The result in the (\ct, \cttilde) parameter plane is shown in \Cref{fig:ctcttilde_ttHcompare} (\emph{right panel}). Here, there is no sign preference in \ct, and the likelihood is completely flat along the entire ellipse of constant $\mu_{t\bar{t}H+tH+tWH}$, as anticipated. The fact that $\mu_{t\bar{t}H+tH+tWH}\simeq 1$ is still preferred --- albeit at a very low significance
--- is due to inclusive, non-$t\bar{t}H$ specific, Higgs rate measurements included in the fit, as well as the inclusion of the LHC Run-1 ATLAS and CMS combined measurements. Both types of measurements maintain some sensitivity to top quark associated Higgs production processes.


\subsection{Discussion}
\label{sec:discussion}

As shown in \cref{sec:res_ttH}, top-quark associated Higgs production plays a key role in constraining the \cp-nature of the Higgs--top-quark interaction. The corresponding constraints can be tightened by taking into account not only rate information, but also kinematic information which partly incorporate \cp-odd observables, as done in the recent ATLAS and CMS analyses~\cite{Sirunyan:2020sum,Aad:2020ivc}. Due to differences in the used models, a comparison of these studies to our results is difficult. The CMS study constrains the \cp-violating phase to be lower than 55 degrees at the $2\,\sigma$ level under the assumption that all Higgs production modes (apart from top-quark associated Higgs production) are constrained to their SM predictions. In our 5D parametrization, this would imply that $\cv = \kg = \kgamma = 1$ is assumed. For comparison, in our results in the 5D parametrization where \cv, \kg and \kgamma\ are treated as free parameters, we obtain an upper limit of 72 degrees (see \cref{sec:res_ctcttilde}). The ATLAS study finds a limit of 43 degrees regardless whether \kg and \kgamma are treated as free parameters or not. If \kg and \kgamma are calculated as a function of the \cp-violating phase, the result could be compared to our result of a $22.5$ degree upper limit in the 2D parameterization. The comparability of the results is less obvious for the case that \kg and \kgamma are treated as free parameters. Since the Monte-Carlo samples used in Ref.~\cite{Aad:2020ivc} are generated at the NLO level employing the ``Higgs Characterization Model'', top-quark associated Higgs production cannot be regarded as being independent of Higgs production via gluon fusion, since top-quark associated Higgs production depends on the Higgs--gluon--gluon operator at NLO (see the discussion in \cref{sec:xs_fits_calc} for more details). Therefore, the results of Ref.~\cite{Aad:2020ivc} can only be compared to our result in an approximate way.

As discussed in \cref{sec:intro}, results as the ones presented in Ref.~\cite{Sirunyan:2020sum,Aad:2020ivc} are in any case difficult to integrate into global fits without further experimental information, since the interpretation of these results depends strongly on the exact model used for deriving the constraints. Also, the resulting limit on \cp-odd couplings from such analyses cannot be directly applied to other models than the model studied in the experimental analyses.

Unfolded differential cross section or STXS measurements can be used to lower the model-dependence. But also for such measurements, the sensitivity of the kinematic distributions to the \cp-character of the Higgs boson can affect the efficiency of the event selection, thus introducing some model-dependence (see \cref{sec:prospects}). In the particular case of the measurement of $t\bar{t}H$ kinematic shapes for \cp\ tests, the subtraction of the $tH$ and $tWH$ background in a model-independent way is delicate, due to the strong dependence of the $tH$ and $tWH$ cross sections on the \cp\ properties of  the Higgs boson.

Total rate measurements are, however, largely model-independent. While total rate measurements for combined top-quark associated Higgs production have already been performed and first shape measurements with the full Run 2 LHC data are on-going, few attempts have been made so far to disentangle $tH$ and $t\bar t H+tWH$ production. However, as discussed above and shown in \cref{fig:mu_tHoverttH}, separate measurements of $tH$ and $t\bar t H+tWH$ production have the potential to constrain further the $(\ct,\cttilde)$ parameter space. While such measurements may not be feasible with current data, they may become possible with upcoming data, as discussed in the next Section.


\section{Future sensitivity to $tH$ production and constraints on \cp}
\label{sec:prospects}


The aim of this section is to evaluate the possibility for separately measuring $tH$ and $t\bar tH+tWH$ production at the LHC with 300 fb$^{-1}$ and at the HL-LHC with 3000 fb$^{-1}$ of data and to discuss the additional constraints brought by this new measurement on the Higgs-\cp\ nature. For this purpose, we evaluate the feasibility of an analysis for the separate measurement of $tH$ and $t\bar tH+tWH$ production that is designed to remain independent of the \cp-character of the top-Yukawa coupling. This allows us to directly compare the measured rates to the theoretical predictions, e.g.\ from the model described in \cref{sec:xs_fits}, and enables their inclusion in a global fit such as the one presented earlier.

The analysis is implemented in \texttt{Delphes}~\cite{deFavereau:2013fsa} and employs a simplified analysis strategy. A corresponding analysis by ATLAS and CMS thus should yield an improvement over the results presented here. The inclusion of the $tH$ signal strength, as proposed here, in a global study of the \cp-properties, along with various other inclusive and differential measurements, would yield the highest sensitivity.

\subsection{Analysis strategy}

Only $H\to\gamma\gamma$ decays are considered in the study, due to their sharp two photon invariant mass ($m_{\gamma\gamma}$) peak signature around the Higgs boson mass ($m_{H}\sim 125$~GeV) at the LHC and equally at the HL-LHC. This allows the subtraction of the background originating from non-Higgs processes, which is typically smoothly falling as a function of $m_{\gamma\gamma}$, with a rather simple fit of the $m_{\gamma\gamma}$ distribution in the experimental data, including a fixed Higgs boson mass. In this work, we assume that the non-Higgs background is subtracted by such a fit. The $H\to\gamma\gamma$ decay channel also benefits from a relatively high decay branching ratio ($\sim 10^{-3}$), e.g. with respect to the four-lepton decay channel ($\sim 10^{-4}$), which allows to apply tighter cuts to select $tH$ events while retaining a sufficient number of events. These properties make $H\to\gamma\gamma$ decays particularly relevant in the context of this phenomenological work, however we encourage the experiments to explore all relevant Higgs boson decay channels.

We consider the splitting of data events in two categories. The first category (1-lepton category) targets leptonic $W$ boson decays in $tH$ events and so requires exactly one electron or muon in the final state. The second category (2-lepton category) targets $t\bar{t}H$ and $tWH$ events in which both $W$ bosons decay leptonically, and thus requires exactly two electrons or muons with opposite sign in the final state. The two categories are orthogonal to each other due to the different lepton multiplicity requirement. The purpose of the 2-lepton category, which is expected to be free of any $tH$ events, is to allow an independent measurement of the $t\bar{t}H+tWH$ production. By fitting the $t\bar{t}H+tWH$ event yields in the 1-lepton and 2-lepton categories simultaneously, the number of $t\bar{t}H+tWH$ events entering the 1-lepton category, which originates mostly from events including one $W$ boson decaying to leptons and the other one to hadrons, can then be predicted accurately.

A particular attention will be given to the robustness of the analysis with respect to the Higgs boson \cp\ properties, which is an essential feature to be able to include the results in a global \cp\ fit such as the one presented earlier in this paper.


\subsection{Technical setup}

Only Higgs boson production modes predicted by the SM which include at least one `prompt' lepton originating from a $Z$ or $W$ boson decay, i.e. $t\bar{t}H$, $tWH$, $tH$ (neglecting the $tH$ $s$-channel contribution) and $VH$ ($V=Z,W$, including the $gg\to ZH$ contribution) production, are considered in the following. This is justified by the fact that the ATLAS and CMS detectors allow for a very strong rejection of leptons which have a non-prompt origin (e.g.~bottom or charm hadron decays, electrons arising from a photon conversion, jets misidentified as electrons, or muons produced from in-flight pion or kaon decays) such that the resulting background can usually be neglected in data analyses after requiring exactly one electron or muon (as in the 1-lepton category) or exactly two electrons or muons with opposite sign (as in the 2-lepton category) in the final state.

$t\bar{t}H$, $tWH$, $tH$ and $VH$ events in $\sqrt{s}=13$ TeV proton-proton collisions are simulated using the same setup as described in \cref{sec:xs_fits_calc}. Monte Carlo (MC) events are produced under four different \cp\ hypotheses for the Higgs--top-quark coupling, following the notation introduced in \cref{sec:model}:
\begin{itemize}
    \item SM Higgs boson: $\ct=\cv=1$, $\cttilde=0$,
    \item pure \cp-odd state: $\cttilde=\cv=1$, $\ct=0$,
    \item \cp-mixed state, benchmark 1: $\ct=\cttilde=0.8$, $\cv=1$,
    \item \cp-mixed state, benchmark 2: $\ct=0.5$, $\cttilde=1.2$, $\cv=0.95$.
\end{itemize}
We assume the $H\to\gamma\gamma$ decay branching ratio to equal the SM prediction for all four \cp\ hypotheses ($\kgamma=1$), i.e.~allowing the existence of higher-dimensional BSM  operators that couple the Higgs boson directly to photons for the non-SM \cp\ cases (see \cref{eq:dim5}). Since only $H\to\gamma\gamma$ decays are considered in this work and the impact of the BSM operators on the $H\to\gamma\gamma$ decay kinematics relevant for the present analysis are known to be small, a different $H\to\gamma\gamma$ decay branching ratio is however simply equivalent to a rescaling of the integrated luminosity at first order. In addition, the Higgs production rate via gluon fusion is assumed to be equal to the SM prediction for all \cp hypotheses ($\kg=1$). These assumptions are necessary to relax the tight constraints from diagrams where the Higgs--top-quark couplings intervene at the loop-level, see \cref{sec:results}. They require in general the presence of corresponding BSM operators and some non-trivial, accidental cancellations (see \cref{eq:dim5}), with \cg and \cgtilde of the order of $\mathcal{O}(\lesssim 0.5)$ (see \cref{eq:ggH}).

The analyses presented in Refs.~\cite{Sirunyan:2020sum,Aad:2020ivc} exclude a pure \cp-odd top-Yukawa coupling by $3.9\,\sigma$ (ATLAS) and $3.2\,\sigma$ (CMS). The first benchmark of a \cp-mixed state defined above lies within the $1\,\sigma$ region of the five parameter fit presented in \cref{sec:results} [(\ct, \cttilde, \cv, \kgamma, \kg) free]. It corresponds to a \cp-violating phase of $45^\circ$ and an absolute rescaling factor for the top-Yukawa coupling of 0.8 with respect to the SM. The second benchmark of a \cp-mixed state is chosen to be further away from the SM point. It lies within the $2\,\sigma$ region of the five parameter fit presented in \cref{sec:results}.

The SM $tH$ and $t\bar{t}H$ samples are also generated at the next-to-leading order (NLO) in QCD with the \texttt{MSTW2018NLO}~\cite{Martin:2009iq} PDF set and \texttt{Pythia 8} for comparison (see discussion of \cref{sec6:lo_nlo_1} below). It should be kept in mind that the SM Higgs boson scenario is the only one out of the four Higgs boson \cp\ hypotheses considered here which can be generated at NLO+PS with the current implementation of the ``Higgs charaterization model'' when $\kg$ is fixed to unity (see discussion in \cref{sec:xs_fits_calc}).

The simulated events are reconstructed with \texttt{Delphes 3.4.2}~\cite{deFavereau:2013fsa} using either the ATLAS or the HL-LHC configuration card provided in the  \texttt{Delphes} repository. The \texttt{Delphes} software allows for an approximate, very fast simulation of the current ATLAS detector~\cite{PERF-2007-01} response (ATLAS card) or of the future ATLAS or CMS HL-LHC detector response (HL-LHC card)~\cite{Cepeda:2019klc,Atlas:2019qfx}, reconstructing typical objects and quantities used in data analyses and accounting for the limited energy resolution, reconstruction efficiency and acceptance of these objects at the LHC and at the HL-LHC. Additional proton-proton interactions occurring during the same bunch crossing (pileup) are assumed to be subtracted from experimental data by dedicated algorithms and are not included in the MC generation. In all the results which follow, the ATLAS card has been modified to reconstruct hadronic jets using the anti-$k_t$ algorithm~\cite{Cacciari:2008gp} with radius parameter $R=0.4$ instead of $R=0.6$, which is the default value in the configuration card. The value $R=0.4$ is more widely used in recent ATLAS data analyses and has the additional advantage to harmonise the jet definition between the ATLAS and the HL-LHC card. All the other parameters of the ATLAS card are left unchanged. The HL-LHC card was used in a recent review of the opportunities for Higgs physics at the HL-LHC~\cite{Cepeda:2019klc,Atlas:2019qfx} so that it is left unchanged. The same proton-proton center of mass energy of $\sqrt{s}=13$ TeV is used in the LHC and HL-LHC scenario and an integrated luminosity of 300~fb$^{-1}$ (3000~fb$^{-1}$) is assumed at the LHC (HL-LHC). The center of mass energy may be increased to $\sqrt{s}=14$~TeV at the HL-LHC~\cite{Cepeda:2019klc,Atlas:2019qfx}, which would enhance Higgs production cross sections by about 20\% (10\%) for $t\bar{t}H$, $tWH$ and $tH$ ($VH$). The choice to use $\sqrt{s}=13$~TeV for all scenarios is therefore conservative.


\subsection{Event selection}

We start by implementing a simplified version of the typical event preselection used in the latest $t\bar{t}H,~H\to\gamma\gamma$ ATLAS and CMS measurements~\cite{Sirunyan:2020sum,Aad:2020ivc}: all events must include at least two photons with $|\eta|<2.5$ and $p_{T}^{\gamma}$ larger than 35~GeV and 25~GeV. In case of the presence of more than two photons in the event, the two photons with highest $p_{T}$ are selected, forming the Higgs boson candidate. The invariant mass of the two selected photons, $m_{\gamma\gamma}$, must be within the range
[105--160]~GeV, and the leading (subleading) photon in $p_T$ has to satisfy $p_{T}^{\gamma}/m_{\gamma\gamma}>0.35~(0.25)$. At least one hadronic jet identified as originating from the hadronization of a $b$-quark ($b$-jet) is required in both categories, and must satisfy $p_{T}^{b}>25$~GeV and $|\eta|<2.0$, and the missing transverse momentum $p_{T}^{miss}$ (as defined in \texttt{Delphes}) must be higher than 25 GeV. The $t\bar{t}H$ event yield obtained after applying this selection to events reconstructed with the ATLAS card, and requiring in addition at least one electron or muon with $p_{T}^{\ell}>15$ GeV, was compared with a recent ATLAS measurement~\cite{ATLAS-CONF-2019-004}. A good agreement between the two numbers (within $10\%$)  was observed after rescaling the $t\bar{t}H$ event yield obtained in this work to the signal strength fitted by ATLAS ($\mu_{t\bar{t}H}=1.4$).

We then split the events according to the lepton multiplicity: the 1-lepton (2-lepton) category includes at least one (two) electron(s) or muon(s) with $p_{T}^{\ell}>15$~GeV and no additional lepton in the full acceptance. The two-lepton category requires in addition that the two leptons have opposite signs and, in case they have the same flavor ($ee$ or $\mu\mu$), that their invariant mass ($m_{\ell\ell}$) falls outside the $Z$ boson mass range, defined as [80--100]~GeV, preventing potential contamination from $ZH$ events.
The selections described above will be referred to as the \emph{1-lepton preselection} and the \emph{2-lepton selection} in the following.

\begin{figure}[tbp!]
  \centering
\includegraphics[width=0.49\textwidth]{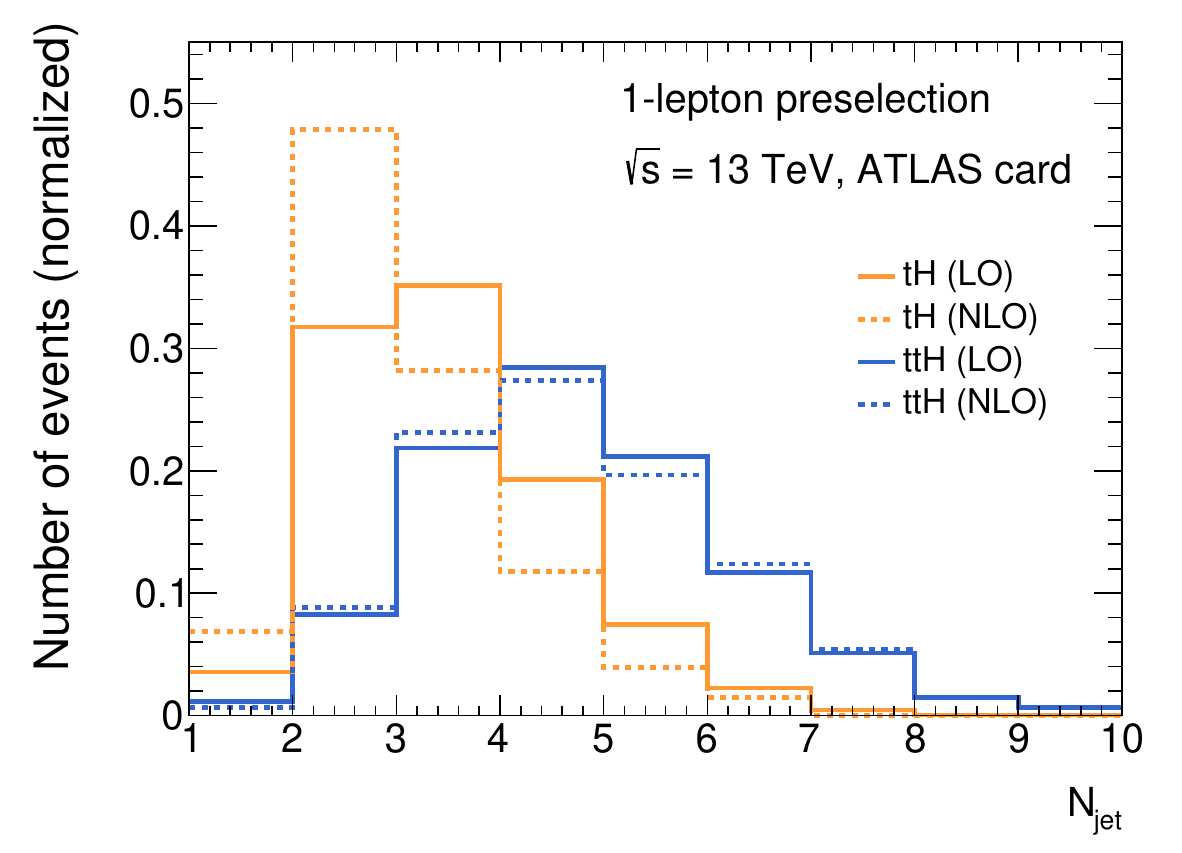}\hfill
\includegraphics[width=0.49\textwidth]{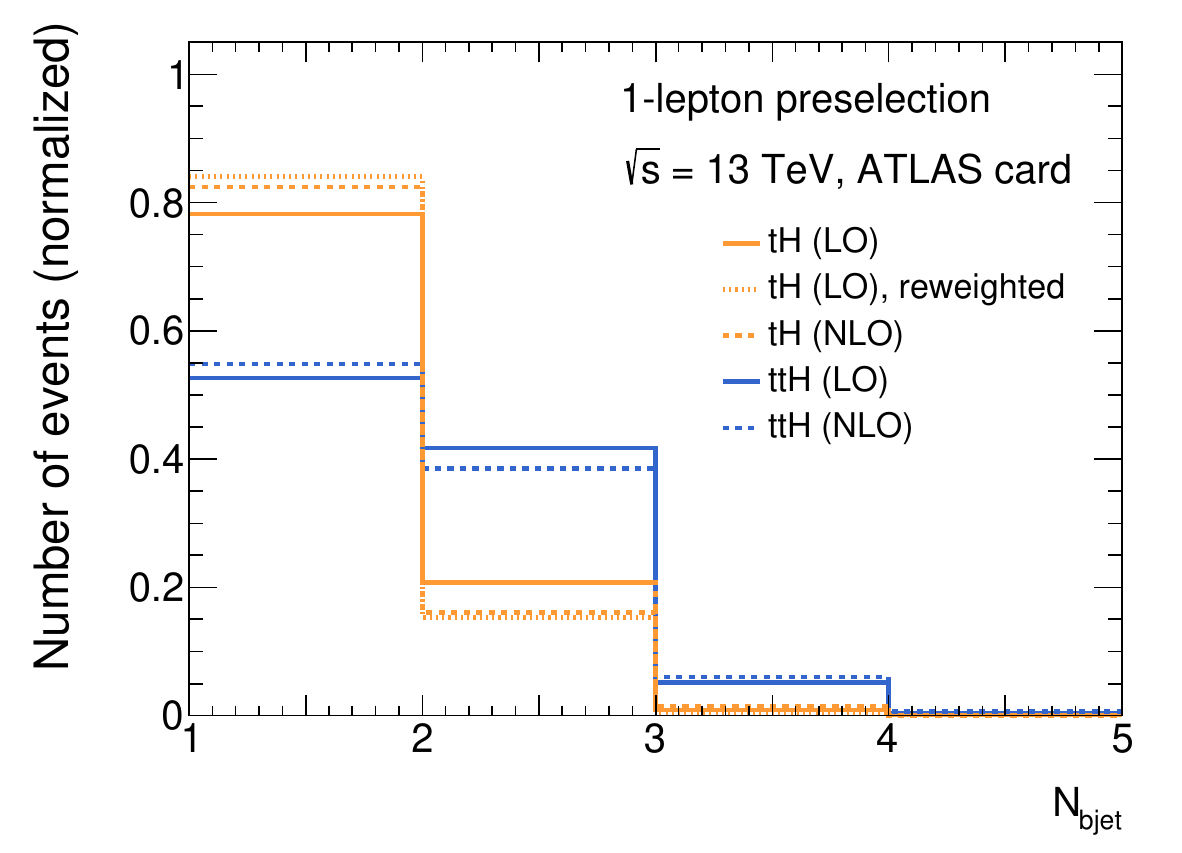}\\
\includegraphics[width=0.49\textwidth]{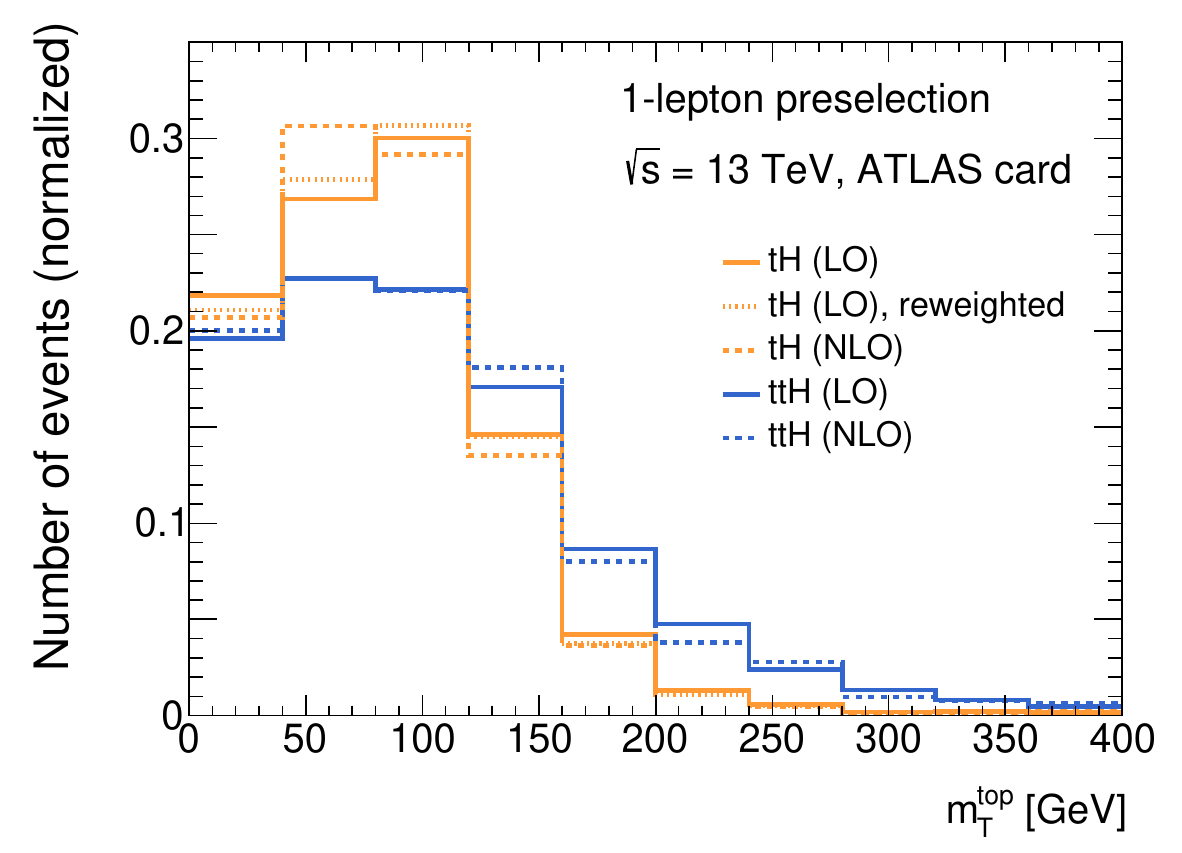}\\
\includegraphics[width=0.49\textwidth]{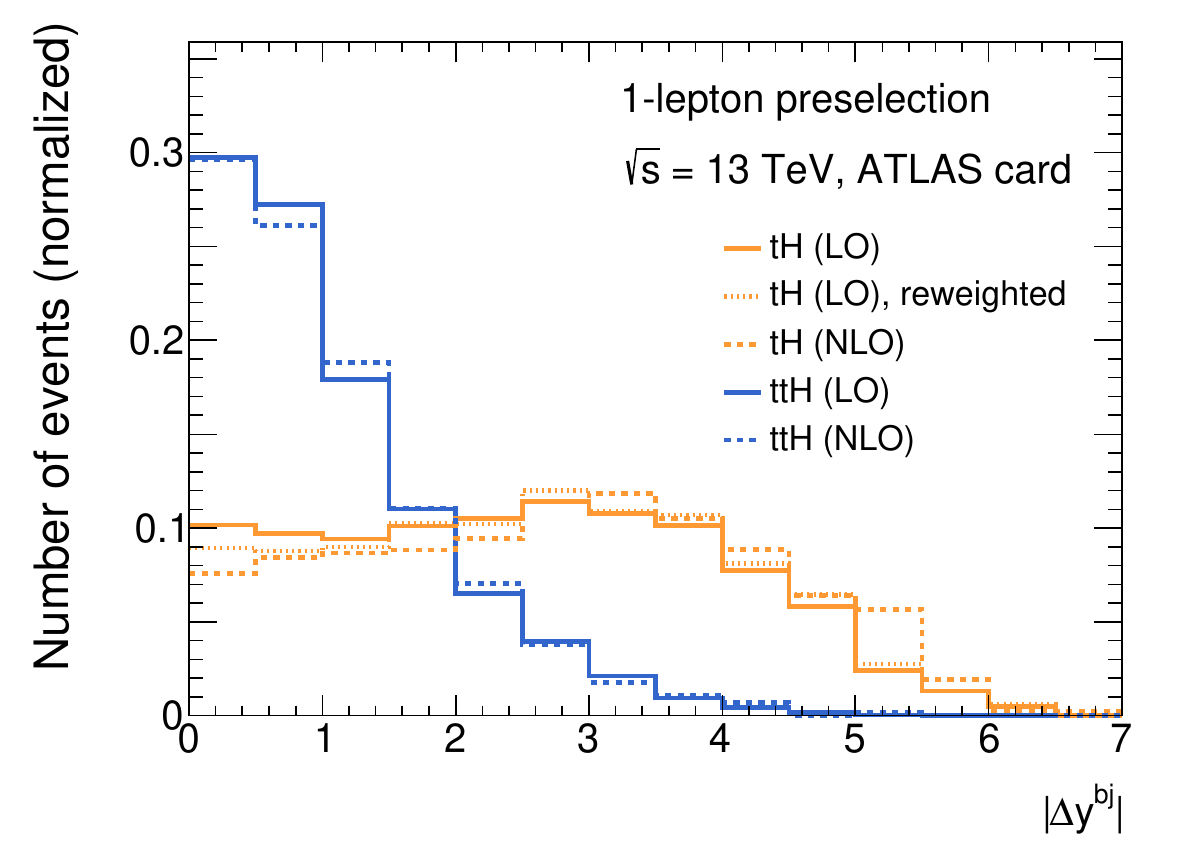}\hfill
\includegraphics[width=0.49\textwidth]{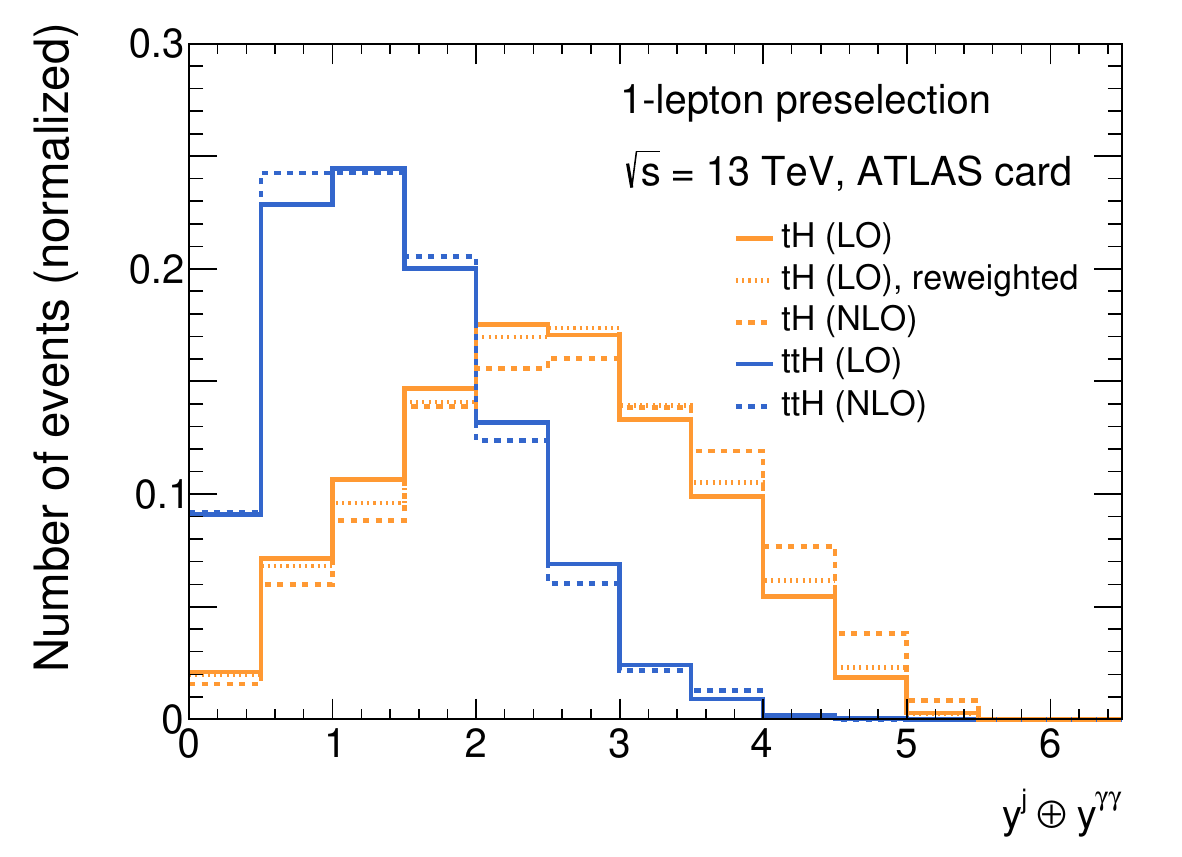}
\caption{Comparison of the shape of (top, left) $N_{jet}$, (top, right) $N_{bjet}$, (middle) $m_{T}^{top}$, (bottom, left) $|\Delta y^{bj}|$ and (bottom, right) $y^{j} \oplus y^{\gamma\gamma}$ at (solid line) LO+PS and (coarse dashed line) NLO+PS in SM (red) $tH$ and (green) $t\bar{t}H$ events passing the \emph{1-lepton preselection}. The shape of $N_{bjet}$, $m_{T}^{top}$, $|\Delta y^{bj}|$ and $y^{j} \oplus y^{\gamma\gamma}$ are also shown for $tH$ events at LO+PS after reweighting the $N_{jet}$ shape to the one observed at NLO+PS (thin dashed line). Events are reconstructed with the \texttt{Delphes} software using the ATLAS card described in the text, similar results are obtained when using the HL-LHC card.}
\label{sec6:lo_nlo_1}
\end{figure}

\begin{figure}[tbp!]
  \centering
\includegraphics[width=0.49\textwidth]{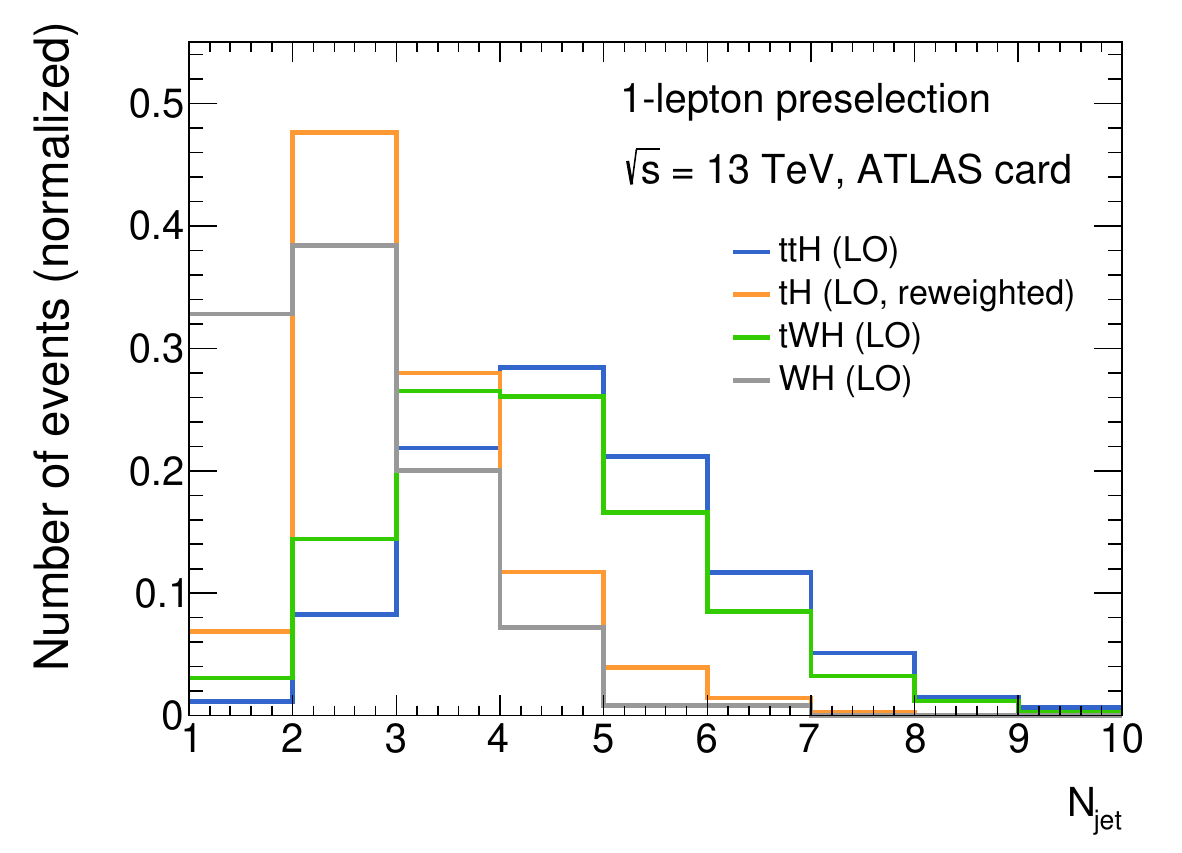}\hfill
\includegraphics[width=0.49\textwidth]{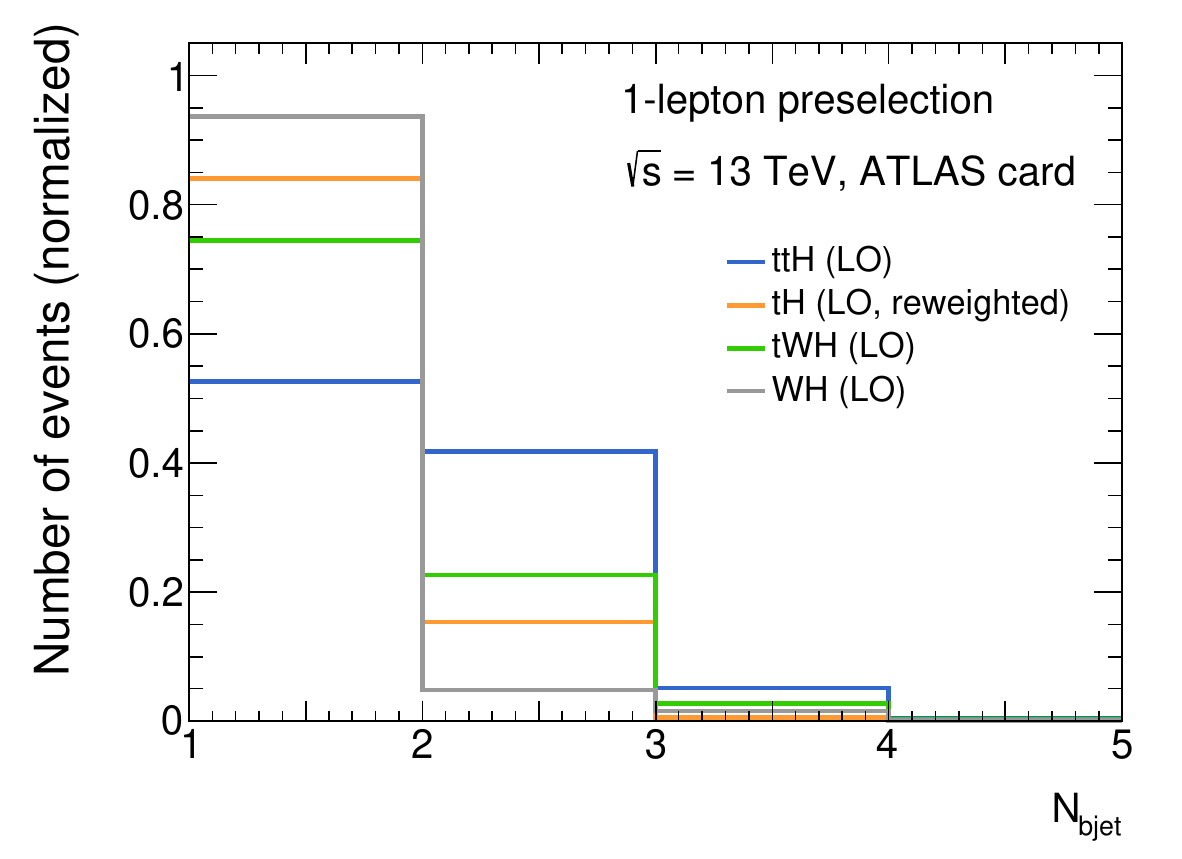}\\
\includegraphics[width=0.49\textwidth]{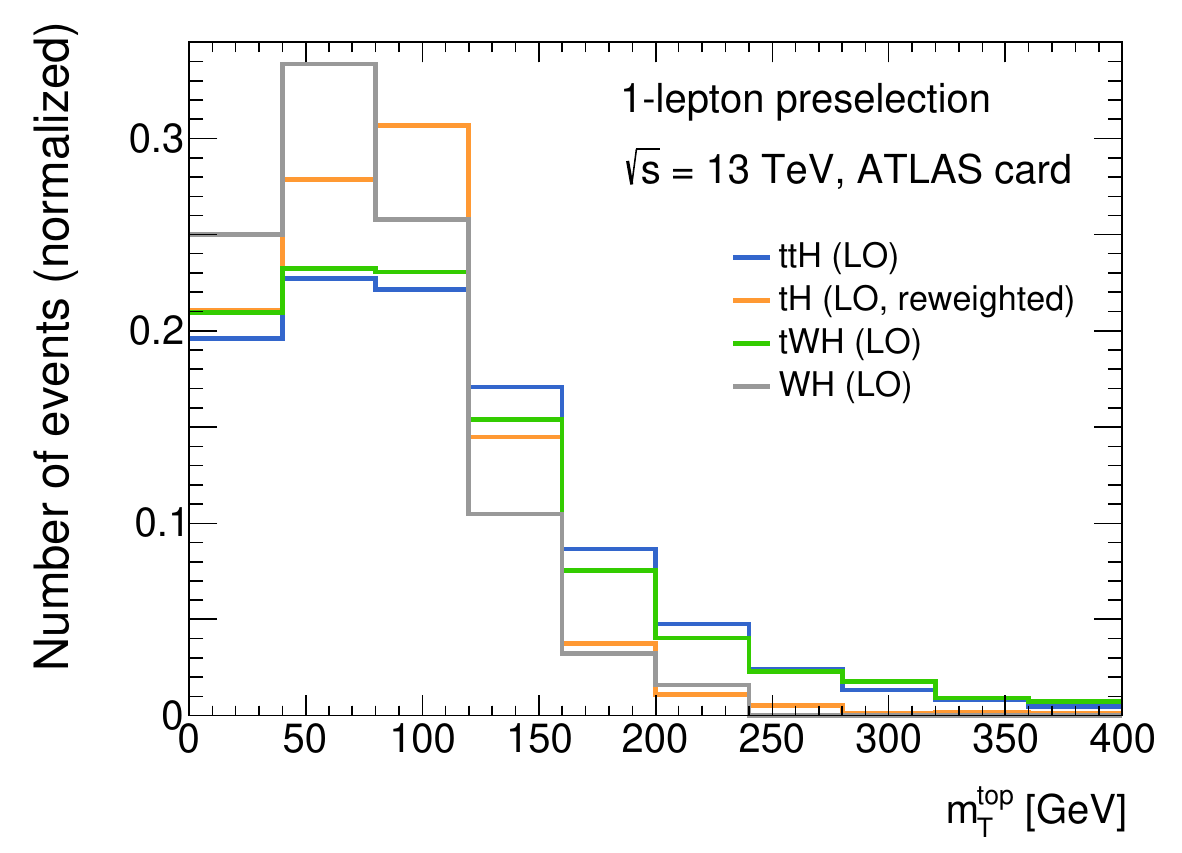}\\
\includegraphics[width=0.49\textwidth]{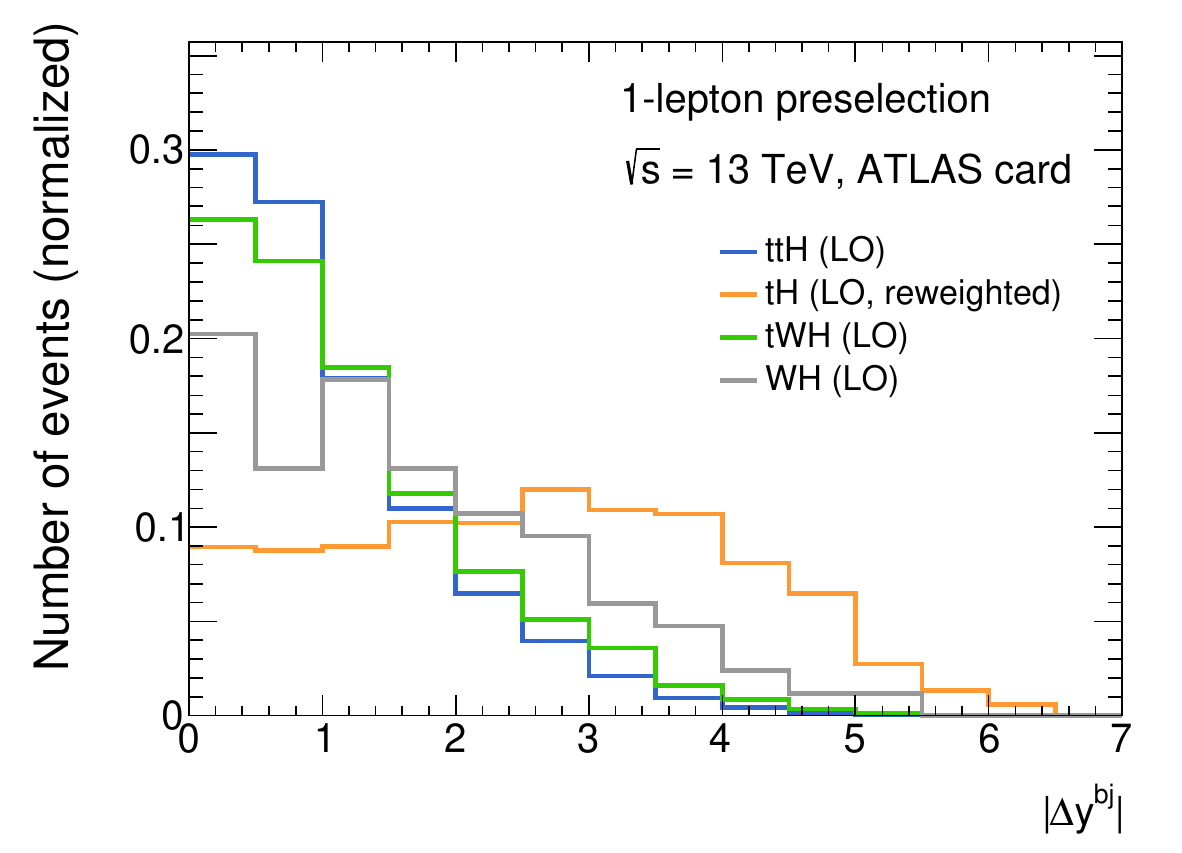}\hfill
\includegraphics[width=0.49\textwidth]{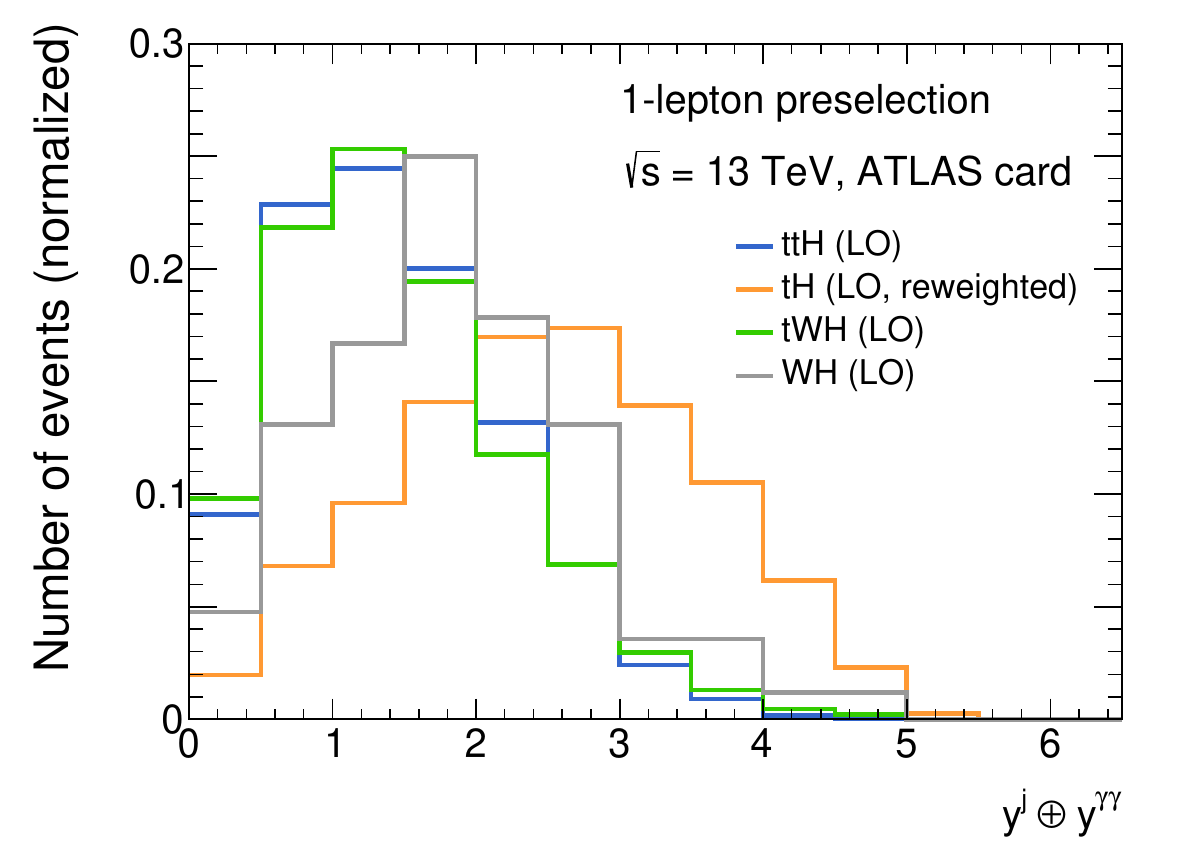}
\caption{Comparison of the shape of (top, left) $N_{jet}$, (top, right) $N_{bjet}$, (middle) $m_{T}^{top}$, (bottom, left) $|\Delta y^{bj}|$ and (bottom, right) $y^{j} \oplus y^{\gamma\gamma}$ at LO+PS in SM $tH$, $t\bar{t}H$, $tWH$ and $VH$ events passing the \emph{1-lepton preselection}. $tH$ events are reweighted to match the NLO+PS $N_{jet}$ shape. Events are reconstructed with the \texttt{Delphes} software using the ATLAS card described in the text, similar results are obtained when using the HL-LHC card.}
\label{sec6:shape}
\end{figure}

In the 1-lepton category, further discriminating variables are considered to enhance the fraction of $tH$ events. In addition to the jet multiplicity ($N_{jet}$) and the $b$-jet multiplicity ($N_{bjet}$), which are both expected to be larger in $t\bar{t}H$ due to the presence of a second top quark, the highest $p_{T}$ $b$-jet is associated to the lepton and the $p_{T}^{miss}$ vector to form the top quark candidate, whose transverse mass is defined as
\begin{align}
m_{T}^{top} = \sqrt{2 p_{T}^{b+\ell} p_{T}^{miss} [1 - \cos\Delta\phi(p_{T}^{b+l},p_{T}^{miss})] },
\end{align}
 where $p_{T}^{b+l}$ corresponds to the vector sum of the $b$-jet and the lepton transverse momenta, and $\Delta\phi(p_{T}^{b+l},p_{T}^{miss})$ corresponds to the difference in azimuthal angle between the missing transverse energy vector and $p_{T}^{b+l}$. The shapes of $N_{jet}$, $N_{bjet}$, and $m_{T}^{top}$ obtained at LO+PS and NLO+PS for SM $t\bar{t}H$ and $tH$ events passing the \emph{1-lepton preselection} are shown in \cref{sec6:lo_nlo_1} (top, middle). We observe a good consistency between the LO+PS and NLO+PS shapes except for $N_{jet}$ in $tH$ events, where the LO+PS calculation predicts a higher jet multiplicity. This feature has already been discussed in Ref.~\cite{Demartin:2015uha} and also appears in the case of single top production alone. It is related to the presence of a $b$ quark in the initial state in the five-flavor scheme (instead of a gluon splitting to a $b\bar{b}$ pair in the four-flavor scheme), which leads to a mismodeling of QCD radiation at leading-order. It was shown in Ref.~\cite{Demartin:2015uha} to be mitigated once higher-order corrections are included (NLO+PS). Since the jet multiplicity is not expected to strongly depend on the \cp\ properties of the Higgs boson, the LO+PS $tH$ sample is reweighted to the $N_{jet}$ shape obtained at NLO+PS in the SM for all four \cp\ hypotheses in the following.\footnote{A similar effect may also affect $tWH$ production. However, since the $tWH$ total rate is much smaller than the $tH$ and $t\bar{t}H$ rates, the effect is neglected and no corrections are included.} The impact of the reweighting on the $N_{bjet}$ and $m_{T}^{top}$ shapes is shown in \cref{sec6:lo_nlo_1} (top right and middle), and found to be small. The discrimination power between the different production modes of $N_{jet}$, $N_{bjet}$ and $m_{T}^{top}$ is found to be largely independent from the Higgs \cp\ properties, and is shown in \cref{sec6:shape} (top, middle) in the SM. The following requirements are applied to enhance the fraction of $tH$ events in the 1-lepton category: $N_{jet}=2$, $N_{bjet}=1$ and $m_{T}^{top}<200$ GeV.

\begin{figure}[tbp!]
  \centering
\includegraphics[width=0.49\textwidth]{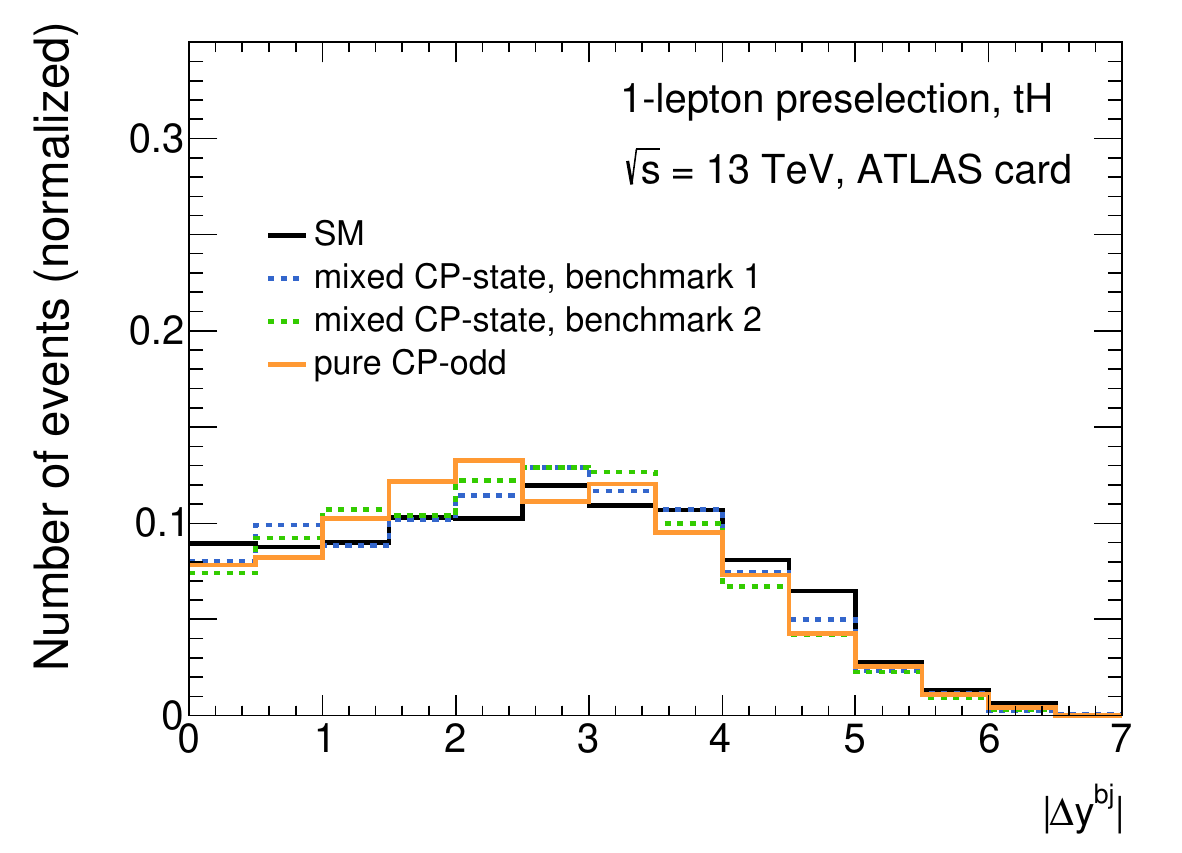}
\includegraphics[width=0.49\textwidth]{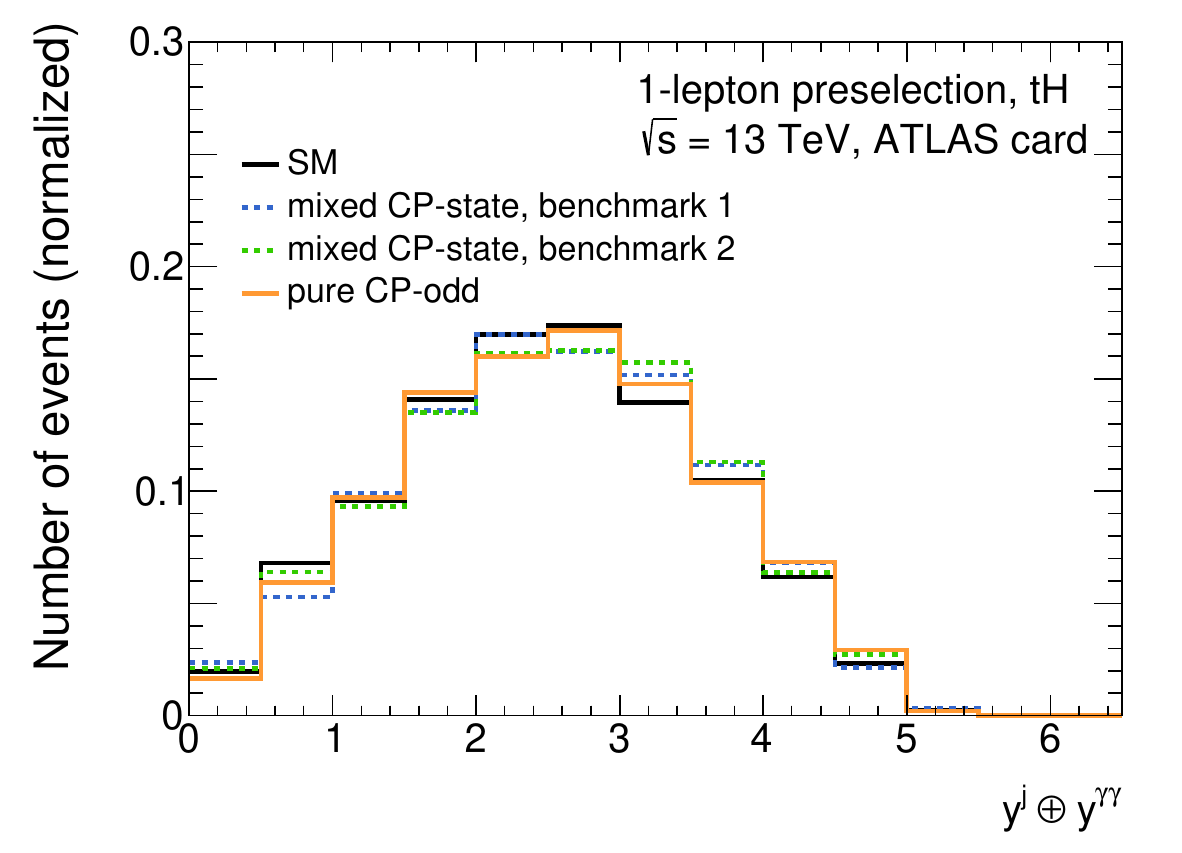}
\includegraphics[width=0.49\textwidth]{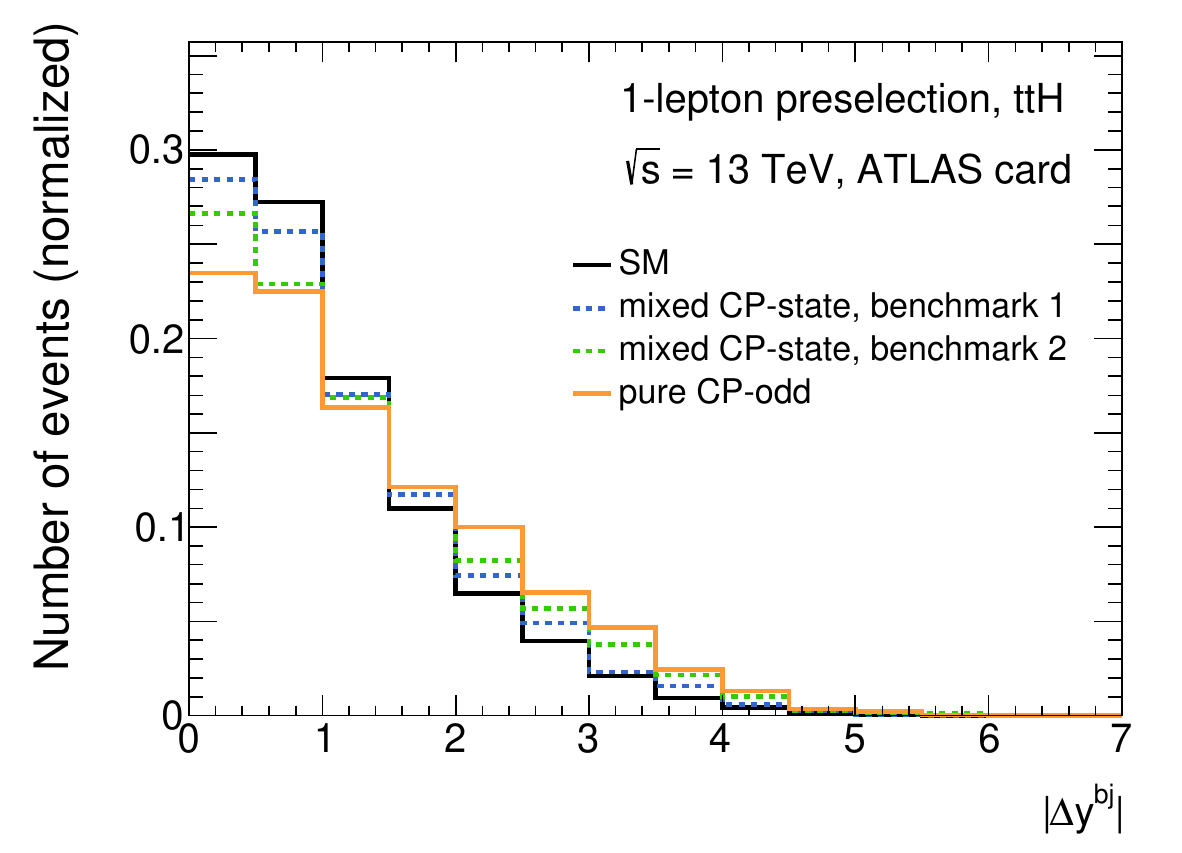}
\includegraphics[width=0.49\textwidth]{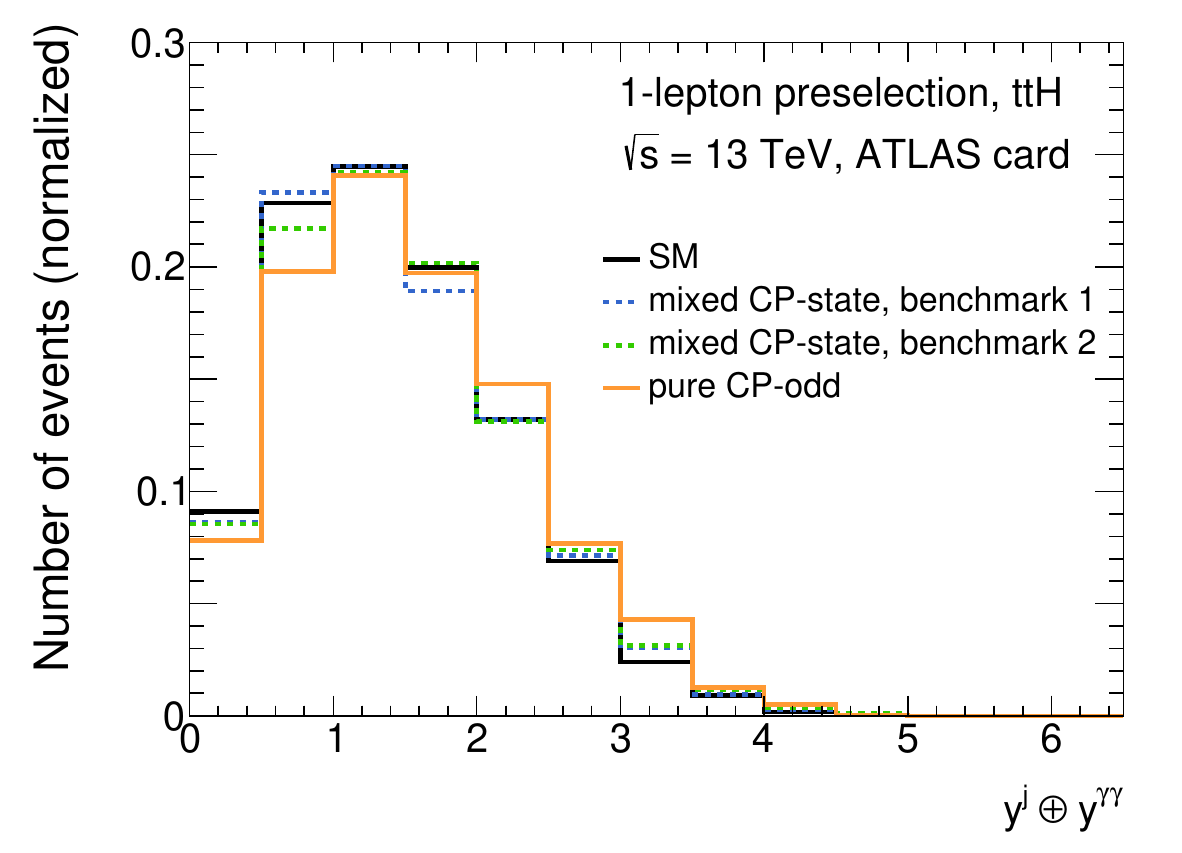}
\caption{Comparison of the shape of (left) $|\Delta y^{bj}|$ and (right) $y^{j} \oplus y^{\gamma\gamma}$ in (top) $tH$ and (bottom) $t\bar{t}H$ production for the four \cp\ hypotheses considered in the study: SM, pure \cp-odd Higgs boson (solid lines), and two benchmarks of \cp-mixed states described in the text (dashed lines). Events are reconstructed with the \texttt{Delphes} software using the modified ATLAS card described in the text, similar results are obtained when using the HL-LHC card.}
\label{sec6:cp}
\end{figure}

\begin{figure}[tbp!]
  \centering
\includegraphics[width=0.49\textwidth]{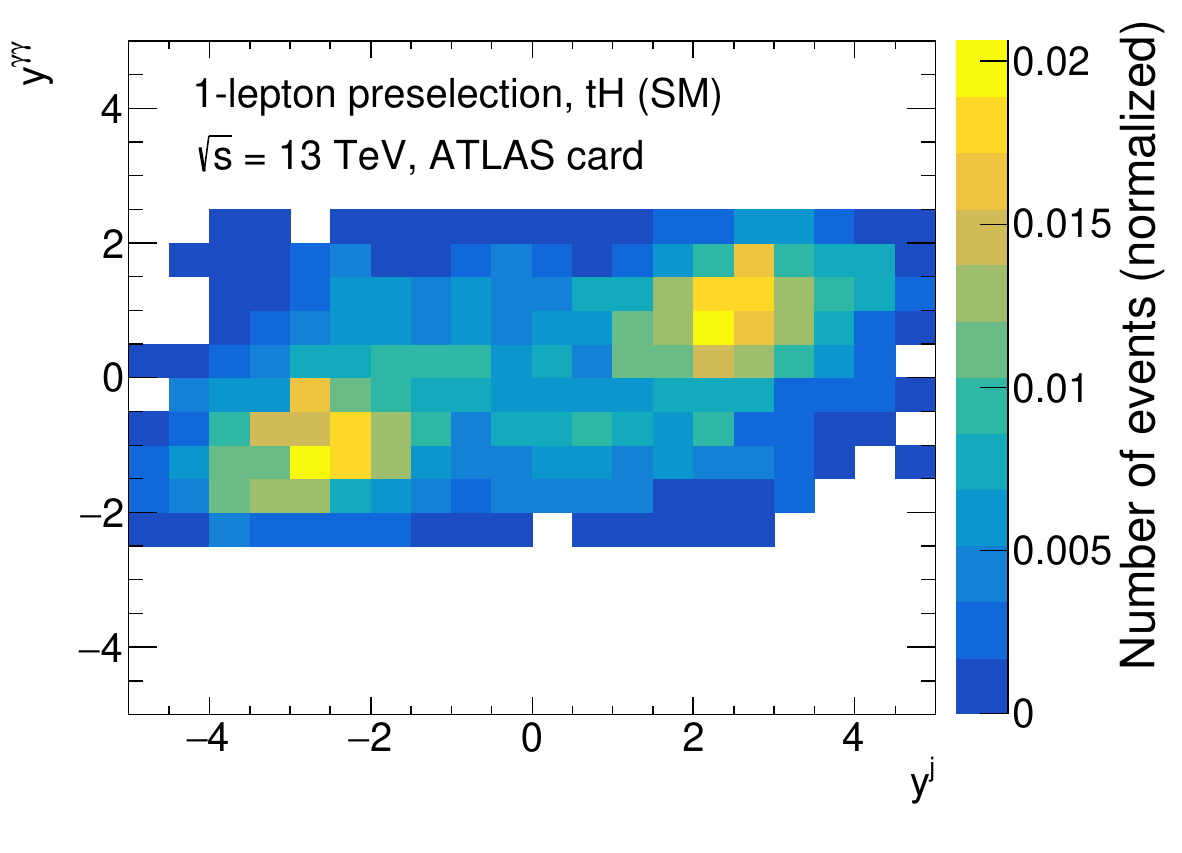}
\includegraphics[width=0.49\textwidth]{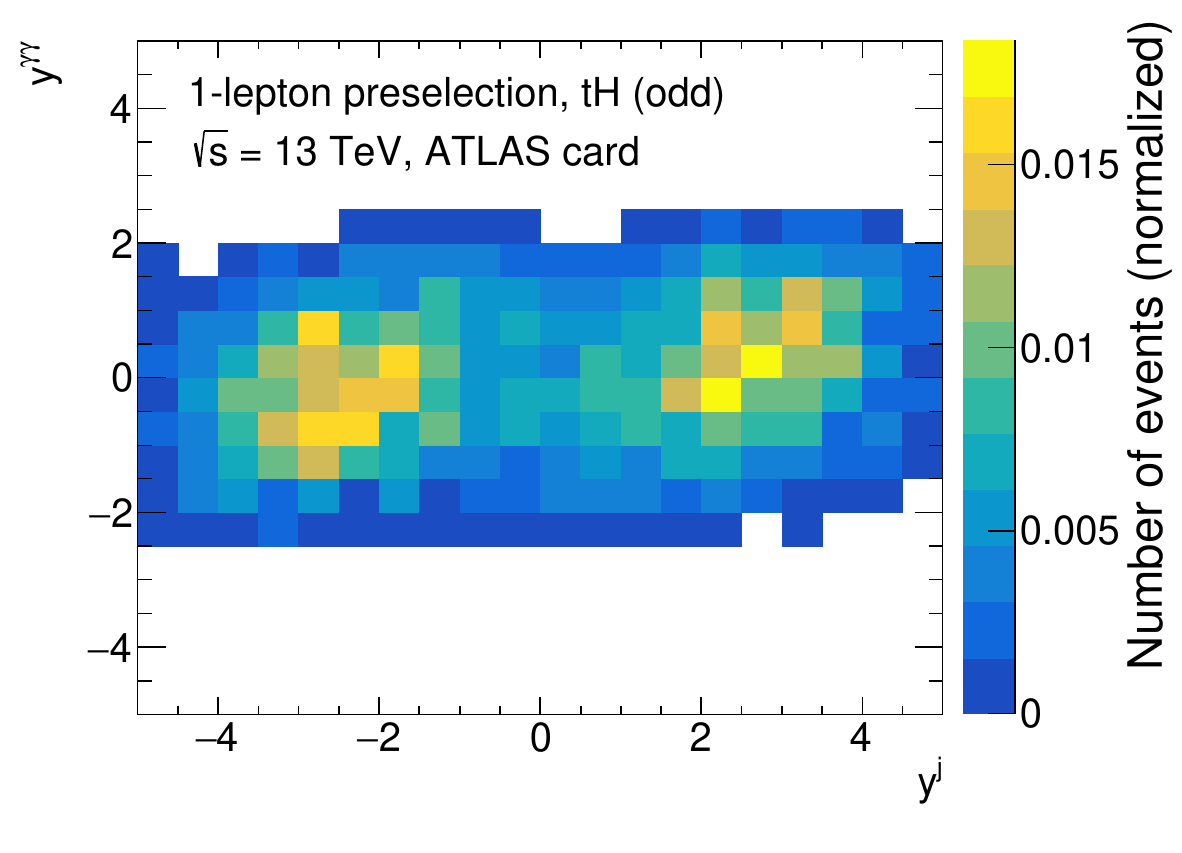}
\includegraphics[width=0.49\textwidth]{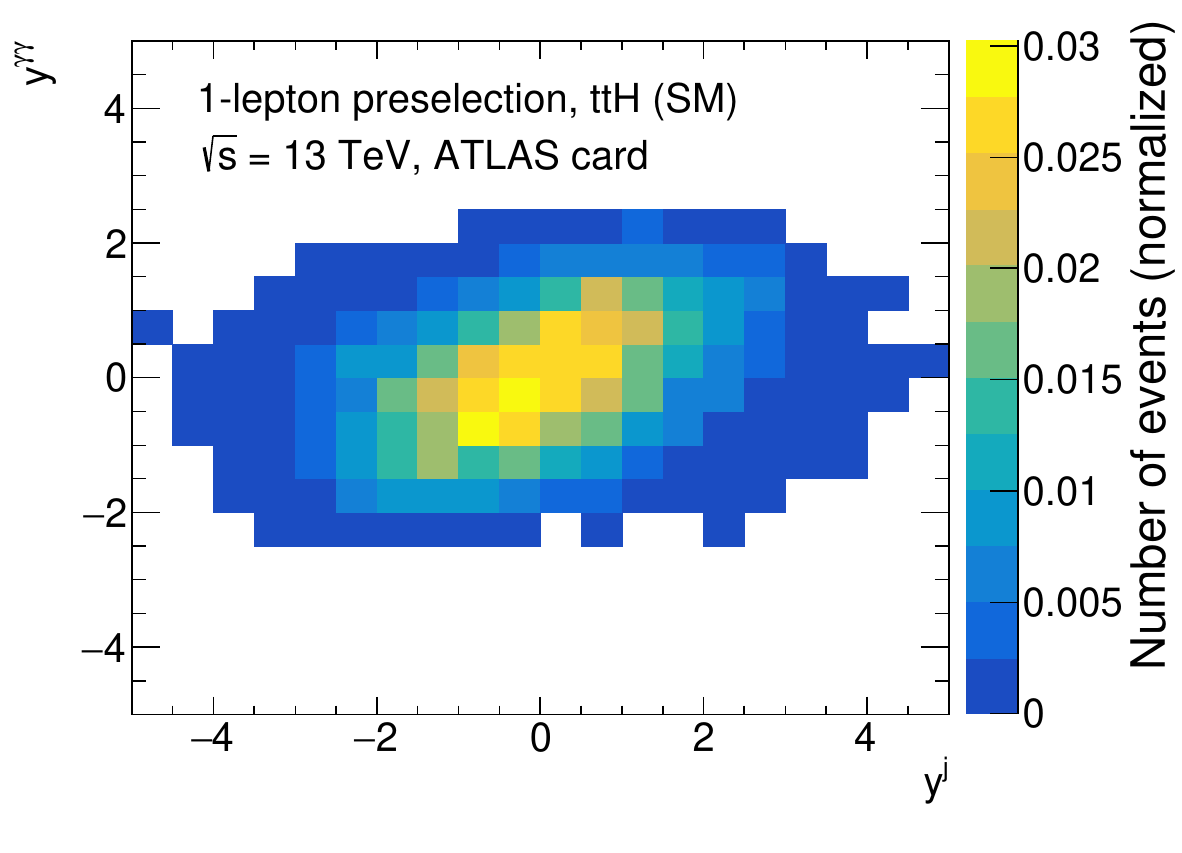}
\includegraphics[width=0.49\textwidth]{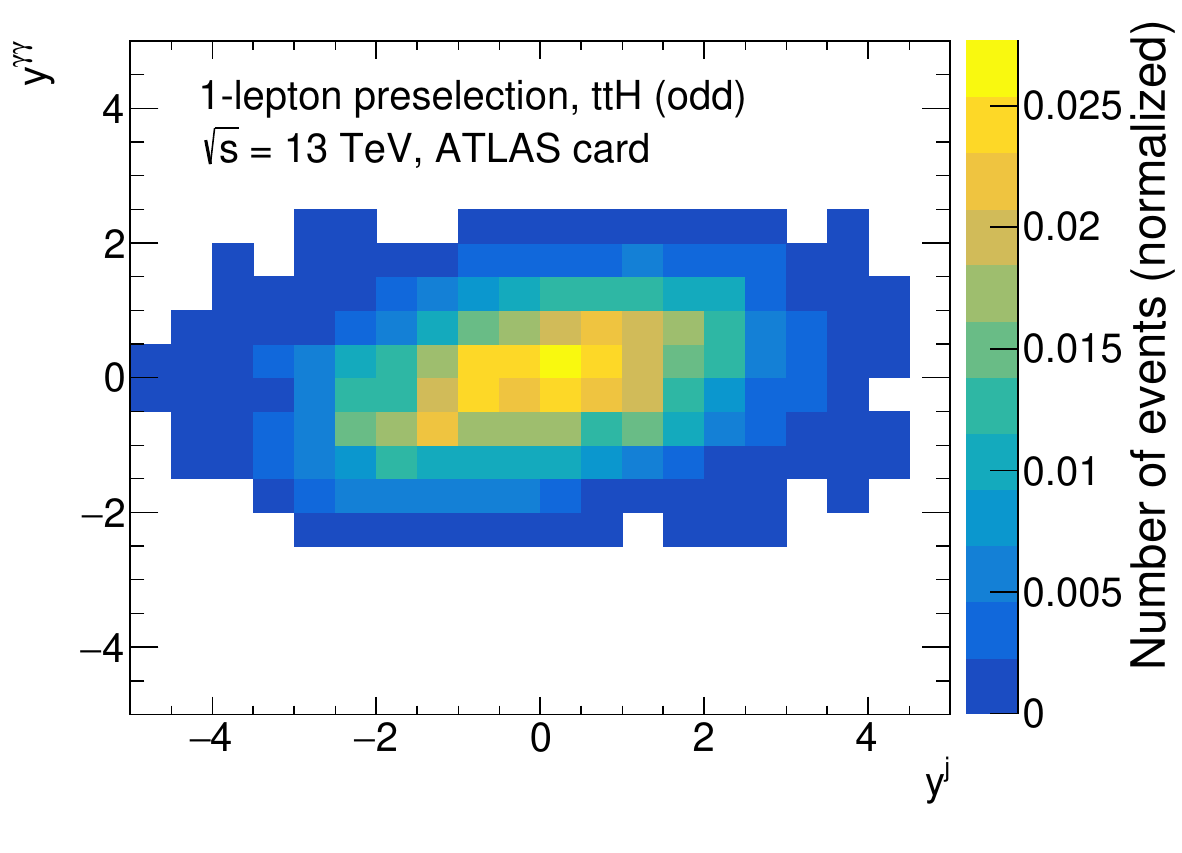}
\caption{Comparison of the correlations between $y^{j}$ and $y^{\gamma\gamma}$ in (top) $tH$ and (bottom) $t\bar{t}H$ production in (left) the SM and (right) for a pure \cp-odd Higgs boson. Events are reconstructed with the \texttt{Delphes} software using the modified ATLAS card described in the text, similar results are obtained when using the HL-LHC card.}
\label{sec6:2d}
\end{figure}

$tH$ events tend to include a high $p_{T}$ jet at large rapidity in the final state, similarly to single top production alone, which is well separated from the selected (central) $b$-jet. To exploit this feature, a first observable that comes to mind is the absolute value of the difference in rapidity\footnote{The pseudo-rapidity ($\eta$) is a good approximation of the rapidity ($y$) for the selected $b$-jet and the next highest $p_{T}$ jet due to the low jet mass, i.e.\ $y^b \sim \eta^b$ and $y^j \sim \eta^j$, but not for the Higgs boson candidate ($y^{\gamma\gamma} \ne \eta^{\gamma\gamma}$), where the effect from the Higgs boson mass is sizable.} between the selected $b$-jet and the next highest $p_{T}$ jet ($|\Delta y^{bj}|$). The shape of $|\Delta y^{bj}|$ is found to be stable after the $N_{jet}$ reweighting of $tH$, as shown in \cref{sec6:lo_nlo_1} (bottom, left), and its discrimination power between the different production modes is found to be large in the SM, as shown in \cref{sec6:shape} (bottom, left). However, when the Higgs boson includes a \cp-odd component, the $b$-jet tends to get closer (further away) in rapidity from that jet in $tH$ ($t\bar{t}H$) events, which reduces significantly the discrimination power and makes it dependent on the \cp\ properties (see \cref{sec6:cp}, left). The full selection efficiency for $t\bar{t}H$ events if requiring e.g.\ $|\Delta y^{bj}| > 2$ varies by $\sim40\%$ depending on the \cp-character of the top-Yukawa coupling. In \cref{sec6:2d}, we evaluate the use of the Higgs rapidity, $y^{\gamma\gamma}$, as an alternative to $y^b$. While $t\bar t H$ events are clustered around the origin in the $(y^j,y^{\gamma\gamma})$ plane, only a few $tH$ events lie close to the origin. Moreover, by comparing the left (SM) and the right plots (pure \cp-odd), we observe that the presence of a \cp-odd component in the Higgs--top-quark interaction leads to a transformation similar to a rotation in the $(y^j,y^{\gamma\gamma})$ plane. Consequently we introduce as an alternative observable to $|\Delta y^{bj}|$ the sum in quadrature of $y^j$ and $y^{\gamma\gamma}$, defined as
\begin{align}
y^{j} \oplus y^{\gamma\gamma} = \sqrt{(y^{j})^2 + (y^{\gamma\gamma})^2}.
\end{align}
The shape of $y^{j} \oplus y^{\gamma\gamma}$ is found to be stable after the $N_{jet}$ reweighting of $tH$, as shown in \cref{sec6:lo_nlo_1} (bottom, right), and allows for a good discrimination power between the different production modes in the SM, as shown in \cref{sec6:shape} (bottom, right). The shape of $y^{j} \oplus y^{\gamma\gamma}$ is stable with respect to the Higgs boson \cp\ properties (see \cref{sec6:cp}, bottom right), and the variation of the full selection efficiency for $t\bar{t}H$ events if requiring e.g. $y^{j} \oplus y^{\gamma\gamma}> 2$ is within $\lesssim 2\%$ in the considered parameter space. Due to this important property, a requirement of $y^{j} \oplus y^{\gamma\gamma}>2$ is included in the 1-lepton selection but no selection on $|\Delta y_{bj}|$ is considered. A summary of the event selection is presented in \cref{sec6:selection}.

We showed in this section that the discriminating observables used to disentangle $tH$ from $t\bar{t}H$ production need to be chosen with care. In particular, the stability of their shape with respect to the \cp\ properties of the Higgs boson is an essential ingredient making it possible to probe $tH$ production in a \cp-independent way. We demonstrated that this property can be achieved and proposed a new observable, $y^{j} \oplus y^{\gamma\gamma}$, which fulfills the conditions.

\begin{table}[!tb]
\small
\begin{center}
\def\arraystretch{1.2}
\begin{tabular}{l|c|c}
\toprule
\textbf{Observable / Selection } & 1-lepton selection & 2-lepton selection \\
\midrule
$N_{\gamma}$ & \multicolumn{2}{c}{$\ge 2$} \\
$m_{\gamma\gamma}$ & \multicolumn{2}{c}{$[105-160]$ GeV} \\
$(p_{T,1}^{\gamma}, p_{T,2}^{\gamma})$ & \multicolumn{2}{c}{$\ge (35, 25)$ GeV} \\
$(p_{T,1}^{\gamma}/m_{\gamma\gamma}, p_{T,2}^{\gamma}/m_{\gamma\gamma})$ & \multicolumn{2}{c}{$\ge (0.35,0.25)$} \\
$N_{bjet}$ & \multicolumn{2}{c}{$\ge 1$} \\
$p_{T}^{miss}$ & \multicolumn{2}{c}{$\ge 25$ GeV} \\
\midrule
$N_{\ell}$ & exactly $1$ & exactly 2 with opposite sign \\
$m_{\ell \ell}$ & -- & $[80,100]$ GeV vetoed if same flavour \\
$N_{jet}$ & exactly 2 & -- \\
$N_{bjet}$ & exactly 1 & -- \\
$m_{T}^{top}$ & $<200$ GeV & -- \\
$y^{j} \oplus y^{\gamma\gamma}$ & $>2$ & -- \\
\bottomrule
\end{tabular}
\end{center}
\caption{Summary of the 1-lepton and 2-lepton selection. The \emph{1-lepton preselection}, not detailed in the table, corresponds to the 1-lepton selection dropping the requirements on $N_{jet}$, $N_{bjet}$, $m_{T}^{top}$ and $y^{j} \oplus y^{\gamma\gamma}$. The object definition follows the one written in the \texttt{Delphes} cards unless mentioned otherwise in the text.}
\label{sec6:selection}
\end{table}



\subsection{Results}

Less than two events in total are expected at the 300~fb$^{-1}$ LHC in the 2-lepton category in all four \cp\ scenarios, indicating that additional data from the HL-LHC is required to successfully implement the strategy being followed in this work. We will therefore focus on the results for the 3000~fb$^{-1}$ HL-LHC in the following. The $gg\to ZH$ contribution is found to be negligible in all scenarios and is not considered further. The $t\bar{t}H$, $tH$, $tWH$, $WH$ and $qq\to ZH$ event yields obtained after applying the different selections described in the previous section are shown in \cref{sec6:yields_hllhc} for the SM and the pure \cp-odd Higgs--top-quark interaction scenarios (the corresponding event yields for the \cp-mixed benchmark scenarios lead to results in between).

\begin{table}[!tb]
\small
\begin{center}
\def\arraystretch{1.2}
\begin{tabular}{l|ccc}
\toprule
\textbf{Selection} & 1-lepton     & 1-lepton      & 2-lepton \\
\textbf{Process}   & preselection & selection     & selection \\
\midrule
SM $ttH$  & 300.4 & 7.2 & 29.3    \\
SM $tH$   & 19.0  & 6.7 & $<0.1$ \\
SM $tWH$  & 8.0   & 0.4 & 0.8 \\
SM $WH$   & 3.6    & 0.4 & $<0.1$ \\
SM $qqZH$ & 0.8    & $<0.1$ & 0.1 \\
\midrule
SM total   & 331.8 & 14.7 & 30.2 \\
\midrule
\cp-odd, $ttH$  & 140.2 & 3.3  & 13.9    \\
\cp-odd, $tH$   & 120.7 & 42.7 & 0.1 \\
\cp-odd, $tWH$  & 37.8  & 1.4  & 3.9   \\
\midrule
\cp-odd total (incl. SM $WH$ + $qqZH$) & 303.1 & 47.8 & 18.0 \\
\bottomrule
\end{tabular}
\end{center}
\caption{Expected number of events at the 3000 fb$^{-1}$ HL-LHC passing the different event selections defined in the text.}
\label{sec6:yields_hllhc}
\end{table}

As expected the fraction of $tH$ ($t\bar{t}H$) events entering the 1-lepton category is enhanced (depleted) in the pure \cp-odd scenario with respect to the SM. $tWH$ production is not negligible at the HL-LHC in the pure \cp-odd scenario, with about four $tWH$ events out of eighteen in total expected in the 2-lepton category, even if it is still largely subdominant with respect to $t\bar{t}H$. For convenience, we will refer in the following to the merged $t\bar{t}H$ and $tWH$ contribution, labeled $t\bar{t}H+tWH$. The $VH$ contribution is very small ($< 5\%$) and independent of the \cp\ properties of the Higgs--top-quark interaction since the $gg\to ZH$ contribution is negligible in all categories and scenarios.

In order to evaluate the sensitivity of the prospect analysis to measure $tH$ production, the total number of events in the 1-lepton and 2-lepton category is included in a dedicated profile likelihood fit defined within the \texttt{HistFitter} software\cite{Baak:2014wma}. Systematic uncertainties are neglected in this fit. Compared to the low count rates, their effect is expected to be subdominant. The $t\bar{t}H+tWH$ and $tH$ signal strengths, labeled as $\mu_{t\bar{t}H+tWH}$ and $\mu_{tH}$, respectively, are free-floating in the fit. The $VH$ signal strength, $\mu_{VH}$, is fixed to unity. Due to the very limited size of the $VH$ contamination in both the 1-lepton and 2-lepton category, $\mu_{VH}$ can be varied by large values without impacting the results. The fit is repeated four times, considering each time an observed dataset built from one of the four different Higgs \cp\ hypotheses introduced earlier and rounded to the closest integer number. The results are summarised in \cref{sec6:fit1}. The selection efficiencies entering the fit results can be considered as independent of the \cp\ properties of the Higgs--top-quark interaction assumed during the hypothesis testing within the statistical uncertainty.

\begin{table}[!tb]
\small
\begin{center}
\def\arraystretch{1.2}
\begin{tabular}{l|ccc}
\toprule
\textbf{Observed data}   &  $\mu_{tH}$ & $\mu_{t\bar{t}H+tWH}$ & $\mu_{tH/(t\bar{t}H+tWH)}$ \\
\midrule
SM          & $1.06\pm 0.6$  & $0.99\pm 0.18$ & $1.07^{+0.82}_{-0.62}$ \\
\cp-odd     & $6.41\pm 1.05$ & $0.60\pm 0.14$ & $10.78^{+3.96}_{-2.82}$ \\
\cp-mixed 1 & $2.29\pm 0.74$ & $0.96\pm 0.18$ & $2.39^{+1.14}_{-0.87}$ \\
\cp-mixed 2 & $3.56\pm 0.87$ & $1.03\pm 0.18$ & $3.47^{+1.33}_{-1.03}$ \\
\bottomrule
\end{tabular}
\end{center}
\caption{Results of the prospect measurement with 3000~fb$^{-1}$ at the HL-LHC described in the text. The observed data sets used for the prospective measurements are built from each of the four \cp\ Higgs scenarios defined earlier in the text. The selection efficiencies entering the fit results can be considered as independent of the \cp\ properties of the Higgs--top-quark interaction within the statistical uncertainty. They deviate insignificantly from unity in the SM scenario due to the rounding to the closest integer number required by Poisson statistics for the observed data.}
\label{sec6:fit1}
\end{table}

If the data is SM-like, an evidence for $tH$ production is obtained with about $2\,\sigma$ significance (corresponding to a $p$-value for the null hypothesis of no $tH$ contribution of about 0.025) while $t\bar{t}H$ is observed with $5\,\sigma$ significance. This translates to an expected upper limit of $\mu_{tH} < 2.21$ at the 95\% confidence level (CL) using the $CL_s$ prescription~\cite{Read:2002hq} if the data is SM-like including the $tH$ signal, and an upper bound of $\mu_{tH} < 1.16$ if there is no $tH$ signal.  This result for discriminating the $tH$ channel from the $t\bar{t}H + tWH$ channel is robust with respect to the \cp\ properties of the Higgs--top-quark interaction assumed during the hypothesis testing owing to the careful selection of observables used to define the 1-lepton and 2-lepton category discussed in the previous section. Our projected result improves on the current strongest limit~\cite{Aad:2020ivc} by about a factor of 5 ($\mu_{tH}<12$ at the 95\% CL), which was set recently by the ATLAS collaboration with the full Run 2 dataset of 139 fb$^{-1}$ using $H\to\gamma\gamma$ decays, under the assumption of a pure \cp-even coupling. The projected exclusion limit obtained in this work is also stronger than the most optimistic projected limit on $tH$ production at the 3000 fb$^{-1}$ HL-LHC documented in Refs.~\cite{Cepeda:2019klc,Atlas:2019qfx}. This illustrates the potential of the proposed strategy to probe $tH$ production at the HL-LHC.

We recall that, unlike in the references quoted above, the results derived from the prospect measurement presented in this work do not consider systematic uncertainties. The ATLAS analysis reporting the current strongest limit on $tH$ production~\cite{Aad:2020ivc} mentions that systematic uncertainties are negligible. They are expected to still play a minor role at the HL-LHC~\cite{Cepeda:2019klc,Atlas:2019qfx}. We report in addition in \cref{sec6:fit1} the results for the measurement of the $tH/(t\bar{t}H+tWH)$  signal strength, $\mu_{tH/(t\bar{t}H+tWH)}$, which is obtained by reparametrizing the free-floating parameters in the fit to $\mu_{tH/(t\bar{t}H+tWH)}$ and $\mu_{t\bar{t}H}$. The $\mu_{tH/(t\bar{t}H+tWH)}$ measurement suffers from a larger statistical uncertainty than the $tH$ measurement but is expected to be more robust with respect to e.g.\ the uncertainty in the photon energy resolution and scale, which should cancel out in the ratio to a large extent and was reported as the leading experimental systematic uncertainties ($\sim 7\%$) in Ref.~\cite{ATLAS-CONF-2019-004}. Current theory uncertainties are expected to be at the $\sim$ 5 to 10\% level for $tH$, $tWH$ and $t\bar{t}H$ production and therefore should be subdominant with respect to the data statistical uncertainty quoted in \cref{sec6:fit1}. We found that $\mu_{tH/\{t\bar{t}H+tWH\}}> 2.57$ is expected to be excluded at the 95\% level if the data is SM-like including $\mu_{tH/(t\bar{t}H+tWH)}=1$, and $\mu_{tH/(t\bar{t}H+tWH)}> 1.14$ if there is no $tH$ signal.

When the observed dataset is built from other \cp\ scenarios, $\mu_{tH}=1$, which is the value expected in the SM, deviates from the best fit $\mu_{tH}$ value by about $5.2\,\sigma$, $1.7\,\sigma$ and $2.9\,\sigma$ in the pure \cp-odd, the first and the second benchmark of a \cp-mixed state scenario, respectively. It should be noted that the hypothesis test between various \cp\ scenarios depends on many other features not explored in this analysis. It should be stressed that our analysis is meant to be included in a global study of the \cp-properties rather than being used alone.

\begin{figure}[!tbp]
  \centering
\includegraphics[width=0.49\textwidth]{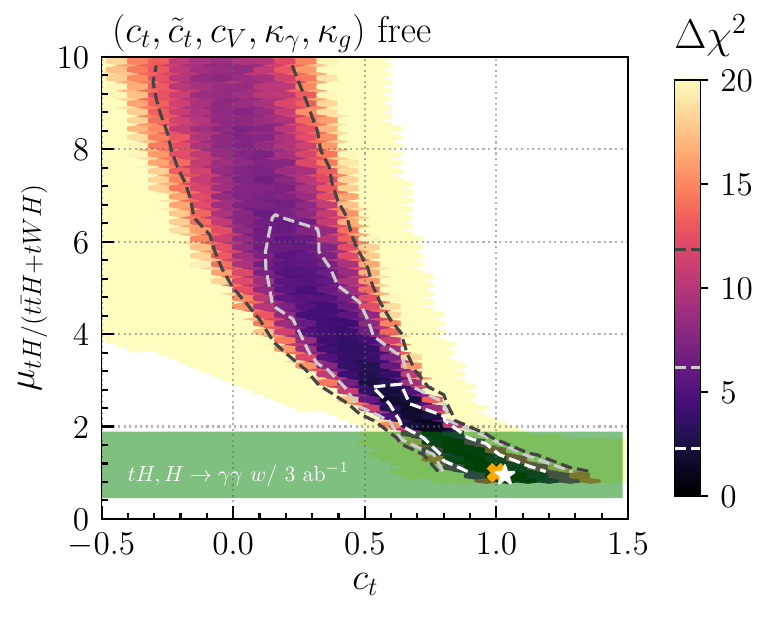}\hfill
\includegraphics[width=0.49\textwidth]{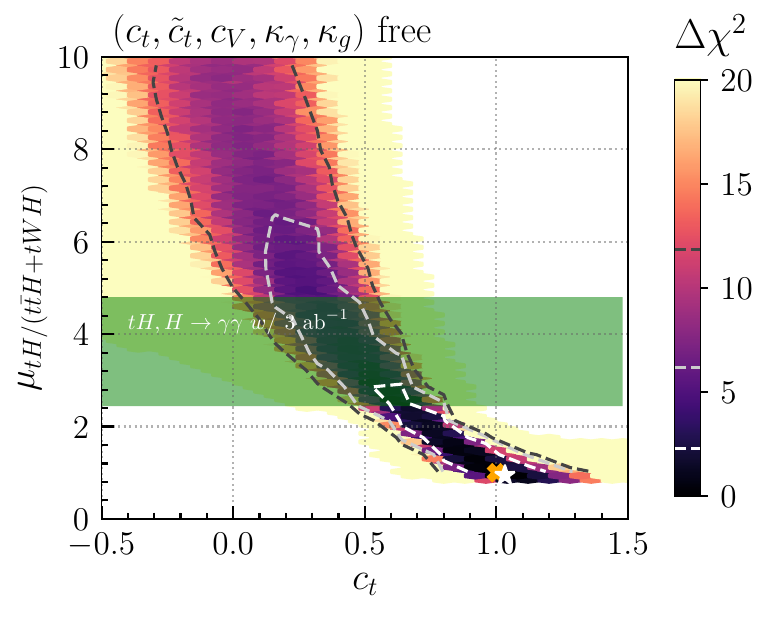}\\
\includegraphics[width=0.49\textwidth]{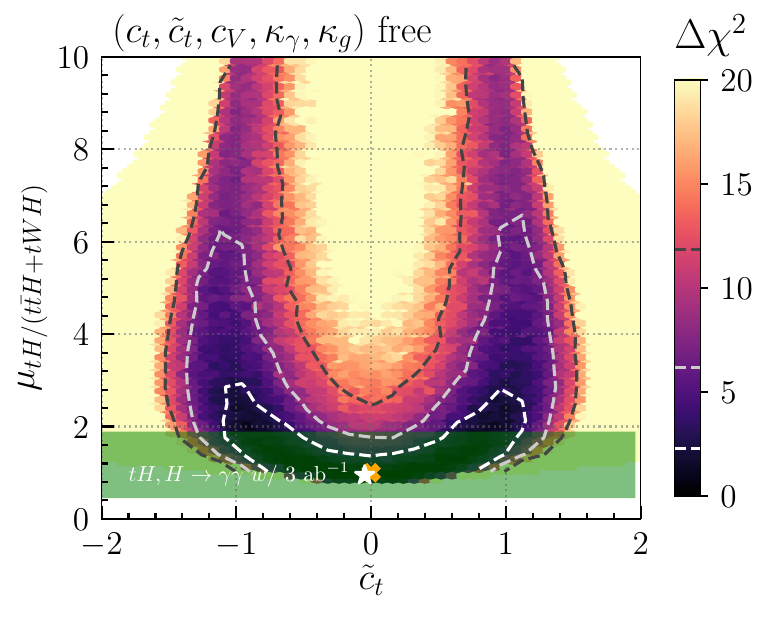}\hfill
\includegraphics[width=0.49\textwidth]{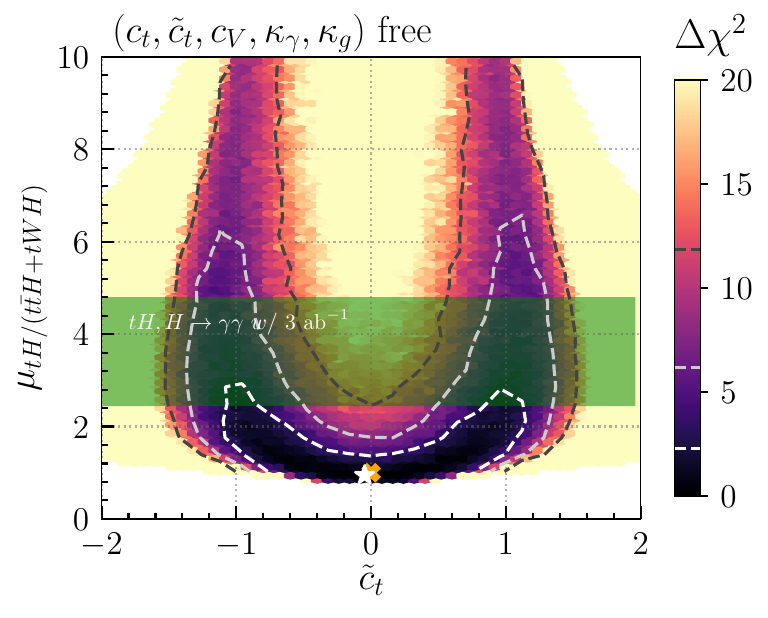}\\
\includegraphics[width=0.49\textwidth]{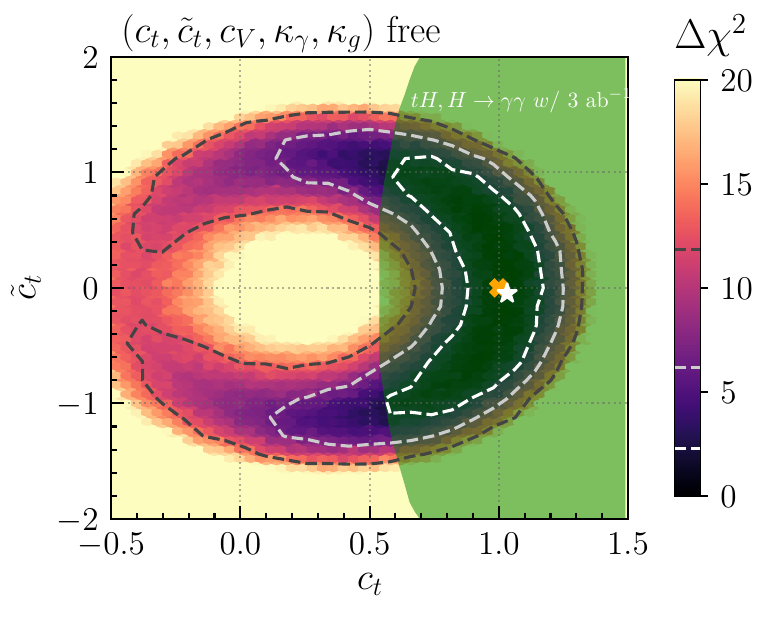}\hfill
\includegraphics[width=0.49\textwidth]{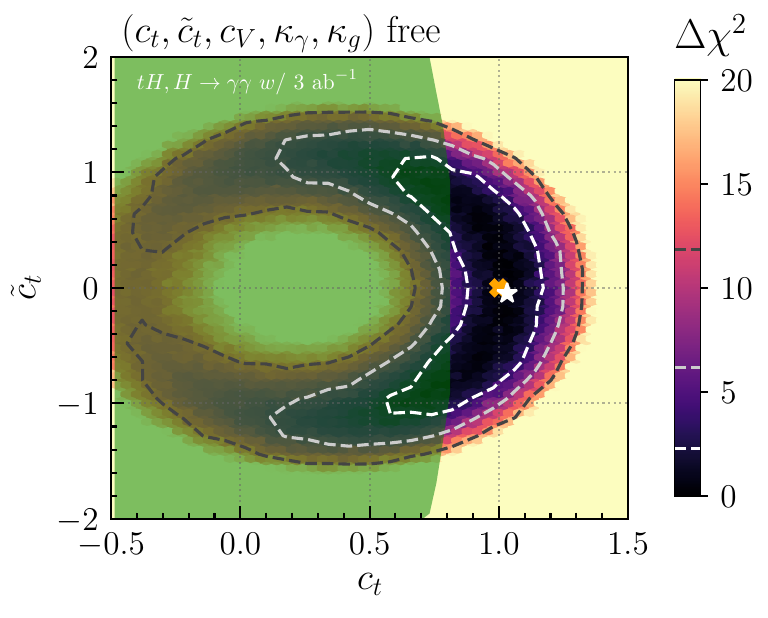}
\caption{Impact of the prospective $\mu_{tH/(t\bar{t}H+tWH)}$ determination of the proposed $tH$ analysis with $3~\mathrm{ab}^{-1}$ of data [the green areas indicate the $1\,\sigma$ precision] on the currently allowed ranges of $\ct$ (\emph{top panels}), $\cttilde$ (\emph{middle panels}), as well as on the (\ct, \cttilde) parameter plane (\emph{bottom panels}). The figures contain the fit results of the 5D parametrization shown in \cref{fig:mu_tHoverttH} (\emph{right panels}) and \cref{fig:ctcttilde} (\emph{bottom right panel}), respectively. We assume the future $\mu_{tH/(t\bar{t}H+tWH)}$ measurement to be consistent either with the SM (\emph{left panels}) or the \cp-mixed 2 benchmark scenario (\emph{right panels}).}
\label{fig:prospects_mu_tHoverttHtWH}
\end{figure}

An independent $tH$ measurement will in any case provide valuable information for determining the nature of BSM physics if a deviation from the SM predictions is observed. Here, we concentrate on illustrating the impact of a prospective HL-LHC determination of $\mu_{tH/(t\bar{t}H+tWH)}$ on the currently allowed ranges in the 5D model parametrization -- (\ct, \cttilde, \cv, \kg, \kgamma) free -- discussed in \Cref{sec:results}. For this purpose, we overlay the anticipated $1\,\sigma$ range of $\mu_{tH/(t\bar{t}H+tWH)}$ from our proposed analysis, \Cref{sec6:fit1}, assuming either a measurement consistent with the SM hypothesis or with the \cp-mixed\,2 benchmark scenario, on the fit results obtained in \Cref{sec:results}. Figure \ref{fig:prospects_mu_tHoverttHtWH} displays this comparison in dependence of \ct (\emph{top panels}), \cttilde (\emph{middle panels}), as well as in the (\ct,\cttilde) parameter plane (\emph{bottom panels}), where the green band indicates the prospective determination of $\mu_{tH/(t\bar{t}H+tWH)}$, assuming the SM hypothesis (\emph{left panels}) or the \cp-mixed\,2 benchmark scenario hypothesis (\emph{right panels}). The fit results are the ones that have been shown above in \cref{fig:mu_tHoverttH} (\emph{right panels}) and \cref{fig:ctcttilde} (\emph{bottom right panel}), respectively.

Assuming a SM-consistent measurement at the HL-LHC, the $\mu_{tH/(t\bar{t}H+tWH)}$ determination will yield a lower limit on \ct while the overall impact on \cttilde will be rather weak. In contrast, if $\mu_{tH/(t\bar{t}H + tWH)}$ is found to be consistent with the \cp-mixed\,2 benchmark scenario at the HL-LHC, a preference for \ct\ values below 1 and non-zero \cttilde will be observed. From the bottom panels in \cref{fig:prospects_mu_tHoverttHtWH} one can see that a SM-consistent future measurement would further constrain the allowed range to positive values of \ct, while a measurement consistent with the \cp-mixed\,2 benchmark scenario would result in a preference for small or even negative \ct values. In the latter case, the combination of the $\mu_{tH/(t\bar{t}H+tWH)}$ measurement with the fit result obtained above based on all other available inclusive and differential Higgs rate measurements (which will be further improved with the accuracies at the HL-LHC) would yield a preference for a parameter region with $\cttilde \ne 0$, i.e.~hinting towards \cp-violation in the top-Yukawa sector.


\section{Conclusions}
\label{sec:conclusions}

Probing the top-Yukawa coupling is crucial to constrain the \cp nature of the Higgs boson discovered at the LHC. In this paper, we used all relevant inclusive and differential Higgs boson rate measurements to derive bounds on a possible \cp-odd component of the top-Yukawa coupling.

We performed this study in the ``Higgs characterization model'', an EFT framework which allows one to study \cp-violating Higgs couplings in a consistent, systematic and accurate way. Within this model, we derived theory predictions for all relevant Higgs production and decay channels. We evaluated the experimental constraints in four different model parameterizations with the parameter space spanning two, three, four or five dimensions. The 2D parameter space, consisting of the \cp-even, \ct, and the \cp-odd, \cttilde, top-Yukawa coupling, is successively enlarged by allowing for new physics contributions to the couplings of the Higgs boson to massive vector bosons, to photons, and to gluons. We found the best-fit points in all four parameterizations to be close to the SM.

We could constrain \cttilde in the 2D parametrization to the the interval $[-0.3,0.3]$ at the $1\,\sigma$ level. In this parametrization, the most constraining measurements arise from Higgs production via gluon fusion and the Higgs decay to di-photons, which are sensitive to the top-Yukawa coupling at the loop level. Allowing for additional freedom in the Higgs coupling to massive vector bosons (i.e.\ not setting \cv to unity) only slightly weakens the constraints. Allowing also \kgamma to vary freely results in a $1\,\sigma$ interval of $[-0.4,0.4]$ for \cttilde. In this parameterization, the strongest constraints originate from Higgs production via gluon fusion. The constraints are significantly weakened if also the Higgs coupling to gluons, \kg, is allowed to vary freely. In this model, \cttilde can lie in the range $[-1.1,1.1]$ at the $1\,\sigma$ level. Measurements of top-associated Higgs production, sensitive to the top-Yukawa coupling at the tree-level, yield the strongest constraints in case of a positive \ct. Moreover, Higgs production in association with a $Z$~boson, using total rate and kinematic information, also constraints the parameter space in particular for negative \ct. We encourage future measurements of kinematic shapes (also for top-quark-associated production) to further constrain the parameter space.

The constraints can also be interpreted in terms of a \cp-violating phase which we find to be $\lesssim 22.5 - 27$ degrees for the first three parametrizations and $\lesssim 72$~degrees for the 5D parameterization at the $2\,\sigma$ level. Similar constraints have been derived in the latest studies reported by ATLAS and CMS~\cite{Sirunyan:2020sum, Aad:2020ivc}. Such results nevertheless cannot be directly applied to other models than the model studied in the experimental analyses, whereas the STXS and total rate measurements included in our global fit are largely model-independent.

The existing constraints place only weak bounds on the possible values of the $tH$ production cross section,  so that there is large room for deviations from the SM in this channel. The measurement of an enhanced rate of Higgs production in association with a single top-quark with respect to the SM would hint at a non-zero \cp-odd top-Yukawa coupling and a lower \cp-even top-Yukawa coupling than predicted by the SM. At the current level of experimental precision, it is, however, hard to disentangle Higgs production in association with a single top-quark and with two top-quarks. Therefore, we proposed a novel analysis strategy for disentangling $tH$ and $t\bar t H+tWH$ production at the HL-LHC, where we compare the number of events in 1- and 2-lepton categories designed such that the selection efficiencies can be considered as independent of the \cp\ properties of the Higgs--top-quark interaction within the statistical uncertainty. Focusing on the Higgs decay to photons, we obtained an expected upper limit of $\mu_{tH} < 2.21$ at the 95\% CL if the data is SM-like including the $tH$ signal, and $\mu_{tH} < 1.16$ if there is no $tH$ signal. These prospective limits may be regarded as a conservative estimate since a dedicated effort including additional decay channels will most likely increase the sensitivity.

If a deviation from the SM in top-associated Higgs production is observed in the future, an independent $tH$ measurement will be important to unravel the nature of this deviation. The prospective impact on the \cp\ properties of the Higgs--top-quark interaction depends on the measured value of the signal strength. In case of a large deviation, the measurement of the $tH$ production cross section will be a crucial ingredient for determining the \cp\ properties of the $t \bar t H$ coupling. On the other hand, if the $tH$ cross section turns out to be SM-like, according to the current projections there will still be significant scope for non-standard \cp\ properties of the Higgs--top-quark interaction. Further experimental input and more precise measurements will be needed in this case to narrow down these possibilities.


\section*{Acknowledgments}
\sloppy{
We thank Lisa Biermann and Martin Mosny for collaboration in an early stage of the project, Marco Zaro for clarifications regarding the ``Higgs charaterization model'' as well as Robert Harlander, Lukas Simon and Jonas Wittbrodt for useful discussion regarding vh@nnlo. H.B., J.K., K.P., T.S.\ and G.W.\ acknowledge support by the Deutsche Forschungsgemeinschaft (DFG, German Research Foundation) under Germany's Excellence Strategy -- EXC 2121 ``Quantum Universe'' – 390833306.
The work of S.H.\ is supported in part by the
Spanish Agencia Estatal de Investigaci{\' o}n (AEI) and the EU Fondo Europeo
de Desarrollo Regional (FEDER) through the project FPA2016-78645-P,
in part by the MEINCOP Spain under contract FPA2016-78022-P,
in part by the “Spanish Red Consolider MultiDark” FPA2017-90566-REDC
and in part by
the AEI through the grant IFT Centro de Excelencia Severo Ochoa SEV-2016-0597.
}


\appendix


 \clearpage
 \section*{Appendix}
 \label{app}

 \section{CP sensitivity in \texorpdfstring{$gg\rightarrow H + 2j$}{gg to H + 2j}}
\label{app:ggH2j}

\begin{figure}\centering
\includegraphics[width=.49\textwidth]{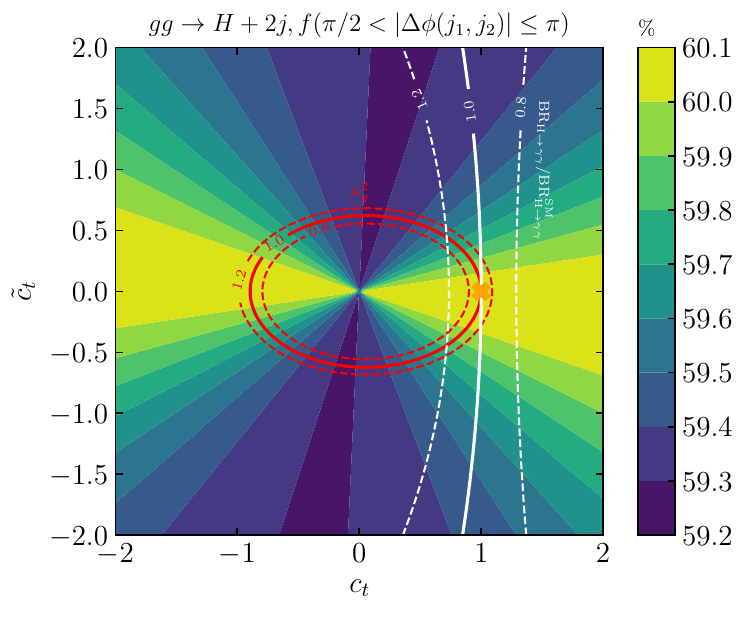}
\includegraphics[width=.49\textwidth]{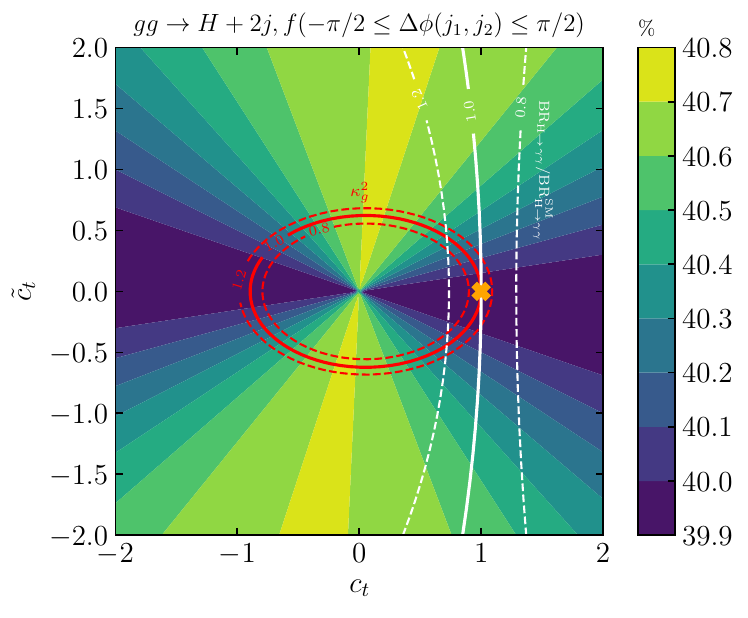}
\caption{Fraction of $gg\rightarrow H+2j$ events falling into the $\Delta\phi(j_1,j_2)$ bin. \textit{Red}: $\kappa_g = 1.0 \pm 0.2$. \textit{White}: $\text{BR}_{H\rightarrow\gamma\gamma}/\text{BR}^\SM_{H\rightarrow\gamma\gamma} = 1.0 \pm 0.2$. The SM is marked by an orange cross.}
\label{fig:ggHj_2j}
\end{figure}

In~\cite{ATLAS:2019jst}, the $\Delta\phi(j_1,j_2)$ distribution in $gg\rightarrow H + 2j$ has been split into four bins. Since the theoretical prediction for the $\Delta\phi_{j_1,j_2}$ distribution in the model considered is symmetric around zero, we merge the four bins used in~\cite{ATLAS:2019jst} into two bins:
\begin{itemize}
\item bin~1: $\pi/2 < |\Delta\phi(j_1,j_2)| \le \pi$,
\item bin~2: $-\pi/2 \le \Delta\phi(j_1,j_2) \le \pi/2$.
\end{itemize}
The corresponding fit formulas are given by
\begin{align}
\mu_{gg\rightarrow H+2j}^\text{bin 1} &= (\ct + \cg)^2 + 0.001 (\ct + \cg)(\cttilde + \cgtilde) + 2.17 (\cttilde + \cgtilde)^2, \\
\mu_{gg\rightarrow H+2j}^\text{bin 2} &= (\ct + \cg)^2 + 0.013 (\ct + \cg)(\cttilde + \cgtilde) + 2.25 (\cttilde + \cgtilde)^2.
\end{align}
The numerical results are presented in \cref{fig:ggHj_2j}. While the fraction of events falling into the region $\pi/2 < |\Delta\phi(j_1,j_2)| \le \pi$ is maximized for $\cttilde\approx 0$ (see \cref{fig:ggHj_2j}, left), the fraction of events in the complementary region of $-\pi/2 \le \Delta\phi(j_1,j_2) \le \pi/2$ reaches its maximal value for $\ct\approx 0$. The overall variation is, however, in the sub-percent level and therefore not resolvable within the current experimental and theoretical precision.


\section{Experimental input}
\label{app:expinput}

\cref{tab:expRun2ATLAS,tab:expRun2CMS} list the LHC Run-2 Higgs signal measurements from ATLAS and CMS, respectively, which are implemented in \texttt{HiggsSignals-2.5.0} and included in our fit. In addition, we include the combined LHC Run-1 results from ATLAS and CMS~\cite{Khachatryan:2016vau} in the fit via \texttt{HiggsSignals}.

\begin{table}
\centering
\footnotesize
\renewcommand{\arraystretch}{1.2}
\begin{tabular}{l c c c c}
\toprule
  Channel  & Luminosity [$\text{fb}^{-1}$]&  \multicolumn{2}{c}{Signal strength $\mu$}  & Reference \\
\midrule
 VBF, $H\to b\bar{b}$             & $30.6$  & \multicolumn{2}{c}{$3.0\substack{+1.7\\-1.8}$ }     & \cite{Aaboud:2018gay}  \\
 $t\bar{t}H$, $H\to b\bar{b}$ ($1\ell$) \ttHmark     & $36.1$  & \multicolumn{2}{c}{$0.67\substack{+0.61\\-0.69}$ }  & \cite{Aaboud:2017rss}  \\
 $t\bar{t}H$, $H\to b\bar{b}$ ($2\ell$) \ttHmark     & $36.1$  & \multicolumn{2}{c}{$0.11\substack{+1.36\\-1.41}$ }  & \cite{Aaboud:2017rss}  \\
 $t\bar{t}H$, multilepton ($2\ell ss$) \ttHmark & $79.9$ &\multicolumn{2}{c}{$0.38\substack{+0.57\\-0.54}$} & \cite{ATLAS:2019nvo} \\
 $t\bar{t}H$, multilepton ($3\ell$) \ttHmark & $79.9$ &\multicolumn{2}{c}{$0.93\substack{+0.58\\-0.52}$} & \cite{ATLAS:2019nvo} \\
 $t\bar{t}H$, multilepton ($4\ell$) \ttHmark & $79.9$ &\multicolumn{2}{c}{$0.52\substack{+0.93\\-0.72}$} & \cite{ATLAS:2019nvo} \\
 $t\bar{t}H$, multilepton ($1\ell+2\tau_h$) \ttHmark & $79.9$ &\multicolumn{2}{c}{$0.30\substack{+1.01\\-0.90}$} & \cite{ATLAS:2019nvo} \\
 $t\bar{t}H$, multilepton ($2\ell+1\tau_h$) \ttHmark & $79.9$ &\multicolumn{2}{c}{$0.49\substack{+0.94\\-0.82}$} & \cite{ATLAS:2019nvo} \\
 $t\bar{t}H$, multilepton ($3\ell+1\tau_h$) \ttHmark & $79.9$ &\multicolumn{2}{c}{$0.43\substack{+1.10\\-0.85}$} & \cite{ATLAS:2019nvo} \\
\midrule
    & &  $\sigma_\text{obs}$ [$\text{pb}$] & $\sigma_{\text{SM}}$ [$\text{pb}$]  &  \\
\midrule
 $gg\to H$, $H\to W^+W^-$ & $36.1$   &  $11.4\substack{+2.2\\-2.1}$  &  $10.4\pm 0.6$ & \cite{Aaboud:2018jqu} \\
 VBF, $H\to W^+W^-$       & $36.1$   &  $0.50\substack{+0.29\\-0.28}$  &  $0.81\pm 0.02$ & \cite{Aaboud:2018jqu} \\
 VBF, $H\to ZZ$ ($p_{T,H}$ high) & $139.0$ & $0.0005\substack{+0.0079\\ - 0.0048}$ & $0.00420 \pm 0.00018$ & \cite{Aad:2020mkp} \\
 VBF, $H\to ZZ$ ($p_{T,H}$ low) & $139.0$ & $0.150\substack{+0.064 \\-0.052}$ & $ 0.1076\substack{+0.034\\-0.0035}$ & \cite{Aad:2020mkp} \\
 $V(\text{had})H$, $H\to ZZ$ & $139.0$ & $0.021\pm 0.035$ & $0.0138\substack{+0.0004\\-0.0006}$ & \cite{Aad:2020mkp} \\
 $V(\text{lep})H$, $H\to ZZ$ & $139.0$ & $0.22\substack{+0.028\\-0.017}$ & $0.0165\substack{+0.0010\\-0.0016}$ &
 \cite{Aad:2020mkp} \\
  $gg\to H$, $H\to ZZ$ ($p_{T,H}$ high) & $139.0$ & $0.038\substack{+0.021\\ -0.016}$ & $0.015\pm 0.004$ &  \cite{Aad:2020mkp} \\
  $gg\to H$, $H\to ZZ$ ($0j,~p_{T,H}$ high) & $139.0$ & $0.63\pm 0.11$ & $0.55\pm 0.04$ &  \cite{Aad:2020mkp} \\
  $gg\to H$, $H\to ZZ$ ($0j,~p_{T,H}$ low) & $139.0$ & $0.170\pm 0.055$ & $0.176\pm 0.025$ &  \cite{Aad:2020mkp} \\
  $gg\to H$, $H\to ZZ$ ($1j,~p_{T,H}$ high) & $139.0$ & $0.009\substack{+0.016\\ -0.012}$ & $0.020\pm 0.004$ &  \cite{Aad:2020mkp} \\
  $gg\to H$, $H\to ZZ$ ($1j,~p_{T,H}$ low) & $139.0$ & $0.05\pm 0.08$ & $0.172\pm 0.025$ &  \cite{Aad:2020mkp} \\
  $gg\to H$, $H\to ZZ$ ($1j,~p_{T,H}$ med.) & $139.0$ & $0.170\pm 0.050$ & $0.119\pm 0.018$ &  \cite{Aad:2020mkp} \\
  $gg\to H$, $H\to ZZ$ ($2j$) & $139.0$ & $0.040\pm 0.075$ & $0.127\pm 0.027$ &  \cite{Aad:2020mkp} \\
  $t\bar{t}H$, $H\to ZZ$ \ttHmark & $139.0$ & $0.025\substack{+0.026\\ -0.017}$ & $0.0154\substack{+0.0010\\-0.0013}$ &  \cite{Aad:2020mkp} \\
  $gg\to H$, $H\to \gamma\gamma$ ($0j$) & $139.0$ & $0.039\pm 0.006$ & $0.0382 \substack{+0.0019 \\ -0.0018}$ & \cite{ATLAS:2019jst} \\
  $gg\to H$, $H\to \gamma\gamma$ ($1j$) & $139.0$ & $0.0162\substack{+0.0031\\-0.0022}$ & $0.0194 \substack{+0.0018 \\ -0.0019}$ & \cite{ATLAS:2019jst} \\
  $gg\to H$, $H\to \gamma\gamma$ ($2j$, $\Delta\Phi_{jj} \in [-\pi,-\tfrac{\pi}{2}]$) & $139.0$ & $0.0023\pm 0.0007$ & $0.0024 \pm 0.0002$ & \cite{ATLAS:2019jst} \\
  $gg\to H$, $H\to \gamma\gamma$ ($2j$, $\Delta\Phi_{jj} \in [-\tfrac{\pi}{2},0]$) & $139.0$ & $0.0011\pm 0.0004$ & $0.0020 \pm 0.0002$ & \cite{ATLAS:2019jst} \\
  $gg\to H$, $H\to \gamma\gamma$ ($2j$, $\Delta\Phi_{jj} \in [0, \tfrac{\pi}{2}]$) & $139.0$ & $0.0014\pm 0.0004$ & $0.0020 \pm 0.0002$ & \cite{ATLAS:2019jst} \\
  $gg\to H$, $H\to \gamma\gamma$ ($2j$, $\Delta\Phi_{jj} \in [\tfrac{\pi}{2},\pi]$) & $139.0$ & $0.0021\pm 0.0007$ & $0.0024 \pm 0.0002$ & \cite{ATLAS:2019jst} \\
  $t\bar{t}H$, $H\to \gamma\gamma$ \ttHmark & $139.0$ & $0.00159\substack{+0.00043\\ -0.00039}$ & $0.00115 \substack{+0.00009 \\ -0.00012}$ & \cite{ATLAS-CONF-2019-004} \\
  VBF, $H\to \tau^+\tau^-$ &  $36.1$         & $0.28\substack{+0.14 \\ -0.13}$ & $0.237\pm 0.006$ & \cite{Aaboud:2018pen} \\
  $gg\to H$, $H\to \tau^+\tau^-$ & $36.1$   & $3.10\substack{+1.90 \\ -1.60}$ & $3.05\pm 0.13$ & \cite{Aaboud:2018pen} \\
  $WH$, $H\to W^+W^-$ & $36.1$   & $0.67\substack{+0.36 \\ -0.30}$ & $0.293\pm 0.007$ & \cite{Aad:2019lpq} \\
  $ZH$, $H\to W^+W^-$ & $36.1$   & $0.54\substack{+0.34 \\ -0.25}$ & $0.189\pm 0.007$ & \cite{Aad:2019lpq} \\
  $WH$, $H\to b\bar{b}$ ($p_{T,V} \in [150, 250]~\mathrm{GeV}$) & $139.0$   & $0.019\pm 0.0121$ & $0.0240\pm 0.0011$ & \cite{Aad:2020jym} \\
  $WH$, $H\to b\bar{b}$ ($p_{T,V} \ge 250~\mathrm{GeV}$) & $139.0$   & $0.0072\pm 0.0022$ & $0.0071\pm 0.0030$ & \cite{Aad:2020jym} \\
  $ZH$, $H\to b\bar{b}$ ($p_{T,V} \in [75, 150]~\mathrm{GeV}$) & $139.0$   & $0.0425\pm 0.0359$ & $0.0506\pm 0.0041$ & \cite{Aad:2020jym} \\
  $ZH$, $H\to b\bar{b}$ ($p_{T,V} \in [150, 250]~\mathrm{GeV}$) & $139.0$   & $0.0205\pm 0.0062$ & $0.0188\pm 0.0024$ & \cite{Aad:2020jym} \\
  $ZH$, $H\to b\bar{b}$ ($p_{T,V} \ge 250~\mathrm{GeV}$) & $139.0$   & $0.054\pm 0.0017$ & $0.0049\pm 0.0005$ & \cite{Aad:2020jym} \\
\bottomrule
\end{tabular}
\caption{ATLAS Higgs rate measurements from LHC Run-2 included in the fit via \texttt{HiggsSignals}. Measurements from dedicated $t\bar{t}H+tH$ analyses are marked with \ttHmark\ and are not included in a specific fit that we performed for comparison (see \cref{sec:res_ttH}).}
\label{tab:expRun2ATLAS}
\end{table}

\begin{table}
\centering
\footnotesize
\def\arraystretch{1.2}
\begin{tabular}{l c c  c c}
\toprule
  Channel  & Luminosity [$\text{fb}^{-1}$]&  \multicolumn{2}{c}{Signal strength $\mu$}  & Reference \\
\midrule
 $pp\to H$, $H\to \mu^+\mu^-$     & $35.9$  & \multicolumn{2}{c}{$1.0\substack{+1.1\\-1.1}$ }     & \cite{Sirunyan:2018hbu}\\
 $WH$, $H\to b\bar{b}$            & $35.9$  & \multicolumn{2}{c}{$1.7\substack{+0.7\\-0.7}$ }     & \cite{Sirunyan:2017elk}\\
 $ZH$, $H\to b\bar{b}$            & $35.9$  & \multicolumn{2}{c}{$0.9\substack{+0.5\\-0.5}$ }     & \cite{Sirunyan:2017elk}\\
 $pp\to H$ (boosted), $H\to b\bar{b}$ & $35.9$ & \multicolumn{2}{c}{$2.3\substack{+1.8\\-1.6}$}   & \cite{Sirunyan:2017dgc}\\
$t\bar{t}H$, $H\to b\bar{b}$ ($1\ell$) \ttHmark    & $35.9 \oplus 41.5$  & \multicolumn{2}{c}{$0.84\substack{+0.52\\-0.50} \oplus 1.84\substack{+0.62 \\-0.56}$}   & \cite{Sirunyan:2018mvw,CMS:2019lcn}\\
$t\bar{t}H$, $H\to b\bar{b}$ ($2\ell$) \ttHmark    & $35.9 \oplus 41.5$  & \multicolumn{2}{c}{$-0.24\substack{+1.21\\-1.12} \oplus 1.62\substack{+0.90 \\-0.85}$}   & \cite{Sirunyan:2018mvw,CMS:2019lcn}\\
$t\bar{t}H$, $H\to b\bar{b}$ (hadr.) \ttHmark    & $41.5$  & \multicolumn{2}{c}{$-1.69\pm 1.43$}   & \cite{CMS:2019lcn}\\
 $t\bar{t}H$, multilepton ($1\ell+2\tau_h$) \ttHmark & $35.9 \oplus 41.5$ & \multicolumn{2}{c}{$-1.52\substack{+1.76\\-1.72} \oplus 1.4\substack{+1.24\\-1.14}$} & \cite{Sirunyan:2018shy,CMS:2018dmv}\\
 $t\bar{t}H$, multilepton ($2\ell ss+1\tau_h$) \ttHmark & $35.9 \oplus 41.5$ & \multicolumn{2}{c}{$0.94\substack{+0.80\\-0.67} \oplus 1.13\substack{+1.03\\-1.11}$} & \cite{Sirunyan:2018shy,CMS:2018dmv}\\
 $t\bar{t}H$, multilepton ($2\ell ss$) \ttHmark & $35.9 \oplus 41.5$ & \multicolumn{2}{c}{$1.61\substack{+0.58\\-0.51} \oplus 0.87\substack{+0.62\\-0.55}$} & \cite{Sirunyan:2018shy,CMS:2018dmv}\\
 $t\bar{t}H$, multilepton ($3\ell+1\tau_h$) \ttHmark & $35.9 \oplus 41.5$ & \multicolumn{2}{c}{$1.34\substack{+1.42\\-1.07} \oplus -0.96\substack{+1.96\\-1.33}$} & \cite{Sirunyan:2018shy,CMS:2018dmv}\\
 $t\bar{t}H$, multilepton ($3\ell$) \ttHmark & $35.9 \oplus 41.5$ & \multicolumn{2}{c}{$0.82\substack{+0.77\\-0.71} \oplus 0.29\substack{+0.82\\-0.62}$} & \cite{Sirunyan:2018shy,CMS:2018dmv}\\
 $t\bar{t}H$, multilepton ($4\ell$) \ttHmark & $35.9 \oplus 41.5$ & \multicolumn{2}{c}{$0.57\substack{+2.29\\-1.57} \oplus 0.99\substack{+3.31\\-1.69}$} & \cite{Sirunyan:2018shy,CMS:2018dmv}\\
\midrule
    &  &  $\sigma_\text{obs}$ [$\text{pb}$] & $\sigma_{\text{SM}}$ [$\text{pb}$]  &  \\
\midrule
 $gg\to H$, $H\to W^+W^-$ ($0j$) & $137.0$ & $0.0423\substack{+0.0063\\ -0.0059}$ & $0.0457\substack{+0.0029\\ -0.0018}$ & \cite{CMS:2019kqw} \\
 $gg\to H$, $H\to W^+W^-$ ($1j$) & $137.0$ & $0.0240\substack{+0.0057\\ -0.0051}$ & $0.0217\substack{+0.0023\\ -0.0022}$ & \cite{CMS:2019kqw} \\
 $gg\to H$, $H\to W^+W^-$ ($2j$) & $137.0$ & $0.0151\substack{+0.0051\\ -0.0046}$ & $0.0100\substack{+0.0020\\ -0.0011}$ & \cite{CMS:2019kqw} \\
 $gg\to H$, $H\to W^+W^-$ ($3j$) & $137.0$ & $0.0050\substack{+0.0045\\ -0.0042}$ & $0.0033\substack{+0.0002\\ -0.0004}$ & \cite{CMS:2019kqw} \\
 $gg\to H$, $H\to W^+W^-$ ($4j$) & $137.0$ & $0.0064\substack{+0.0039\\ -0.0034}$ & $0.0018\substack{+0.0001\\ -0.0002}$ & \cite{CMS:2019kqw} \\
 VBF, $H\to ZZ$                  & $137.1$ & $0.279\substack{+0.211\\-0.162}$     & $0.450\pm 0.010$ & \cite{CMS:2019chr}\\
 $gg/b\bar{b}\to H$, $H\to ZZ$   & $137.1$ & $5.328\pm 0.611$     & $5.550\substack{+0.600\\-0.650}$ & \cite{CMS:2019chr}\\
 $VH$, $H\to ZZ$                 & $137.1$ & $0.305\substack{+0.243\\-0.194}$     & $0.270\pm 0.010$ & \cite{CMS:2019chr}\\
 $t\bar{t}H,tH$, $H\to ZZ$ \ttHmark                 & $137.1$ & $0.0078\pm 0.0552$     & $0.060\substack{+0.011\\-0.012}$ & \cite{CMS:2019chr}\\
 $gg\to H$, $H\to \gamma\gamma$ ($0j$) & $77.4$ & $0.072 \pm0.0122$ & $0.0610\substack{+0.0037\\-0.0031}$ & \cite{CMS:1900lgv} \\
 $gg\to H$, $H\to \gamma\gamma$ ($1j,~p_{T,H}$ high) & $77.4$ & $0.0029\substack{+0.0017\\-0.0012}$ & $0.0017\pm 0.0002$ &\cite{CMS:1900lgv} \\
 $gg\to H$, $H\to \gamma\gamma$ ($1j,~p_{T,H}$ low) & $77.4$ & $0.021\substack{+0.0090\\-0.0075}$ & $0.015\pm 0.0015$& \cite{CMS:1900lgv} \\
 $gg\to H$, $H\to \gamma\gamma$ ($1j,~p_{T,H}$ med.) & $77.4$ & $0.0076\pm 0.0040$ & $0.010\pm 0.001$& \cite{CMS:1900lgv} \\
 $gg\to H$, $H\to \gamma\gamma$ ($2j$) & $77.4$ & $0.0084\substack{+0.0066\\ -0.0055}$ & $0.011\pm 0.002$& \cite{CMS:1900lgv} \\
 $gg\to H$, $H\to \gamma\gamma$ (BSM) & $77.4$ & $0.0029\pm 0.00104$ & $0.0013\pm 0.0003$ &\cite{CMS:1900lgv} \\
 VBF, $H\to \gamma\gamma$ & $77.4$ & $0.0091\substack{+0.0044\\-0.0033}$ & $0.0011\pm 0.002$ &\cite{CMS:1900lgv} \\
 $t\bar{t}H$, $H\to \gamma\gamma$ \ttHmark & $137.0$  & $0.00155\substack{+0.00034\\-0.00032}$  & $0.00113\substack{+0.00008\\-0.00011}$     & \cite{Sirunyan:2020sum} \\
 $V(\text{had})H$, $H\to \tau^+\tau^-$ & $77.4$ & $-0.0433\substack{+0.057\\-0.054}$ & $0.037\pm 0.001$ &\cite{CMS:2019pyn} \\
 VBF, $H\to \tau^+\tau^-$ & $77.4$ & $0.114\substack{+0.034\\-0.033}$ & $0.114\pm 0.009$ &\cite{CMS:2019pyn} \\
 $gg\to H$, $H\to \tau^+\tau^-$ ($0j$) & $77.4$ & $-0.680\substack{+1.292\\-1.275}$ & $1.70\pm 0.10$ &\cite{CMS:2019pyn} \\
 $gg\to H$, $H\to \tau^+\tau^-$ ($1j,~p_{T,H}$ high) & $77.4$ & $0.108\substack{+0.071\\-0.061}$ & $0.060\pm 0.010$ &\cite{CMS:2019pyn} \\
 $gg\to H$, $H\to \tau^+\tau^-$ ($1j,~p_{T,H}$ low) & $77.4$ & $-0.139\substack{+0.562\\-0.570}$ & $0.410\pm 0.060$ &\cite{CMS:2019pyn} \\
 $gg\to H$, $H\to \tau^+\tau^-$ ($1j,~p_{T,H}$ med.) & $77.4$ & $0.353\substack{+0.437\\-0.420}$ & $0.280\pm 0.040$ &\cite{CMS:2019pyn} \\
 $gg\to H$, $H\to \tau^+\tau^-$ ($2j$) & $77.4$ & $0.0987\substack{+0.1911\\-0.1806}$ & $0.210\pm 0.050$& \cite{CMS:2019pyn} \\
 $gg\to H$, $H\to \tau^+\tau^-$ ($1j,~p_{T}^{j_1}> 200$ GeV) & $77.4$ & $0.0199\substack{+0.0145\\-0.0148}$ & $0.0141\pm 0.0004$ &\cite{CMS:2019pyn} \\
 $gg\to H$, $H\to \tau^+\tau^-$ (Rest) & $77.4$ & $-0.195\substack{+0.506\\-0.491}$ & $0.184\pm 0.005$ &\cite{CMS:2019pyn} \\
\bottomrule
\end{tabular}
\caption{CMS Higgs rate measurements from LHC Run-2 included in the fit via \texttt{HiggsSignals}. Measurements from dedicated $t\bar{t}H+tH$ analyses are marked with \ttHmark\ and are not included in a specific fit that we performed for comparison (see \cref{sec:res_ttH}).}
\label{tab:expRun2CMS}
\end{table}

These measurements are implemented in \texttt{HiggsSignals} along with correlation matrices and detailed  information about the composition of the signal in terms of the various relevant Higgs boson production and decay processes. If this information is not provided by the experiments, the signal is assumed to be composed of the relevant Higgs processes with equal acceptances, and only correlations of the luminosity uncertainty (within one experiment) and theoretical rate uncertainties are taken into account. For details, see Ref.~\cite{Bechtle:2013xfa}.

An important point for our study arises, however, in the implementation of measurements targeting Higgs production in association with top quarks. While current experimental studies mainly focus on $t\bar{t}H$ production, which is the dominant top associated production mode in the SM, these searches are often also sensitive to $tH$ and $tWH$ production. These processes are, due to their rather small rate in the SM, either neglected or assumed to be equal to the SM prediction when performing the $t\bar{t}H$ measurements, and no detailed information about their potential contribution to the signal is quoted.\footnote{Currently, the only measurement quoting the $tH$ contribution to the signal is the $t\bar{t}H, H\to \gamma\gamma$ analysis by CMS, see Tab.~1 of Ref.~\cite{Sirunyan:2020sum}.}

In the current \texttt{HiggsSignals} implementation, we treat $t\bar{t}H$ measurements without public information about the $tH$ and $tWH$ contribution as follows: (\emph{i}) the signal acceptances of $t\bar{t}H$ and $tWH$ are assumed to be identical; (\emph{ii}) the $tH$ contribution is neglected in the $2$-lepton category of $t\bar{t}H$, $H\to b\bar{b}$ analyses, and in all other observables the $tH$ signal acceptance is assumed to be identical to the $t\bar{t}H$ acceptance. The first point (\emph{i}) is certainly a reasonable approximation as the final states of $tWH$ and $t\bar{t}H$ are experimentally very similar. In contrast, $tH$ production can only lead to one lepton (at parton level), which motivates our choice in the second point (\emph{ii}). However, the approximation of equal acceptances of $tH$ and $t\bar{t}H$ in the remaining measurements  is clearly a simplification. Given the multivariate nature of most of the recent data analyses, it is difficult to improve upon this.

\begin{figure}[ht]
    \centering
  \includegraphics[width=0.49\textwidth]{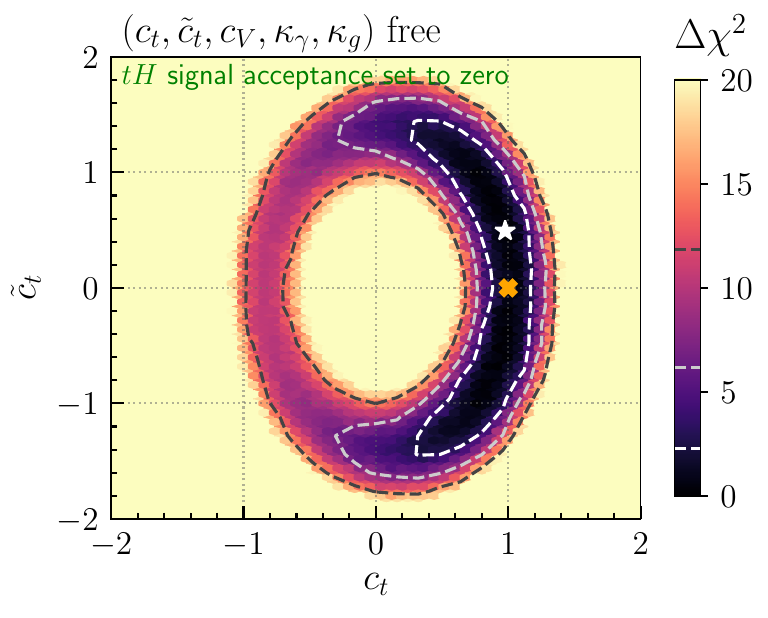}\hfill
  \includegraphics[width=0.49\textwidth]{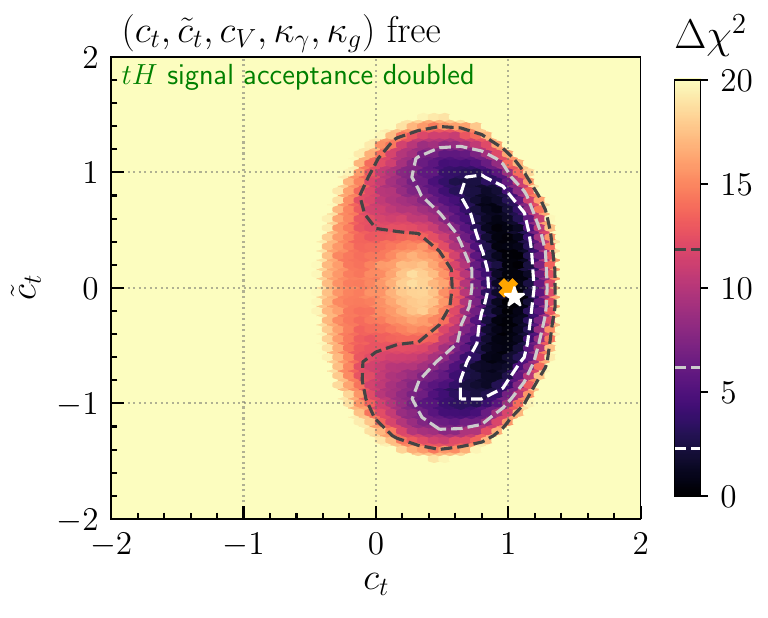}
  \caption{Fit results in the (\ct, \cttilde) parameter plane of the 5D parametrization for two variations of the assumed signal acceptance of $tH$ production in all included Higgs boson signal rate measurements. In the \emph{left panel} we completely neglected the $tH$ production,  in the \emph{right panel} we doubled the $tH$ signal acceptance. The corresponding result for the default $tH$ signal acceptances is shown in \cref{fig:ctcttilde} (\emph{bottom right panel}).}
  \label{fig:ctcttilde_tHvariation}
\end{figure}

In order to quantify the impact of this assumption on our fit and the associated uncertainty, we repeat the fit in the 5D parametrization -- (\ct, \cttilde, \cv, \kg, \kgamma) free -- for two variations of the $tH$  signal acceptance in \emph{all} implemented measurements. In the first variation, we set the $tH$ acceptance to zero, i.e.\ we  neglect entirely the contribution from $tH$ production, and in the second variation we double the $tH$ acceptance. One should keep in mind, though, that this variation is clearly very drastic, as in reality the $tH$ acceptance will certainly not be under- or over-estimated in the same way in all observables. The results in the (\ct, \cttilde) parameter plane for these two variations are shown in \cref{fig:ctcttilde_tHvariation}. These results should be compared with \cref{fig:ctcttilde} (bottom, right). The impact of the variation is largest in the parameter region with an enhanced $tH$ cross section, i.e.\ the negative \ct range and/or regions with large \cttilde.


\section{Additional information about the best-fit points}
\label{app:bfpoint}

The $F$-test is a possibility to quantify how much a model \textbf{B} with more free parameters improves over the description of the data compared to model \textbf{A} with less free parameters. If \textbf{B} contains \textbf{A}, as in extensions of the SM, \textbf{B} will always provide a better or equal fit to the data than \textbf{A}. The $F$-test weighs the higher complexity of \textbf{B} over \textbf{A} against the improvement in the fit. Here, model \textbf{A} is a more restricted model with number of parameters, $n_\text{par,\textbf{A}}$, smaller than the number of parameters of model \textbf{B}, $n_\text{par,\textbf{B}}$. Model~\textbf{A} is furthermore nested within model \textbf{B}, i.e., it can be obtained for a specific parameter choice of model~\textbf{B}. The test statistic $f$ is then calculated as
\begin{align}
f = \frac{\chi^2_\textbf{A}}{\nu_\textbf{A}} \cdot
\left(\frac{\chi^2_\textbf{B}}{\nu_\textbf{B}}\right)^{-1},
\end{align}
with the number of degrees of freedom $\nu_\textbf{A,B} = n_\text{obs} - n_\text{par,\textbf{A,B}}$, and
 $n_\text{obs}$ the number of measurements. The $\chi^2$ values refer to the minimal $\chi^2$ value found in the parameter space (i.e.~the best-fit value). In our case, we consider the SM as the restricted model \textbf{A}. As the SM is contained within all models we consider, the minimal $\chi^2$ in the models is smaller (or equal) to the $\chi^2$ in the SM.

 The cumulative probability $F$ quantifies the significance of the $\chi^2$ improvement found in the more general model, while accounting for the larger number of parameters. It is found by integrating the test statistic $f$ from zero to the $f$-value determined by the data via the fit. The null hypothesis, which is that model~\textbf{B} does not provide a significantly better fit to the data than model~\textbf{A} (i.e.~the SM), can be rejected, for instance, at the $68\%$ ($95\%$) confidence level if $F > 0.68~(0.95)$. In the last column of \Cref{tab:BFpoints} we give the calculated cumulative probability of this $F$-test for all considered model fits. We see that none of the considered models is favored over the SM even at the $68\%$~C.L. In fact, as all considered models feature $F < 0.50$, for the current data the relative simplicity of the SM can be considered as preferable over the minimally better fit of the SM extensions fitted here.

\begin{table}
\centering
\def\arraystretch{1.2}
\small
\begin{tabular}{l cccccc  c c}
\toprule
\multirow{2}{*}{Fit observables}  & \multicolumn{6}{c}{Best-fit parameter point} & \multirow{2}{*}{$\chi_\text{BF}^2/n_\text{obs}$} & $F$-test\\
\cmidrule{2-7}
 &  \ct & \cttilde & \cv & \kg  & \kgamma & \kggzh &   & (w.r.t.~SM) \\
\midrule
all   &  $0.991$   & $-0.009$    &  --  & --  &  -- & --   &  $84.32/106$ & $46\%$    \\
all   & $0.996 $   & $0.004$   & $0.997$ & --  &  -- & --   &  $84.30/106$     & $45\%$\\
all   &  $0.997$  & $-0.015$    & $0.989$ & --  & $1.001$   & --   &  $84.28/106$   & $43\%$ \\
all   &  $1.032$  & $-0.048$   & $0.991$   & $0.993$ & $0.994$  & --   & $84.08/106$   & $41\%$   \\
all, no shape~mod.   & $1.038$   & $0.065$   & $0.999$     & $0.989$    &  $1.001$   & --   &  $84.06/106$    & $41\%$ \\
no $t\bar{t}H$ obs.,  &  \multirow{2}{*}{$1.285$}   & \multirow{2}{*}{$0.667$}   & \multirow{2}{*}{$0.989$} & \multirow{2}{*}{$0.993$} &  \multirow{2}{*}{$0.979$}  & \multirow{2}{*}{--}   & \multirow{2}{*}{$61.55/77$} & \multirow{2}{*}{$46\%$}\\
 no shape~mod. & & & & & & & & \\
all, no shape~mod.   & $1.032$   & $0.075$   & $0.996$  & $0.991$    & $1.004$  & $0.948$  &  $84.01/106$  & $39\%$\\
no $t\bar{t}H$ obs.,   &  \multirow{2}{*}{$0.323$}  & \multirow{2}{*}{$-1.629$}   & \multirow{2}{*}{$0.998$} & \multirow{2}{*}{$0.990$}  & \multirow{2}{*}{$0.978$}    &  \multirow{2}{*}{$0.988$}  &  \multirow{2}{*}{$61.49/77$} & \multirow{2}{*}{$43\%$} \\
 no shape~mod. & & & & & & & & \\
\bottomrule
\end{tabular}
\caption{Results for the best-fit (BF) points for all performed fits: The first column indicates the experimental input for the fit, the middle columns give the parameter values for the free fit parameters (-- signifies fixed/derived parameters), and the two last columns give the total $\chi^2$ over the number of measurements, $n_\text{obs}$, and the $F$-test value  (see text for definition), using the SM as the reference model. The SM $\chi^2$ values are $84.45$ ($64.10$) for the full observable set (observable set without dedicated $t\bar{t}H$ analyses).}
\label{tab:BFpoints}
\end{table}



\newpage

\bibliographystyle{JHEP.bst}
\bibliography{bibliography}{}

\end{document}